\documentclass[a4paper,11pt]{article}
\usepackage{moreverb}
\usepackage[round]{natbib}
\usepackage{times,epsfig,makeidx,amsfonts,amsmath,dcolumn,enumerate,multirow,graphicx,amssymb,colortbl,bm,url}
\usepackage{algorithm}
\usepackage[]{algorithmic}
\usepackage{mathtools,upgreek}
\usepackage{afterpage,lscape}
\usepackage{paralist}
\usepackage{xr}
\usepackage[margin=1in]{geometry}
\usepackage[outdir=./]{epstopdf}

\newtheorem{remark}{Remark}

\begin{document}

\title{Objective priors for the number of degrees of freedom of a multivariate $t$ distribution and the $t$-copula}
\author{Cristiano Villa$^a$$^*$  and Francisco J. Rubio$^b$$^{**}$
\\ \small $^{a}$ School of Mathematics, Statistics and Actuarial Science, University of Kent, UK.
\\ \small $^{b}$ London School of Hygiene \& Tropical Medicine, London, WC1E 7HT, UK.
\\ \small $^{*}$ E-mail:  C.Villa-88@kent.ac.uk
\\ \small $^{**}$ E-mail:  Francisco.Rubio@lshtm.ac.uk}

\date{}
\maketitle
\begin{abstract}
An objective Bayesian approach to estimate the number of degrees of freedom $(\nu)$ for the multivariate $t$ distribution and for the $t$-copula, when the parameter is considered discrete, is proposed. Inference on this parameter has been problematic for the multivariate $t$ and, for the absence of any method, for the $t$-copula. An objective criterion based on loss functions which allows to overcome the issue of defining objective probabilities directly is employed. The support of the prior for $\nu$ is truncated, which derives from the property of both the multivariate $t$ and the $t$-copula of convergence to normality for a sufficiently large number of degrees of freedom. The performance of the priors is tested on simulated scenarios \footnote{The R codes and the replication material are available as a supplementary material of the electronic version of the paper} and on real data: daily logarithmic returns of IBM and of the Center for Research in Security Prices Database.
\end{abstract}


\section{Introduction}\label{sec:intro}
One way to model multivariate quantities is through a multivariate probability distribution and, due to its simplicity and appealing properties, the multivariate Normal distribution represents the most popular choice. However, due to the ``lightness'' of the tails, the Normal distribution does not properly represent the probability of occurrence of rare events. In other words, the multivariate Normal distribution is not the best choice to model data sets which contain outliers. An alternative is represented by the multivariate $t$ distribution, whose expression is presented in Section \ref{sec:preliminaries}; in fact, this distribution has a shape parameter (\textit{i.e.}~the number of degrees of freedom) that controls the tail behaviour allowing to capture heavier tails than those of the Normal distribution. The appropriateness of the $t$ distribution (univariate or multivariate) to deal with outliers has been thoroughly discussed in the literature \citep{West84,Langeetal89}, and it has been applied in numerous contexts, such as medicine \citep{L94}, finance and biology \citep{FS99}, portfolio optimisation \citep{K04}, financial engineering \citep{R11}, among many others.

An alternative method for extending a distribution to the multivariate case, and model a set of variables, consists of using a copula distribution \citep{N07}. The idea is to use a multivariate probability distribution (\textit{i.e.}~the \emph{copula}), whose marginals are uniform densities on $[0,1]$, to represent the dependence between the variables. The $t$-copula \citep{DM05}, which is formally presented in Section \ref{sec:preliminaries}, represents a popular choice in applied statistics as, in comparison to the Normal copula for example, it allows for capturing a wider variety of tail dependencies between the corresponding marginal distributions \citep{N09}. The use of copulas has attracted great attention in financial applications \citep{G09}, where the tail dependence is a common feature of many quantities, such as stock returns \citep{Hatetal04}.

Whether it is the case of the multivariate $t$ or of the $t$-copula, inference about the number of degrees of freedom represents a key task, as the parameter governs the ``heaviness'' of the tails. The Bayesian framework requires a prior distribution to be assigned to the parameter, representing the initial uncertainty about its true value. In the case sufficient information is available, the prior can be elicited by translating the information into a probability distribution. However, there are several cases where prior information is not sufficient (or it is deliberately not considered), nevertheless the Bayesian approach is still pursued. The collection of methodologies and techniques to define prior distributions under conditions of minimal prior information, including their derivation through formal rules \citep{WassKass1996}, goes under the name of \emph{Objective Bayes}, see \cite{Berger2006}. It is in the above sense that the term ``objective'' or ``objective prior'' has to be interpreted throughout this work.

In the univariate scenario, several prior distributions have been proposed for the degrees of freedom parameter of the Student-$t$ distribution. In particular, \cite{L94} presents the expression for the Jeffreys prior (further studied in \citealp{F08}) as well as other heuristic priors; \cite{JS10} proposed a proper prior with the same tail behaviour as that of the Jeffreys prior; while \cite{RS15} introduce a noninformative prior based on a measure of kurtosis. Of particular interest for this work is the prior introduced in \cite{VW14}, as the prior for the number of degrees of freedom we discuss here is based on the result proposed by the authors. \cite{VW14} discuss a discrete prior distribution which is truncated from above. The general idea is to assign a \emph{worth} to each parameter value by objectively measuring the loss in information in removing the parameter value when it is the true one. More details about the method are discussed in Section \ref{sec:ObjPriors}.

In the multivariate case, little attention has been paid to the study of priors for the degrees of freedom. To the best of our knowledge, \cite{L94} represents the only reference addressing this problem. \cite{L94} presents the expression for the Jeffreys prior of the degrees of freedom, and briefly discusses some heuristic prior choices. Although $t$-copula models have been implemented in a Bayesian framework, the choice of the prior for the degrees of freedom has been mainly done from an informal perspective, such as the use of uniform priors on a bounded interval \citep{S12}.

In this paper, we address the problem of estimating the number of degrees of freedom of the multivariate $t$ distribution and of the $t$-copula. In particular, we approach the task by considering the Bayesian framework in the presence of minimal prior information. In Section \ref{sec:preliminaries}, we describe the multivariate $t$ distribution and the $t$-copula. In Section \ref{sec:priors}, we present the proposed priors and introduce weakly informative priors for the remaining parameters. In Section \ref{sec:simulation}, we present a thorough simulation study where we illustrate the frequentist properties of the proposed priors. In Section \ref{sec:applications}, we present some financial applications of the proposed Bayesian models using real data. Finally, Section \ref{sec:discussion} contains some points for discussion and final remarks.

\section{The multivariate $t$ distribution and the $t$-copula}\label{sec:preliminaries}
The $d$-variate $t$ probability density function with $\nu>0$ degrees of freedom (see \citealp{L94} and \citealp{K04} for an extensive review of this model) is given by
\begin{eqnarray}\label{PDFMVT}
f_d({\bf x}\mid \bm{\mu},\bm{\Sigma},\nu) &=& \dfrac{\Gamma\left(\dfrac{\nu+d}{2}\right)}{\Gamma\left(\dfrac{\nu}{2}\right)\sqrt{(\pi\nu)^d\vert\bm{\Sigma}\vert}} \left(1+\dfrac{({\bf x}-\bm{\mu})^{\top}\bm{\Sigma}^{-1}({\bf x}-\bm{\mu})}{\nu}\right)^{-\frac{\nu+d}{2}},
\end{eqnarray}
where ${\bf x}\in\mathbb{R}^d$, $\bm{\mu}\in\mathbb{R}^d$ is the location (vector) parameter and $\bm{\Sigma}\in\mathbb{R}^{d\times d}$ is the positive definite scale matrix. Similarly to the univariate case, the parameter $\nu$ controls the heaviness of the tails of the density, with particular cases of $\nu=1$, where the distribution coincides with a multivariate Cauchy density, and of $\nu\rightarrow\infty$, where the distribution converges to a multivariate Normal density. As is discussed in Section \ref{sec:ObjPriors}, the convergence property of the multivariate $t$ is exploited to truncate the prior on $\nu$. In fact, for a sufficiently large value of the number of degrees of freedom, the difference (in terms of practically any distance between probability measures) between a multivariate $t$ and a multivariate Normal will be sufficiently small to consider all the $t$ densities as virtually the same.

A copula, say $C$, is a distribution function in dimension $d$ defined over the support $[0,1]^d$ with uniformly distributed marginals. As per Sklar's theorem, we can write a multivariate distribution function $F$ with marginals $F_1,\ldots,F_d$ as
\begin{eqnarray*}
F(x_1,\ldots,x_d) = C\left(F_1(x_1),\ldots,F_d(x_d)\right),
\end{eqnarray*}
for some copula $C$. This idea is often used to construct multivariate distributions by \textit{joining} any set of univariate distribution functions by means of a copula $C$ \citep{N07}. In this paper, we focus on the case where $C$ is the $t$-copula and the marginal distributions, $F_1,\dots,F_d$, are given by univariate $t$ densities (although our results apply to any marginal distributional assumptions). The $t$-copula is defined as (see \citealp{DM05})
\begin{eqnarray}\label{tcopcdf}
C_d^t({\bf u}\mid {\bf R},\nu) = \int_{-\infty}^{t^{-1}_{\nu}(u_1)}\cdots\int_{-\infty}^{t^{-1}_{\nu}(u_d)} \dfrac{\Gamma\left(\dfrac{\nu+d}{2}\right)}{\Gamma\left(\dfrac{\nu}{2}\right)\sqrt{(\pi\nu)^d\vert{\bf R}\vert}} \left(1+\dfrac{{\bf x}^{\top}{\bf R}^{-1}{\bf x}}{\nu}\right)^{-\frac{\nu+d}{2}} d{\bf x},
\end{eqnarray}
where ${\bf u}=(u_1,\dots,u_d)\in[0,1]^d$, ${\bf R}$ is a correlation matrix, and $t^{-1}_{\nu}$ denotes the quantile function associated to a Student-$t$ variate with $\nu>0$ degrees of freedom. The corresponding density function is given by \citep{DM05}
\begin{eqnarray}\label{tcoppdf}
c_d^t({\bf u}\mid {\bf R},\nu) = \dfrac{f_d(t^{-1}_{\nu}(u_1),\dots,t^{-1}_{\nu}(u_d)\vert {\bf 0},{\bf R},\nu)}{\prod_{j=1}^d f_1(t^{-1}_{\nu}(u_j)\vert \nu)}.
\end{eqnarray}
\begin{remark}\label{remark1}
	Analogously to the multivariate $t$ distribution, the $t$-copula converges pointwise to the Normal copula for increasing values of the number of degrees of freedom. That is, $C_{\nu,d}^t\rightarrow C_d^N$ for $\nu\rightarrow\infty$.
\end{remark}

\section{Prior distributions and inference}\label{sec:priors}
The inference on the parameters of the distributions, that is the multivariate $t$ and the $t$-copula, is performed with a Bayesian approach. Thus, for the multivariate $t$ we adopt the prior structure
\begin{eqnarray}\label{overallmultitprior}
\pi(\bm{\mu},\bm{\Sigma},\nu) = \pi(\nu \mid \bm{\mu},\bm{\Sigma})\pi(\bm{\mu},\bm{\Sigma}),
\end{eqnarray}
while for the $t$-copula, we have
\begin{equation}\label{overallcopulaprior}
\pi(\nu,\bm{R}) = \pi(\nu \mid \bm{R})\pi(\bm{R}).
\end{equation}
As the aim of this paper is to outline an objective approach, we work under the assumptions that little or no information about any parameter is known. Hence, priors $\pi(\bm{\mu},\bm{\Sigma})$ and $\pi(\bm{R})$ are chosen to be minimally informative. The prior on $\nu$ is considered to be dependent on the other parameters in general. As we will see in the following sections: (i) for the multivariate $t$ distribution, the proposed objective prior for $\nu$ is independent of $(\mu,\Sigma)$, and (ii) for the $t$-copula, the proposed objective prior for $\nu$ depends on the correlation matrix $\bm{R}$ (although we show that, for practical purposes, the dependence is negligible).

\subsection{Objective prior for $\nu$}\label{sec:ObjPriors}
In this section, we present the proposed objective priors for the number of degrees of freedom for the multivariate $t$ and the $t$-copula. As mentioned in Section \ref{sec:intro}, the literature related to the above problem is scarce. In particular, the here proposed objective prior for the $t$-copula case is, to the best of our knowledge, the sole available. For what it concerns the multivariate $t$, there are fundamentally three options \citep{L94}. \cite{A67} proposed a prior of the form $\pi(\nu)\propto(\nu+1)^{-3/2}$, for $\nu\geq1$. Jeffreys prior, obtained by applying the Jeffreys rule \citep{J57}, has the following form
$$\pi(\nu) \propto \left\{\psi\left(\frac{\nu}{2}\right)-\psi\left(\frac{\nu+d}{2}\right)-\frac{2d(\nu+d+4)}{\nu(\nu+d)(\nu+d+2)}\right\}^{1/2},$$
where $d$ is the dimension of the multivariate density, $\psi(x)=d^2/\{d^2\log\Gamma(x)\}$ is the trigamma function, and $\Gamma(\cdot)$ is the Gamma function. Finally, the third objective prior we consider is discussed in \cite{RR77}, and it has the form $\pi(\nu)\propto\nu^{-2}$, for $\nu\geq1$. For the simulation study presented in Section \ref{sec:simulation} we will compare the loss-based prior we propose in this paper with the above three options.

The principle to derive the objective prior for the number of degrees of freedom is the same for both the multivariate $t$ and the $t$-copula. In particular, we will apply the criterion based on loss in information introduced in \cite{VW15}.

We make two important assumptions about the parameter space for the number of degrees of freedom $\nu$. First, $\nu$ can only take positive integer values and, second, the parameter space is truncated at a value $\nu_{\max}$, which typically is 30. The first assumption originates from the fact that it is unlikely to have a sufficient number of observations which would allow to discern between (univariate or multivariate) $t$ distributions with a difference in the number of degrees of freedom smaller than one \citep{JPR04}. In support of the above assumption, we can see in Table \ref{table:KL} that the Kullback--Leibler divergence between distributions with discrete consecutive $\nu$, for dimensions $d=1,2,3$, gets small already for $\nu>5$.
The second assumption is based on the property of the $t$ density, for any dimension, to converge to a Normal density for $\nu\rightarrow+\infty$, of the same dimension. Although this is an approximation, and as such, devoid of an unequivocal value, it is common practice to consider the approximation as satisfactory for $\nu\approx30$; see, for example, \cite{C56}. The property applies to the $t$-copula as well \citep{Eetal01}.
As such, on the basis of the above two assumptions, we consider the parameter space for $\nu$ discrete and truncated at $\nu_{\max}=30$, where the model identified by $\nu_{\max}$ will represent the multivariate Normal distribution or the Normal (\textit{i.e.}~Gaussian) copula. A thorough discussion on the motivations leading to a discrete and truncated parameter space for the number of degrees of freedom can be found in \cite{VW14}; although the discussion made by the authors refers to the univariate $t$ density, the conclusions can be sensibly extended.

The key idea is to assign a \emph{worth} to each model identified by a value of $\nu$ by objectively measuring what is lost if that specific model is removed (\textit{i.e.}~not considered), and it is the true model. In Bayesian analysis it is well known that, if a model is misspecified, the posterior will asymptotically accumulate on the model which happens to be the most similar to the true one, where the similarity is ``measured'' through the Kullback--Leibler divergence \citep{Berk66}. In other words, the Kullback--Leibler divergence between the model identified by $\nu$ and the nearest one represents the loss in information one would incur in not considering that specific model (assumed to be the true one). Thus, $-D_{KL}\Big(f(\cdot\mid \nu)\mid \mid f(\cdot\mid \nu^\prime)\Big)$ represents the loss in keeping $\nu$ when $\nu'$ is the true value.

The prior distribution on the number of degrees of freedom is then constructed by linking the above loss to $\pi(\nu)$ by means of the self-information loss function. This particular kind of loss function measures the loss in information intrinsic to a probability statement. That is, if $P(A)$ is the probability that event $A$ is true, then $-\log P(A)$ is the self-information loss of $P(A)$. Therefore, if $f(\cdot\vert\nu)$ represents a sampling distribution with parameter value $\nu$, we equate the two measures of the loss in information at $\nu$, obtaining
\begin{eqnarray}\label{tprior}
-\log \pi(\nu) &=& -D_{KL}\Big(f(\cdot \mid \nu)\mid\mid f(\cdot\mid\nu^\prime)\Big), \notag \\
\pi(\nu) &\propto& \exp\left\{\min_{\nu^\prime\neq\nu}D_{KL}\Big(f(\cdot\mid\nu)\mid\mid f(\cdot\mid\nu^\prime)\Big)\right\}-1,
\end{eqnarray}
where the ``$-1$'' results from the process of bringing the two loss measures on the same scale (see \citealp{VW15}, equation (3), for a thorough discussion). In detail, let us set $u_1(\nu)=\log\pi(\nu)$ and let the minimum divergence from $\nu$ be represented by $u_2(\nu)$.  We want $u_1(\nu)$ and $u_2(\nu)$ to be matching utility functions; though as it stands $-\infty < u_1 \leq 0$ and $0 \leq u_2 < \infty$, and we want $u_1=-\infty$ when $u_2=0$. The scales are matched by taking exponential transformations; so $\exp(u_1)$ and $\exp(u_2)-1$ are on the same scale. Hence, we have
\begin{equation}\label{priorscale1}
e^{u_1(\nu)} = \pi(\nu) \propto e^{g\{u_2(\nu)\}}.
\end{equation}
By setting $g(u)=\log(e^u-1)$ in \eqref{priorscale1}, we derive \eqref{tprior}. The next two sections will detail the derivation of the prior for, respectively, the multivariate $t$ distribution and the $t$-copula.

It is important to note that the discretisation of the parameter space for $\nu$ can be made arbitrarily denser (\textit{e.g.}~including non-integer values). In Section \ref{sec:nonintegerprior} we show that the method to derive the loss-based prior is not affected by how the intervals between consecutive values of $\nu$ are chosen and that the prior mass on $\nu$ is determined by minimising the Kullback--Leibler divergence as in \eqref{tprior}. In particular, we show how the loss-based prior performs when the chose a support for $\nu$ of half-integers.

\subsubsection{Multivariate $t$}\label{sec:tprior}
Let $f_d({\bf x}\mid \bm{\mu},\bm{\Sigma},\nu)$ be a multivariate $t$, of dimension $d$, with location vector $\bm{\mu}$, scale matrix $\bm{\Sigma}$ and $\nu$ degrees of freedom. The aim is to define an objective prior for the parameter $\nu$. For simplicity in the notation, we will write $f_{d,\nu}=f_d({\bf x}\mid \bm{\mu},\bm{\Sigma},\nu)$, for $\nu=1,\ldots,\nu_{\max-1}$, and $f_{d,\nu_{\max}}=N_d({\bf x}\mid \bm{\mu},\bm{\Sigma})$, with
$$N_d({\bf x}\mid \bm{\mu},\bm{\Sigma}) = \frac{1}{\sqrt{(2\pi)^d|\bm{\Sigma}|}}\exp\left\{-\frac{1}{2}({\bf x}-\bm{\mu})^\top\bm{\Sigma}^{-1}({\bf x}-\bm{\mu})\right\},$$
where in this case $\bm{\mu}$ is the vector of means and $\bm{\Sigma}$ is the covariance matrix.
The prior for $\nu$ here discussed depends on the Kullback--Leibler divergence between two multivariate densities. In particular, for $\nu=1,\ldots,\nu_{\max-1}$, the prior is based on the Kullback--Leibler divergence between two multivariate $t$ densities which differ only in the number of degrees of freedom.
The divergence between two $d$-variate $t$ densities, $f_{d,\nu}$ and $f_{d,\nu^{\prime}}$, is given by
\begin{eqnarray}\label{DKLt}
&&D_{KL}(f_{d}(\cdot\mid \bm{\mu},\bm{\Sigma},\nu)\mid\mid f_{d,}(\cdot\mid \bm{\mu},\bm{\Sigma},\nu^{\prime})) = D_{KL}(f_{d}(\cdot\mid {\bf 0},{\bf I},\nu)\mid\mid f_{d}(\cdot\mid  {\bf 0},{\bf I},\nu^{\prime})) \nonumber\\
&=&\int_{{\mathbb R}^n} f_{d}({\bf x}\mid {\bf 0},{\bf I},\nu) \log \dfrac{f_{d}({\bf x}\mid {\bf 0},{\bf I},\nu)}{f_{d}({\bf x}\mid {\bf 0},{\bf I},\nu^{\prime})} \,d{\bf x}\nonumber\\
&=& \int_{{\mathbb R}^n} K(d,\nu)\left(1+\dfrac{{\bf x}^{\top}{\bf x}}{\nu}\right)^{-\frac{\nu+d}{2}} \log \dfrac{K(d,\nu)\left(1+\dfrac{{\bf x}^{\top}{\bf x}}{\nu}\right)^{-\frac{\nu+d}{2}}}{K(d,\nu^{\prime})\left(1+\dfrac{{\bf x}^{\top}{\bf x}}{\nu^{\prime}}\right)^{-\frac{\nu^{\prime}+d}{2}}} \,d{\bf x}\nonumber\\
&=& \log \dfrac{K(d,\nu)}{K(d,\nu^{\prime})} -\dfrac{\nu+d}{2}{\mathbb E}_{d,\nu}\left[ \log\left(1+\dfrac{{\bf x}^{\top}{\bf x}}{\nu}\right)\right] + \dfrac{\nu^{\prime}+d}{2}{\mathbb E}_{d,\nu}\left[ \log\left(1+\dfrac{{\bf x}^{\top}{\bf x}}{\nu^{\prime}}\right)\right],
\end{eqnarray}
where
$$K(d,\nu)=\dfrac{\Gamma\left(\dfrac{\nu+d}{2}\right)}{\Gamma\left(\dfrac{\nu}{2}\right)\sqrt{(\pi\nu)^d}},$$
and ${\mathbb E}_{d,\nu}$ represents the expected value with respect to $f_{d}(\cdot \mid {\bf 0},{\bf I},\nu)$. \cite{K04}, page 23, present a tractable expression for the first expectation in \eqref{DKLt}. More specificallly,
\begin{eqnarray*}
{\mathbb E}_{d,\nu}\left[ \log\left(1+\dfrac{{\bf x}^{\top}{\bf x}}{\nu}\right)\right]  =  \Psi\left(\dfrac{\nu+d}{2}\right) - \Psi\left(\dfrac{\nu}{2}\right),
\end{eqnarray*}
where $\Psi$ is the digamma function. For the second expectation in \eqref{DKLt}, we use Lemma 2 in \cite{Z99} to obtain the following expression after a change of variables in terms of spherical coordinates
\begin{eqnarray*}
{\mathbb E}_{d,\nu}\left[ \log\left(1+\dfrac{{\bf x}^{\top}{\bf x}}{\nu^{\prime}}\right)\right] = K(d,\nu) \dfrac{\pi^{\frac{d}{2}}}{\Gamma\left(\frac{d}{2}\right)}  \int_0^{\infty} \left(1+\dfrac{t}{\nu}\right)^{-\frac{\nu+d}{2}} t^{\frac{d}{2}-1}\log\left(1+\dfrac{t}{\nu^{\prime}}\right) dt,
\end{eqnarray*}
which only requires one-dimensional numerical integration, regardless of the dimension $d$. Table \ref{table:KL} shows the KL divergences for $\nu=1,\dots,30$. As one would expect, the minimum divergence from $f_{d,\nu}$ will either be $f_{d,\nu-1}$ or $f_{d,\nu+1}$, as this generates the smallest perturbation in the density yielding a relatively similar distribution.
\begin{table}[ht]
\small
\begin{center}
$\begin{array}{ccccccc}
\hline
& \multicolumn{2}{c}{d=1}  & \multicolumn{2}{c}{d=2}  & \multicolumn{2}{c}{d=3} \\
\hline
\nu & D_{KL}(f_{\nu}\| f_{\nu-1}) & D_{KL}(f_{\nu}\| f_{\nu+1}) & D_{KL}(f_{\nu}\| f_{\nu-1}) & D_{KL}(f_{\nu}\| f_{\nu+1})& D_{KL}(f_{\nu}\| f_{\nu-1}) & D_{KL}(f_{\nu}\| f_{\nu+1})\\
\hline
1 & \text{--} & 1.131\times 10^{-1} & \text{--} & 1.416\times 10^{-1} & \text{--} & 1.552\times 10^{-1} \\
  2 & 6.210\times 10^{-2} & 1.917\times 10^{-2} & 7.944\times 10^{-2} & 2.733\times 10^{-2} & 8.851\times 10^{-2} & 3.208\times 10^{-2} \\
  3 & 1.364\times 10^{-2} & 5.897\times 10^{-3} & 1.956\times 10^{-2} & 9.139\times 10^{-3} & 2.313\times 10^{-2} & 1.129\times 10^{-2} \\
  4 & 4.700\times 10^{-3} & 2.412\times 10^{-3} & 7.283\times 10^{-3} & 3.961\times 10^{-3} & 9.021\times 10^{-3} & 5.087\times 10^{-3} \\
  5 & 2.047\times 10^{-3} & 1.170\times 10^{-3} & 3.353\times 10^{-3} & 2.005\times 10^{-3} & 4.307\times 10^{-3} & 2.654\times 10^{-3} \\
  6 & 1.033\times 10^{-3} & 6.364\times 10^{-4} & 1.764\times 10^{-3} & 1.127\times 10^{-3} & 2.332\times 10^{-3} & 1.529\times 10^{-3} \\
  7 & 5.768\times 10^{-4} & 3.761\times 10^{-4} & 1.018\times 10^{-3} & 6.838\times 10^{-4} & 1.378\times 10^{-3} & 9.459\times 10^{-4} \\
  8 & 3.473\times 10^{-4} & 2.366\times 10^{-4} & 6.289\times 10^{-4} & 4.394\times 10^{-4} & 8.680\times 10^{-4} & 6.179\times 10^{-4} \\
  9 & 2.215\times 10^{-4} & 1.563\times 10^{-4} & 4.097\times 10^{-4} & 2.955\times 10^{-4} & 5.749\times 10^{-4} & 4.213\times 10^{-4} \\
  10 & 1.479\times 10^{-4} & 1.075\times 10^{-4} & 2.785\times 10^{-4} & 2.061\times 10^{-4} & 3.962\times 10^{-4} & 2.975\times 10^{-4} \\
  11 & 1.025\times 10^{-4} & 7.632\times 10^{-5} & 1.959\times 10^{-4} & 1.483\times 10^{-4} & 2.821\times 10^{-4} & 2.162\times 10^{-4} \\
  12 & 7.326\times 10^{-5} & 5.570\times 10^{-5} & 1.419\times 10^{-4} & 1.094\times 10^{-4} & 2.064\times 10^{-4} & 1.610\times 10^{-4} \\
  13 & 5.375\times 10^{-5} & 4.161\times 10^{-5} & 1.052\times 10^{-4} & 8.252\times 10^{-5} & 1.546\times 10^{-4} & 1.224\times 10^{-4} \\
  14 & 4.033\times 10^{-5} & 3.172\times 10^{-5} & 7.973\times 10^{-5} & 6.342\times 10^{-5} & 1.180\times 10^{-4} & 9.475\times 10^{-5} \\
  15 & 3.084\times 10^{-5} & 2.460\times 10^{-5} & 6.151\times 10^{-5} & 4.956\times 10^{-5} & 9.173\times 10^{-5} & 7.451\times 10^{-5} \\
  16 & 2.399\times 10^{-5} & 1.937\times 10^{-5} & 4.821\times 10^{-5} & 3.929\times 10^{-5} & 7.237\times 10^{-5} & 5.941\times 10^{-5} \\
  17 & 1.894\times 10^{-5} & 1.546\times 10^{-5} & 3.833\times 10^{-5} & 3.155\times 10^{-5} & 5.786\times 10^{-5} & 4.796\times 10^{-5} \\
  18 & 1.515\times 10^{-5} & 1.250\times 10^{-5} & 3.085\times 10^{-5} & 2.563\times 10^{-5} & 4.682\times 10^{-5} & 3.915\times 10^{-5} \\
  19 & 1.227\times 10^{-5} & 1.021\times 10^{-5} & 2.511\times 10^{-5} & 2.104\times 10^{-5} & 3.830\times 10^{-5} & 3.227\times 10^{-5} \\
  20 & 1.004\times 10^{-5} & 8.420\times 10^{-6} & 2.065\times 10^{-5} & 1.743\times 10^{-5} & 3.163\times 10^{-5} & 2.685\times 10^{-5} \\
  21 & 8.291\times 10^{-6} & 7.007\times 10^{-6} & 1.714\times 10^{-5} & 1.457\times 10^{-5} & 2.636\times 10^{-5} & 2.252\times 10^{-5} \\
  22 & 6.909\times 10^{-6} & 5.879\times 10^{-6} & 1.434\times 10^{-5} & 1.227\times 10^{-5} & 2.214\times 10^{-5} & 1.903\times 10^{-5} \\
  23 & 5.803\times 10^{-6} & 4.969\times 10^{-6} & 1.209\times 10^{-5} & 1.041\times 10^{-5} & 1.873\times 10^{-5} & 1.619\times 10^{-5} \\
  24 & 4.910\times 10^{-6} & 4.229\times 10^{-6} & 1.027\times 10^{-5} & 8.886\times 10^{-6} & 1.595\times 10^{-5} & 1.386\times 10^{-5} \\
  25 & 4.182\times 10^{-6} & 3.622\times 10^{-6} & 8.775\times 10^{-6} & 7.633\times 10^{-6} & 1.367\times 10^{-5} & 1.194\times 10^{-5} \\
  26 & 3.584\times 10^{-6} & 3.120\times 10^{-6} & 7.544\times 10^{-6} & 6.593\times 10^{-6} & 1.179\times 10^{-5} & 1.034\times 10^{-5} \\
  27 & 3.089\times 10^{-6} & 2.702\times 10^{-6} & 6.521\times 10^{-6} & 5.725\times 10^{-6} & 1.022\times 10^{-5} & 8.999\times 10^{-6} \\
  28 & 2.677\times 10^{-6} & 2.352\times 10^{-6} & 5.666\times 10^{-6} & 4.995\times 10^{-6} & 8.899\times 10^{-6} & 7.869\times 10^{-6} \\
  29 & 2.332\times 10^{-6} & 2.056\times 10^{-6} & 4.947\times 10^{-6} & 4.378\times 10^{-6} & 7.786\times 10^{-6} & 6.911\times 10^{-6} \\
  30 & 2.040\times 10^{-6} & 1.806\times 10^{-6} & 4.338\times 10^{-6} & 3.853\times 10^{-6} & 6.843\times 10^{-6} & 6.095\times 10^{-6} \\
\hline
\end{array}$
\caption{Comparison of the Kullback--Leibler divergence for contiguous $\nu$ values in dimension $d=1,2,3$. For simplicity in the notation, we have written $f_{d,\nu}$ as $f_\nu$.}
\label{table:KL}
\end{center}
\end{table}
For $\nu=\nu_{\max}$, the minimum Kullback--Leibler divergence is given by

\begin{eqnarray}\label{DKLN}
D_{KL}\Big(N_d({\bf x}\mid{\bf {\bf 0}},{\bf I})\mid\mid f_d({\bf x}\mid {\bf 0},{\bf I},\nu_{\max-1})\Big) &=& \int_{{\mathbb R}^n} N_d({\bf x}\mid{\bf {\bf 0}},{\bf I})\log\left\{\frac{N_d({\bf x}\mid{\bf {\bf 0}},{\bf I})}{f_d({\bf x}\mid {\bf 0},{\bf I},\nu_{\max-1})}\right\}\,d{\bf x} \notag \\
&=& \log\left\{\frac{1}{(2\pi)^{d/2}K(\nu_{\max-1},d)}\right\}-\frac{d}{2} \notag \\
&+& \frac{\nu_{\max-1}+d}{2}\mathbb{E}_{d}\left\{\log\left(1+\frac{{\bf x}^{\top}{\bf x}}{\nu_{\max-1}}\right)\right\},
\end{eqnarray}
where we have used that $D_{KL}\left(N_d(\cdot\mid \bm{\mu},\bm{\Sigma})\mid\mid f_d(\cdot\mid \bm{\mu},\bm{\Sigma},\nu)\right) = D_{KL}(N_d(\cdot\mid {\bf 0},{\bf I})\mid\mid f_d(\cdot\mid  {\bf 0},{\bf I},\nu))$. The expectation in \eqref{DKLN} can be reduced to one-dimensional integration using again a change of variable in terms of spherical coordinates:
\begin{eqnarray*}
\mathbb{E}_{d}\left\{\log\left(1+\frac{{\bf x}^{\top}{\bf x}}{\nu_{\max-1}}\right)\right\} =  \dfrac{1}{2^{\frac{d}{2}}\Gamma\left(\frac{d}{2}\right)}  \int_0^{\infty} e^{-\frac{t}{2}} t^{\frac{d}{2}-1}\log\left(1+\dfrac{t}{\nu_{\max-1}}\right) dt.
\end{eqnarray*}
Therefore, the calculation of the divergences involved in the construction of the proposed prior only require one-dimensional numerical integration, regardless of the dimension of the sampling model. This makes the construction of the proposed prior scalable to high dimensions.

\begin{figure}[h]
\begin{center}
\begin{tabular}{c c}
\includegraphics[width=6cm, height=5cm]{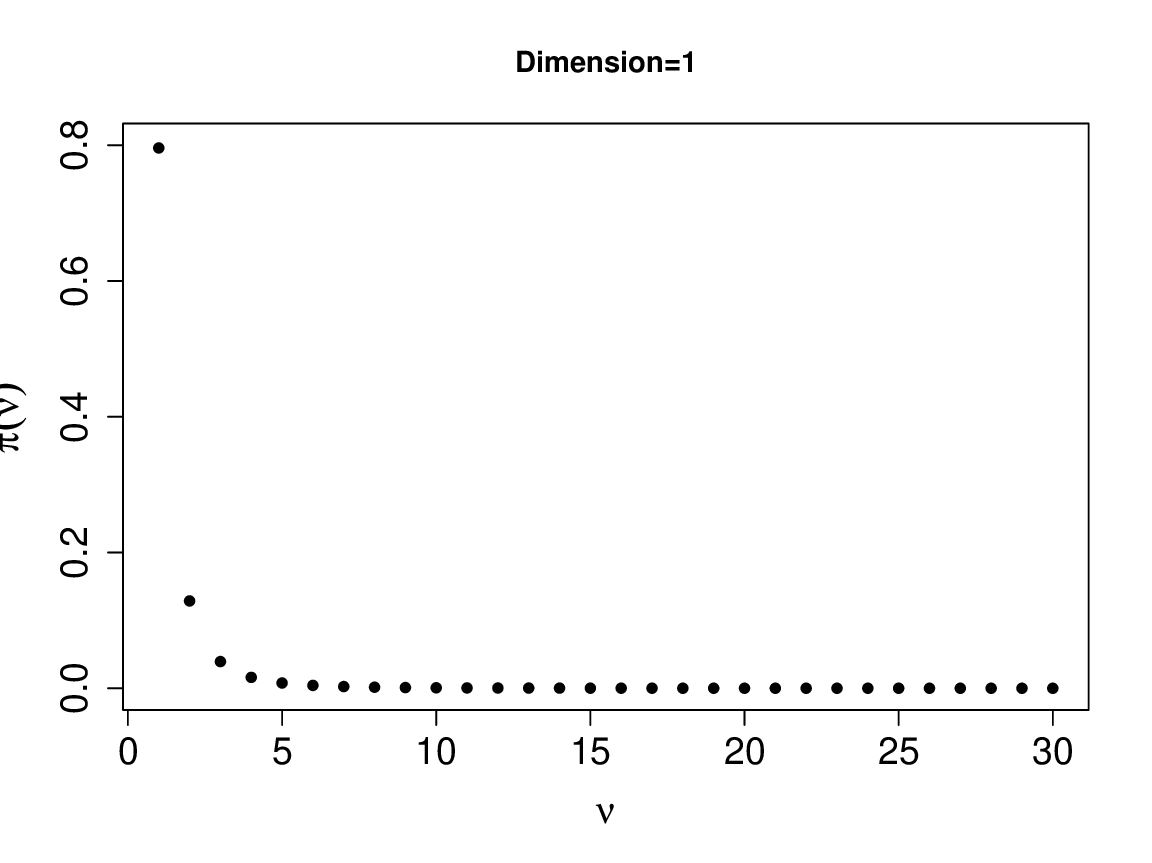}&
\includegraphics[width=6cm, height=5cm]{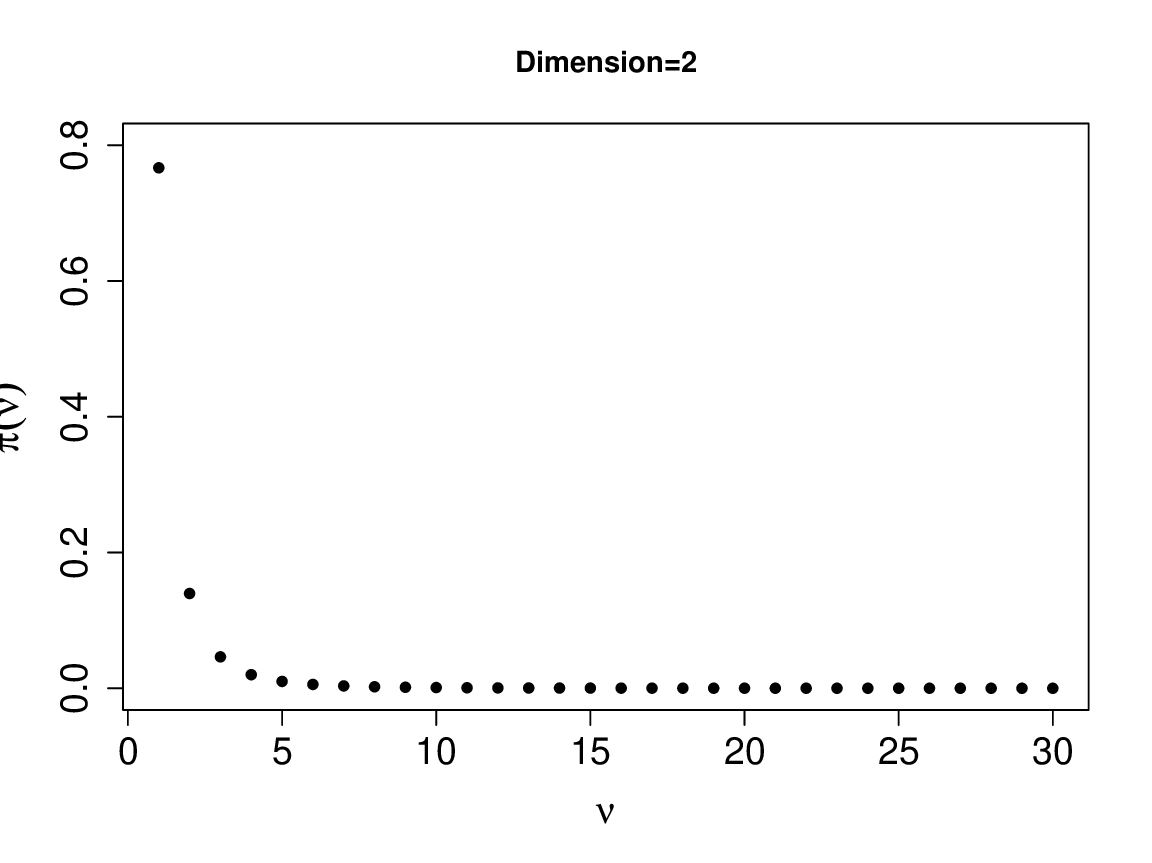}\\
(a) & (b) \\
\includegraphics[width=6cm, height=5cm]{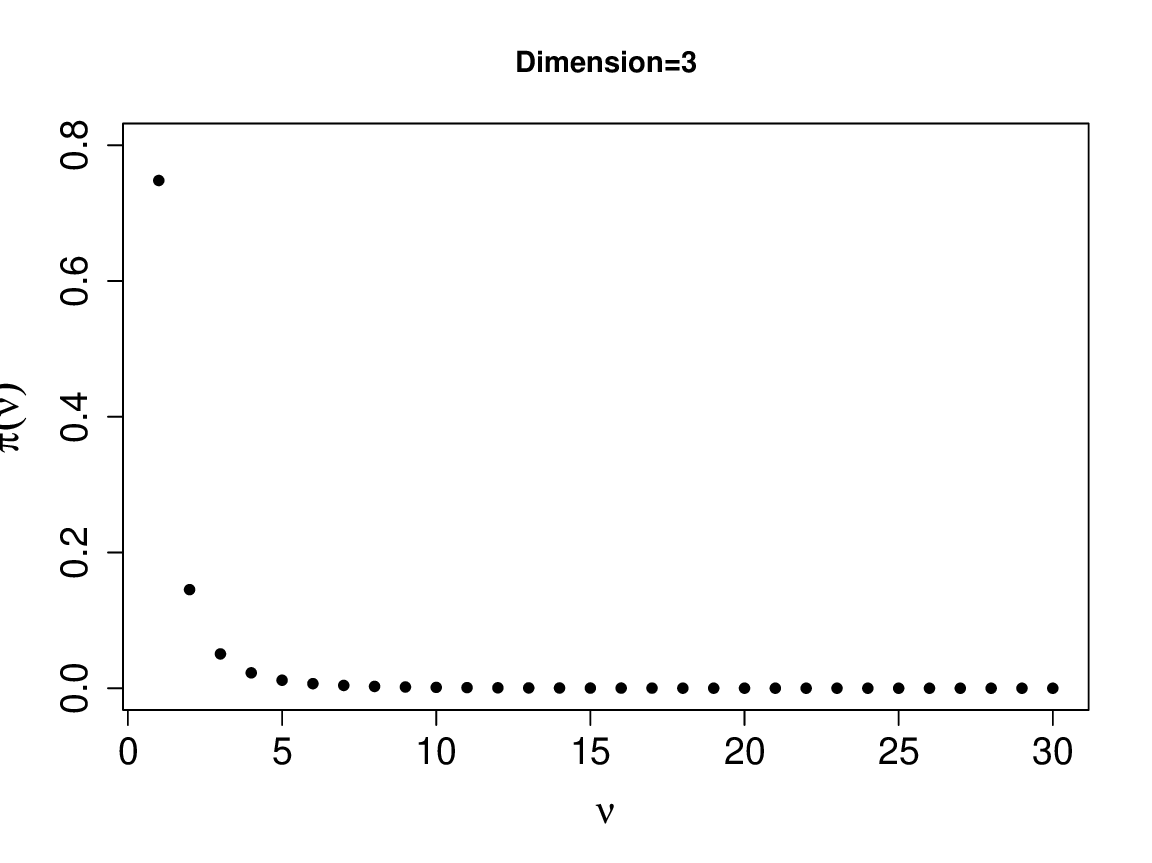}&
\\
(c) & \\
\end{tabular}
\end{center}
\caption{Loss-based prior for the multivariate $t$, $\pi(\nu|\bm{0},\bm{I})$: (a) $d=1$; (b) $d=2$; (c) $d=3$.}
\label{fig:tprior}
\end{figure}

As anticipated, from Table 1 we see that the Kullback--Leibler divergence becomes very small already for moderate values of $\nu$. Furthermore, we note that the nearest density to $f_{d,\nu}$ is always $f_{d,\nu+1}$. Thus, by applying the result in \eqref{tprior}, we have the prior for $\nu$, given $\bm{\mu}$ and $\bm{\Sigma}$, as
\begin{equation}\label{tprior_1}
\pi(\nu\mid\bm{\mu},\bm{\Sigma}) \propto \exp\left\{D_{KL}(f_{d,\nu}\mid\mid f_{d,\nu+1})\right\}-1, \nonumber
\end{equation}
for $\nu=1,\ldots,\nu_{\max-2}$,
\begin{equation}\label{tprio_3}
\pi(\nu\mid\bm{\mu},\bm{\Sigma}) \propto \exp\left\{D_{KL}(f_{d,\nu}\mid\mid f_{d,\nu-1})\right\}-1, \nonumber
\end{equation}
for $\nu=\nu_{\max-1}$, and
\begin{equation}\label{tprior_2}
\pi(\nu\mid\bm{\mu},\bm{\Sigma}) \propto \exp\left\{D_{KL}(N_d\mid\mid f_{d,\nu_{\max-1}})\right\}-1, \nonumber
\end{equation}
for $\nu=\nu_{\max}$. Figure \ref{fig:tprior} shows the induced priors.

\subsubsection{$t$-Copula}\label{sec:tcoppriors}
The Kullback--Leibler divergence between two $d$-variate $t$-copulas, $c_d(\cdot\mid {\nu,{\bf R}})$ and $c_d(\cdot\mid {\nu^{\prime},{\bf R}})$, is given by
\begin{eqnarray}\label{KLCop}
D_{KL}(c_d(\cdot\mid {\nu,{\bf R}})\mid\mid c_d(\cdot\mid {\nu^{\prime},{\bf R}})) &=& \int_{[0,1]^d}  c_d({\bf u}\mid {\nu,{\bf R}}) \log \dfrac{c_d({\bf u}\mid {\nu,{\bf R}})}{c_d({\bf u}\mid {\nu^{\prime},{\bf R}})}\,d{\bf u}.
\end{eqnarray}
This divergence depends on the degrees of freedom $\nu$ and $\nu^{\prime}$ as well as on the correlation matrix ${\bf R}$.

Our aim is to construct a prior for $(\nu,{\bf R})$ by using the decomposition $\pi(\nu,{\bf R})= \pi(\nu\mid{\bf R})\pi({\bf R})$. The prior $\pi(\nu\mid{\bf R})$ will be obtained as in the Multivariate $t$ case (\textit{i.e.}~applying the result in \eqref{tprior}), for each value of the correlation matrix ${\bf R}$, while for the prior $\pi({\bf R})$ we employ independent $\mbox{Beta}(1/2,1/2)$ priors for each of the entries of this matrix. For a more extensive discussion on the choice of priors for correlation parameters, we refer the reader to \cite{S04}.

Each time we evaluate the log-posterior, we need to calculate the prior $\pi(\nu\mid {\bf R})$, which requires the calculation of the $\nu_{\max}$ Kullback--Leibler divergences. In order to have a tractable approximation in the bivariate case, we propose discretising the range of values of $\rho\in(-1,1)$ into intervals of size $0.05$: $(-1,-0.975)\cup(-0.975,-0.925) \cup \dots \cup (0.925,0.975)\cup (0.975,1)$. We have checked the variability of the Kullback--Leibler divergences within these intervals and found that this step-size produces an accurate approximation to the prior using either endpoints. Note that this discretisation only relates to the conditional prior $\pi(\nu\vert {\bf R})$, while there is no approximation on the marginal prior $\pi({\bf R})$.

\begin{figure}[h]
\begin{center}
\begin{tabular}{c c}
\includegraphics[width=6cm, height=5cm]{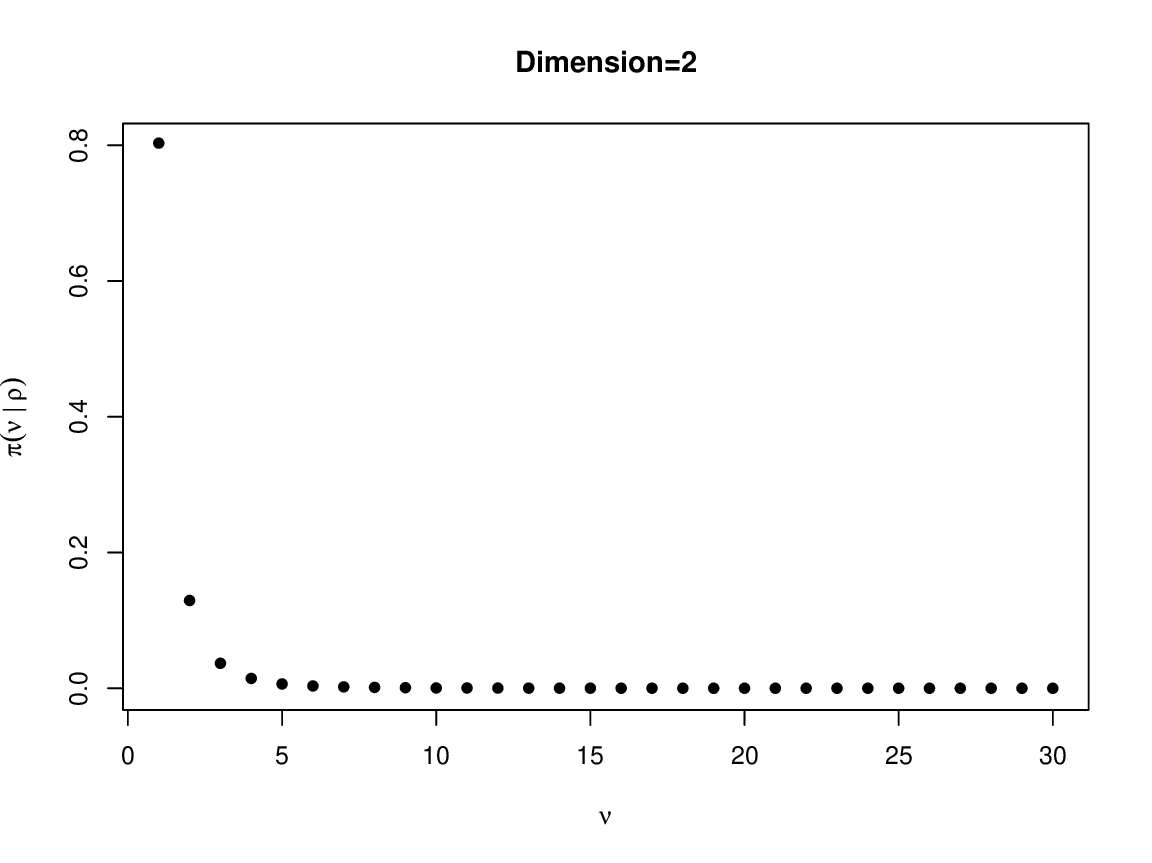}&
\includegraphics[width=6cm, height=5cm]{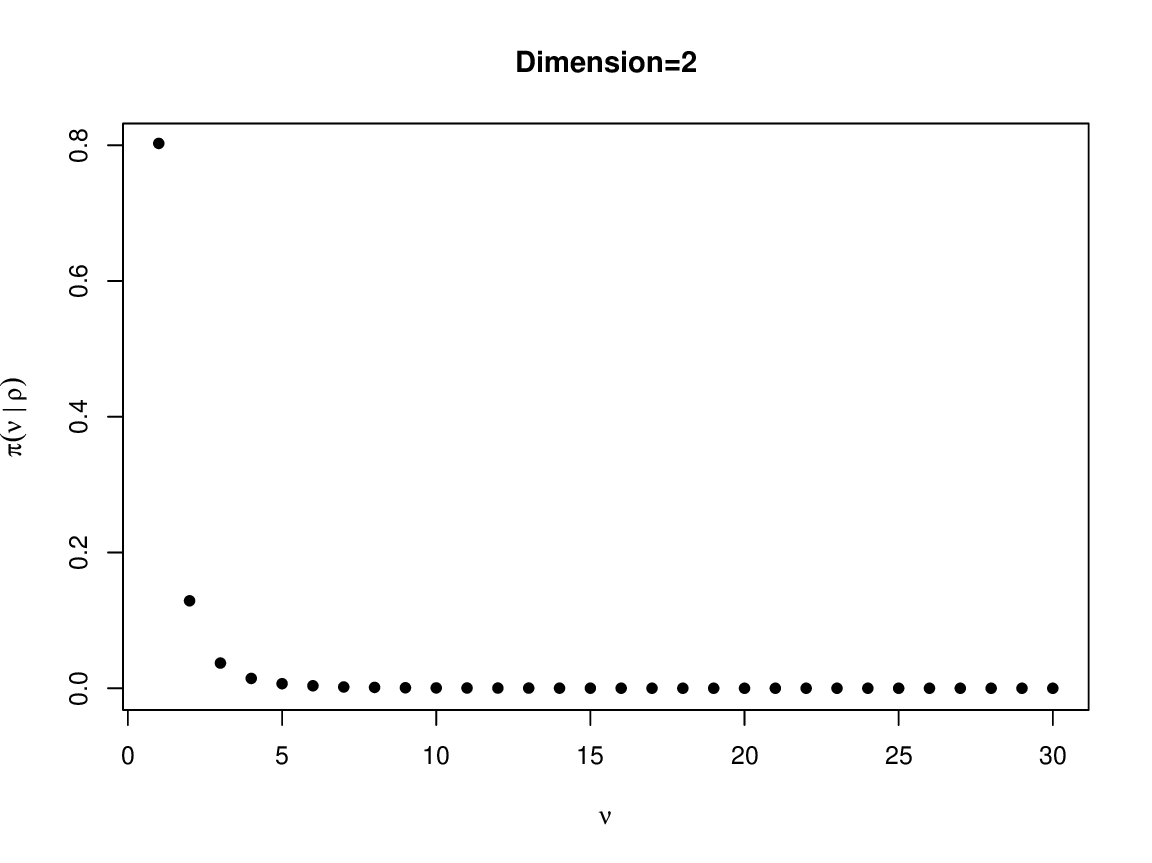}\\
(a) & (b) \\
\includegraphics[width=6cm, height=5cm]{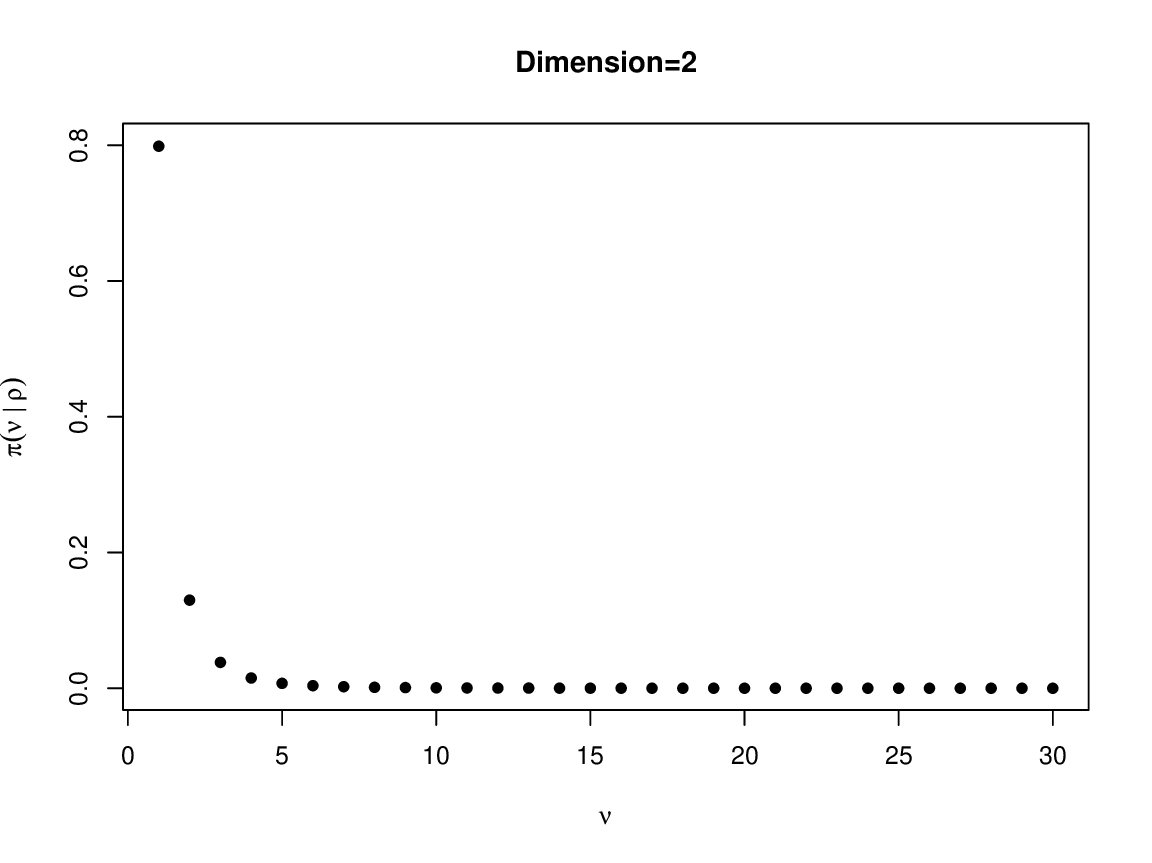}&
\includegraphics[width=6cm, height=5cm]{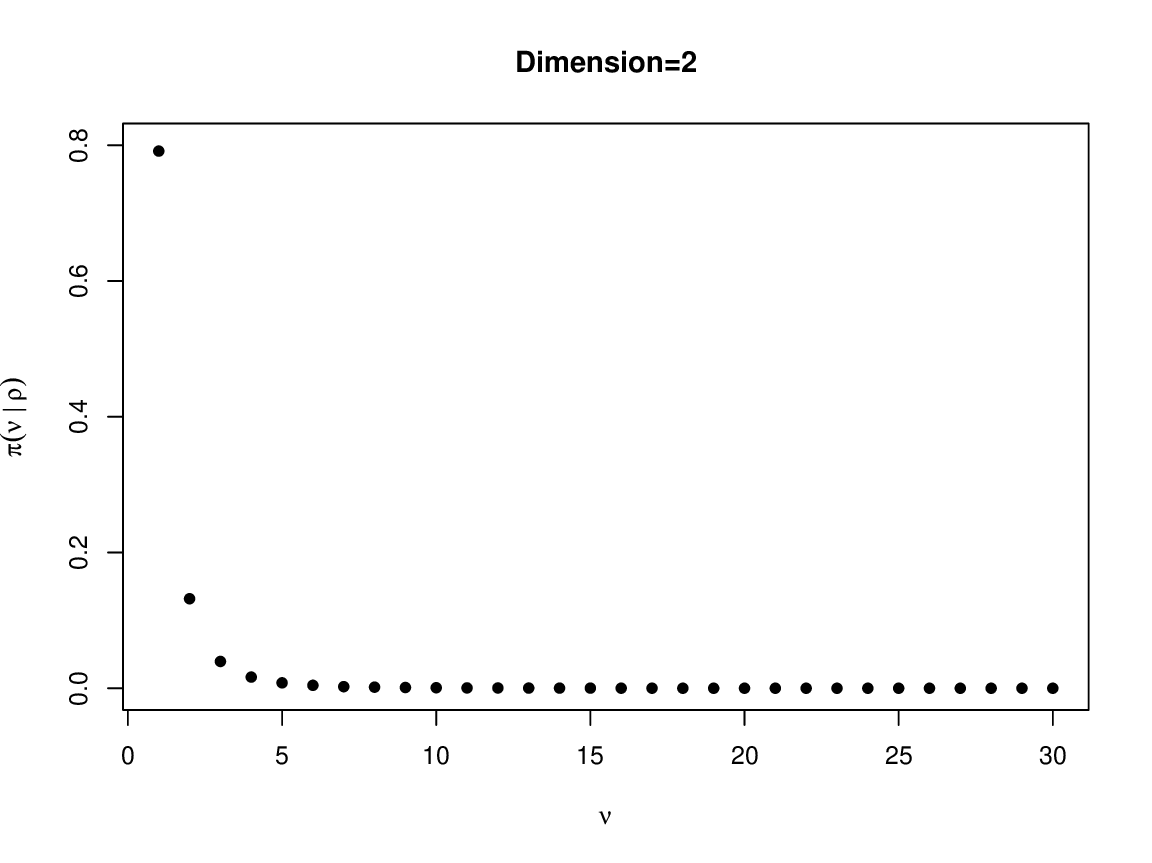}\\
(c) & (d)\\
\end{tabular}
\end{center}
\caption{ $\pi(\nu\mid \rho)$: (a) $\rho=0$; (b) $\rho=0.25$; (c) $\rho=0.5$; (d) $\rho=0.75$.}
\label{fig:tcopcondprior}
\end{figure}

We approximate the Kullback-Leibler divergences using a Monte Carlo approximation to \eqref{KLCop}:
\begin{eqnarray*}
D_{KL}(c_d(\cdot\mid {\nu,{\bf R}})\mid\mid c_d(\cdot\mid {\nu^{\prime},{\bf R}})) &\approx& \dfrac{1}{N}\sum_{j=1}^N   \log \dfrac{c_d({\bf u}_j\mid {\nu,{\bf R}})}{c_d({\bf u}_j\mid {\nu^{\prime},{\bf R}})},
\end{eqnarray*}
where ${\bf u}_1,\dots,{\bf u}_N$ are $d$-variate samples from $c_d(\cdot\mid {\nu,{\bf R}})$. Figure \ref{fig:tcopcondprior} shows the priors obtained for four choices of $\rho$ in the bivariate case ($d=2$) using $N=10^7$ Monte Carlo simulations (the large number of simulations is chosen to improve accuracy). The figure indicates that the conditional value of $\rho$ has negligible influence on the shape of the prior. Thus, in our examples we restrict to $\rho=0$, which greatly simplifies sampling from the posterior distribution. A second approach for approximating the Kullback--Leibler divergences consist of using importance sampling. As the importance function we can employ the copula with the smallest degrees of freedom $\min\{\nu,\nu^{\prime}\}$, which implies heavier tails as desired. We employ the latter method, with $N=5\times 10^7$ Monte Carlo simulations, to approximate the prior probabilities for the $2-$variate $t$-copula with $\rho=0$. Table \ref{table:tcopvalues} shows the values of this prior for $\nu=1,\ldots,30$.

\begin{table}[ht]
\centering
$\begin{array}{ccccccc}
  \hline
\nu & 1 & 2 & 3 & 4 & 5 & 6 \\
  \text{Prob.} & 0.804 & 0.129 & 0.0368 & 0.014 & 0.007 & 0.004 \\
   \hline
  \nu & 7 & 8 & 9 & 10 & 11 & 12 \\
  \text{Prob.} & 0.002 & 1.28\times10^{-3} & 8.05\times10^{-4} & 5.33\times10^{-4} & 3.58\times10^{-4} & 3.05\times10^{-4} \\
   \hline
 \nu & 13 & 14 & 15 & 16 & 17 & 18 \\
  \text{Prob.} & 2.06\times10^{-4} & 1.60\times10^{-4} & 1.30\times10^{-4} & 9.52\times10^{-5} & 6.79\times10^{-5} & 6.04\times10^{-5} \\
   \hline
  \nu & 19 & 20 & 21 & 22 & 23 & 24 \\
 \text{Prob.} & 4.55\times10^{-5} & 3.44\times10^{-5} & 2.19\times10^{-5} & 2.39\times10^{-5} & 2.06\times10^{-5} & 2.31\times10^{-5} \\
  \hline
  \nu & 25 & 26 & 27 & 28 & 29 & 30 \\
  \text{Prob.} & 1.81\times10^{-5} & 1.91\times10^{-5} & 1.28\times10^{-5} & 2.05\times10^{-5} & 7.85\times10^{-6} & 2.78\times10^{-6} \\
   \hline
\end{array}$
\caption{Loss-based prior $\pi(\nu\mid\rho=0)$ for the bivariate $t$-copula.}
\label{table:tcopvalues}
\end{table}


In our simulation study and in the real data analysis we have limited our work to the case of a bivariate copula. This choice is dictated by the fact that most applications where copula functions are employed refer to bi-dimensional problems. However, the proposed prior can be applied to any dimensional size, and \cite{KKC17} discuss a few methods to approximate the Kullback--Leibler divergence between two copulas, even for relatively large dimensions, which are computationally more efficient than Monte Carlo approximation or importance sampling.

\subsection{Prior distributions for the parameters different from $\nu$}\label{sec:otherprios}
For the multivariate $t$ distribution, as prior on the location vector and the scale matrix, as in \eqref{overallmultitprior}, we use the independence-Jeffreys prior
$$\pi(\bm{\mu},\bm{\Sigma}) = \frac{1}{|\bm{\Sigma}|^{3/2}}.$$
We refer the reader to Theorem 1 in \cite{FS99} for a proof of the propriety of the corresponding posterior for the parameters, which only imposes the condition of the sample size: $ n \geq 3$. For other dimensions $d$, Theorem 1 in \cite{FS99} implies that the posterior distribution, under the (marginal) prior structure $\pi(\bm{\mu},\bm{\Sigma}) = \vert \bm{\Sigma} \vert^{-\frac{d+1}{2}}$, is proper provided that $n \geq d + 1$. We can also consider a more general prior structure $\pi(\bm{\mu},\bm{\Sigma}) = \vert \bm{\Sigma} \vert^{-\frac{d+q+1}{2}}$, $q\geq 0$, where the condition for the propriety of the posterior becomes $n \geq d + q + 1$. This structure contains the Jeffreys prior for $(\bm{\mu},\bm{\Sigma})$, which is obtained for $q=1$.

The $t$-copula illustrations are limited to the bivariate case, both in the simulation study and in the real data analysis. As such, the prior in \eqref{overallcopulaprior} becomes
$$\pi(\mu_1,\mu_2,\sigma_1,\sigma_2,\nu_1,\nu_2,\nu,\rho) = \pi(\mu_1)\pi(\mu_2)\pi(\sigma_1)\pi(\sigma_2)\pi(\nu_1)\pi(\nu_2)\pi(\nu,\rho).$$
The minimally informative priors for the location parameters of the marginal $t$ densities are Normal distributions with zero mean and standard deviation 100. That is, $\pi(\mu_j)\sim N(0,100^2)$, for $j=1,2$. This choice allows to ensure that the yielded posteriors are proper and, given the large variance, ensure a representation of minimal prior information. Alternatively, one could use uniform priors on a large compact set. To reflect vague prior information for the scale parameters, we choose half-Cauchy densities for the scale parameters $\pi(\sigma_j)$ \citep{RS15}. The prior distributions for the number of degrees of freedom of the marginal densities, $\pi(\nu_j)$, are based on losses and correspond to the one derived in \cite{VW14}. The joint prior $\pi(\nu,\rho)$ is decomposed as $\pi(\nu\mid \rho)\pi(\rho)$, where $\pi(\nu\mid\rho)$ is the prior defined in Section \ref{sec:tcoppriors}, and $\pi(\rho)$ is a Beta density on $(1+\rho)/2$ with parameters $(1/2,1/2)$.

\subsection{Posterior distribution}\label{sec:posterior}
The joint posterior distributions for all parameters are
$$\pi(\bm{\mu},\bm{\Sigma},\nu\mid\bm{x}) \propto L_d^t(\bm{\mu},\bm{\Sigma},\nu\mid\bm{x})\pi(\nu\mid\bm{\mu},\bm{\Sigma})\pi(\bm{\mu},\bm{\Sigma}),$$
and
$$\pi(\nu,\bm{R}\mid\bm{x})\propto L_d^c(\nu,\bm{R}\mid\bm{x})\pi(\nu\mid\bm{R})\pi(\bm{R}),$$
where $L_d^t$ and $L_d^c$ are the likelihood functions for, respectively, the multivariate $t$ model and the $t$-copula model. In both cases, the posterior distributions are analytically intractable and have to be approximated by using Monte Carlo methods. The details of the algorithms implemented are in Appendix A.

\section{Simulation Study}\label{sec:simulation}
In this section we present the results of the simulation studies performed for the multivariate $t$ distribution and for the $t$ copula. In particular, we analyse the frequentist performances of the respective yielded posterior distributions, focusing on the coverage on the 95\% posterior credible interval and on the relative square-rooted mean squared error (MSE) from the posterior median.

\subsection{Multivariate $t$}\label{sec:simulmultit}
The loss-based prior for the number of degrees of freedom of a multivariate $t$ density has been thoroughly studied by computing the frequentist performances of the yielded posterior. The simulation study includes a comparison of the proposed objective prior with the three options available in the literature, introduced in Section \ref{sec:ObjPriors}. Namely, the Anscombe prior, the Jeffreys prior and the Relles \& Rogers prior. Simulations from the posterior distribution associated to the proposed loss-based priors are obtained using a Markov Chain Monte Carlo (MCMC) algorithm in which continuous parameters are sampled using a Random Walk Metropolis with Normal proposals, while the discrete parameter (the degrees of freedom) is sampled directly using the corresponding posterior probabilities in each iteration (formally, a block Metropolis within Gibbs sampler). For the alternative priors, simulations from the posterior distributions are obtained using the t-walk algorithm \citep{CF10}. In all the simulation scenarios, $N=500$ posterior samples are obtained using a burn-in period of $1000$ iterations, and a thinning period of $10$ iterations ($6000$ iterations in total).

The study consisted in replicating 250 times the derivation of the posterior distribution for $\nu$, under different initial choices, and computing the coverage of the $95\%$ credible interval and the MSE from the median. This has been performed by considering the proposed prior and the three objective alternatives available in the literature. We have considered multivariate $t$ densities of dimension $d=2$ and $d=3$, with zero mean for each component and covariance matrix equal to the identity matrix, so to reflect unit scale for each component and linear independence. The generated samples are of size $n=50$, $n=100$ and $n=250$, so to consider scenarios with little information from the data as well as with large information. The prior for $(\bm{\mu},\bm{\Sigma})$ is the independence-Jeffreys (see Section \ref{sec:otherprios}).

\begin{figure}[h!]
\begin{center}
\begin{tabular}{c c}
\includegraphics[width=6cm, height=5cm]{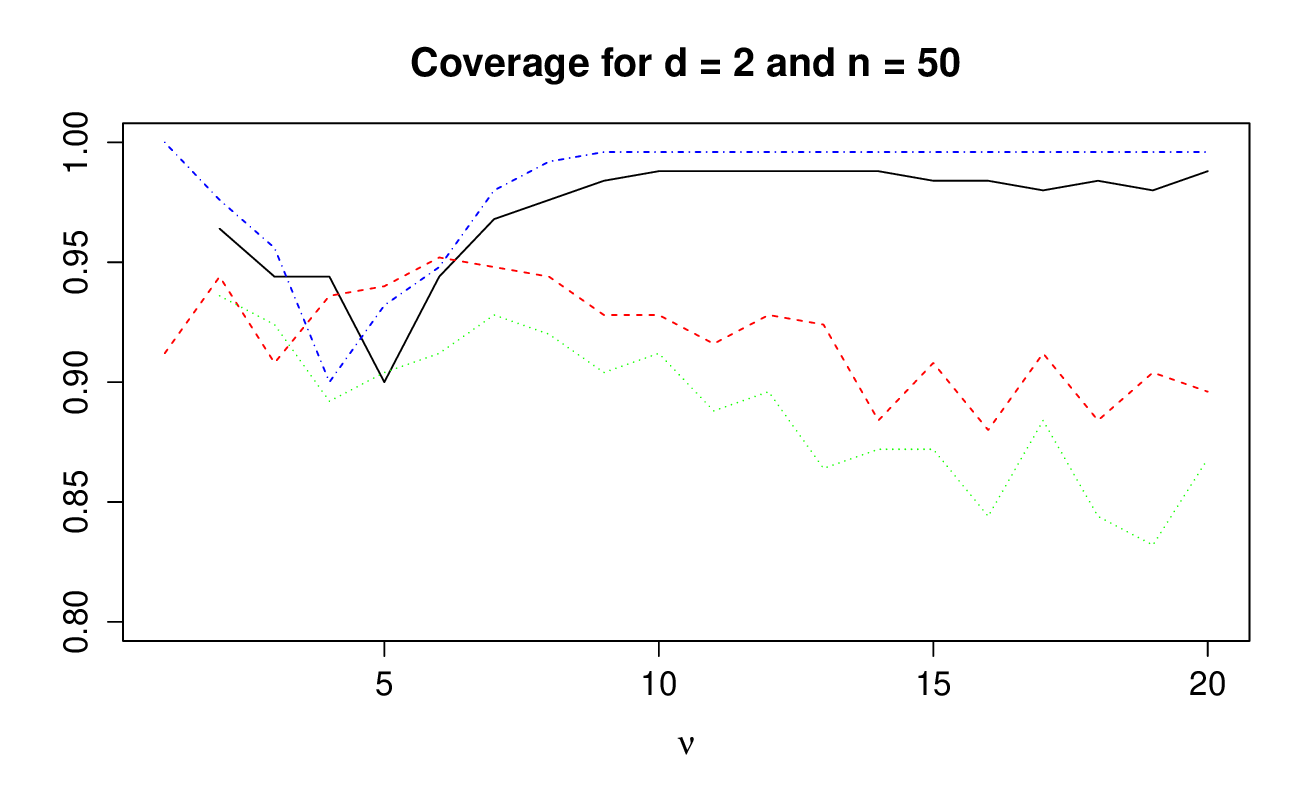}&
\includegraphics[width=6cm, height=5cm]{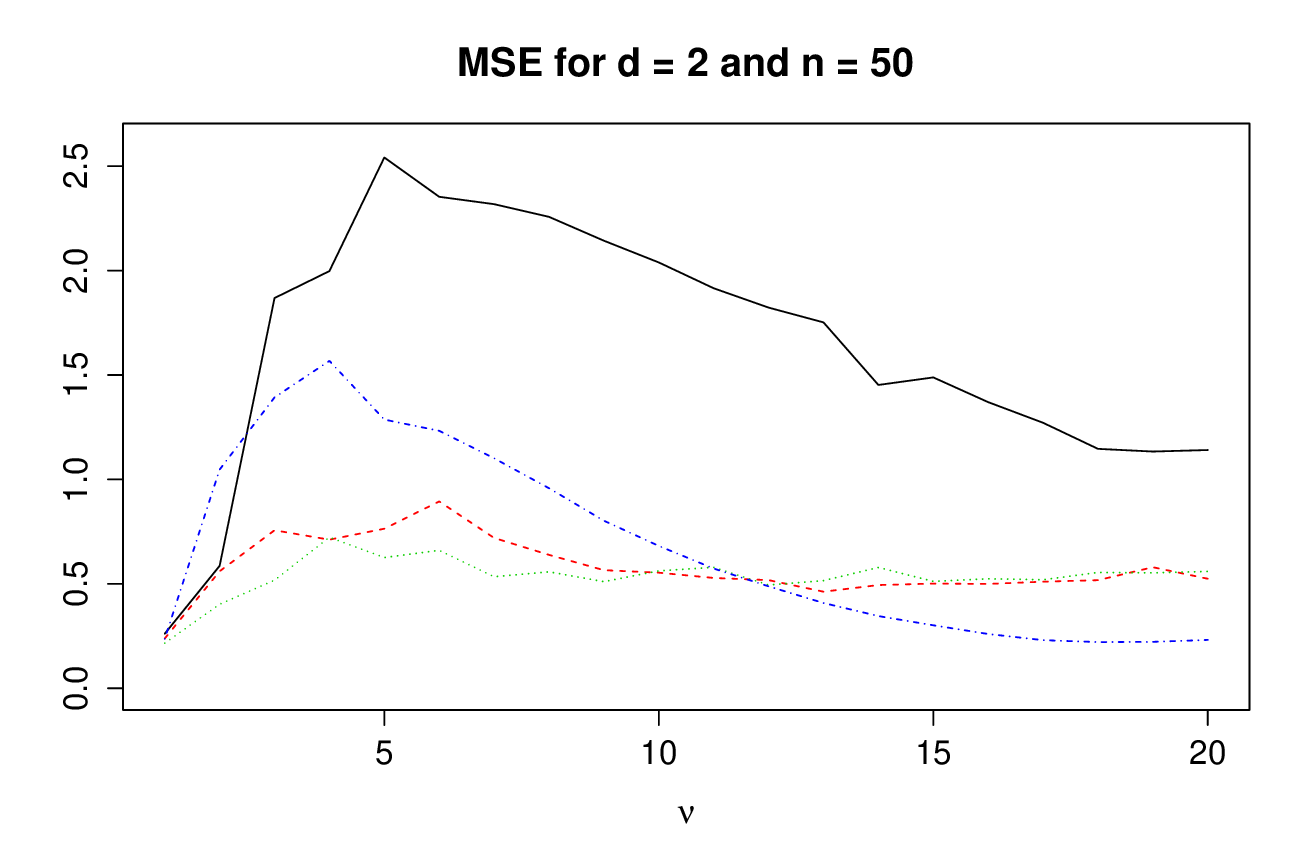}\\
(a) & (b) \\
\includegraphics[width=6cm, height=5cm]{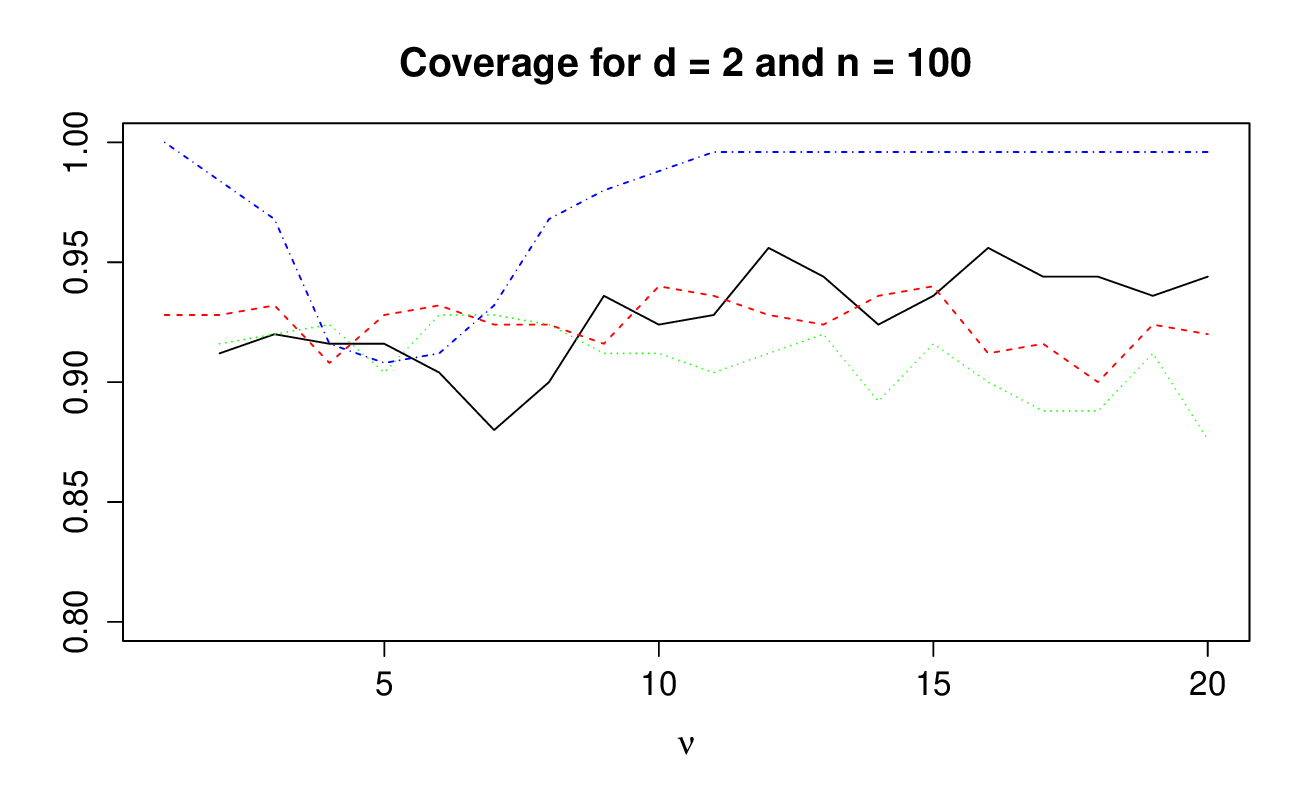}&
\includegraphics[width=6cm, height=5cm]{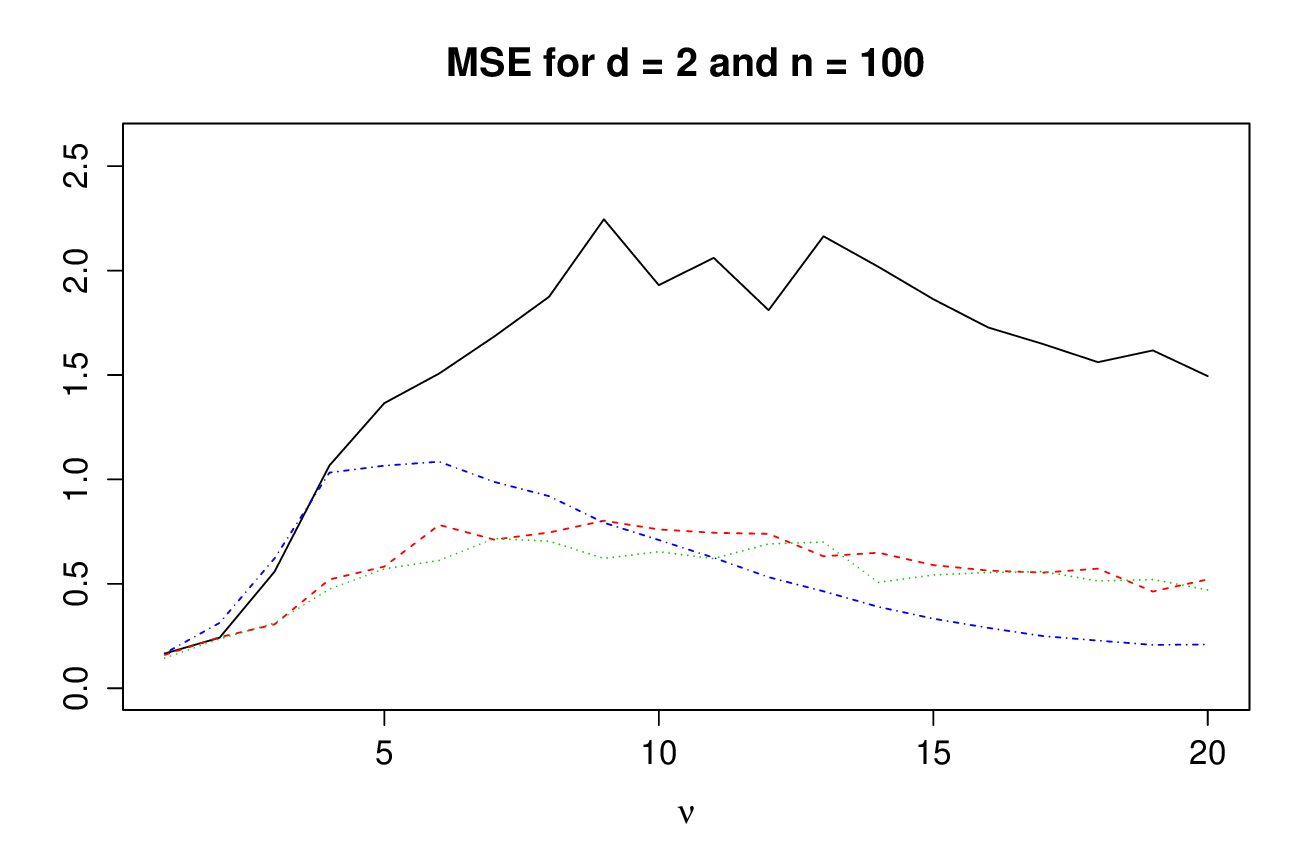}\\
(c) & (d)\\
\includegraphics[width=6cm, height=5cm]{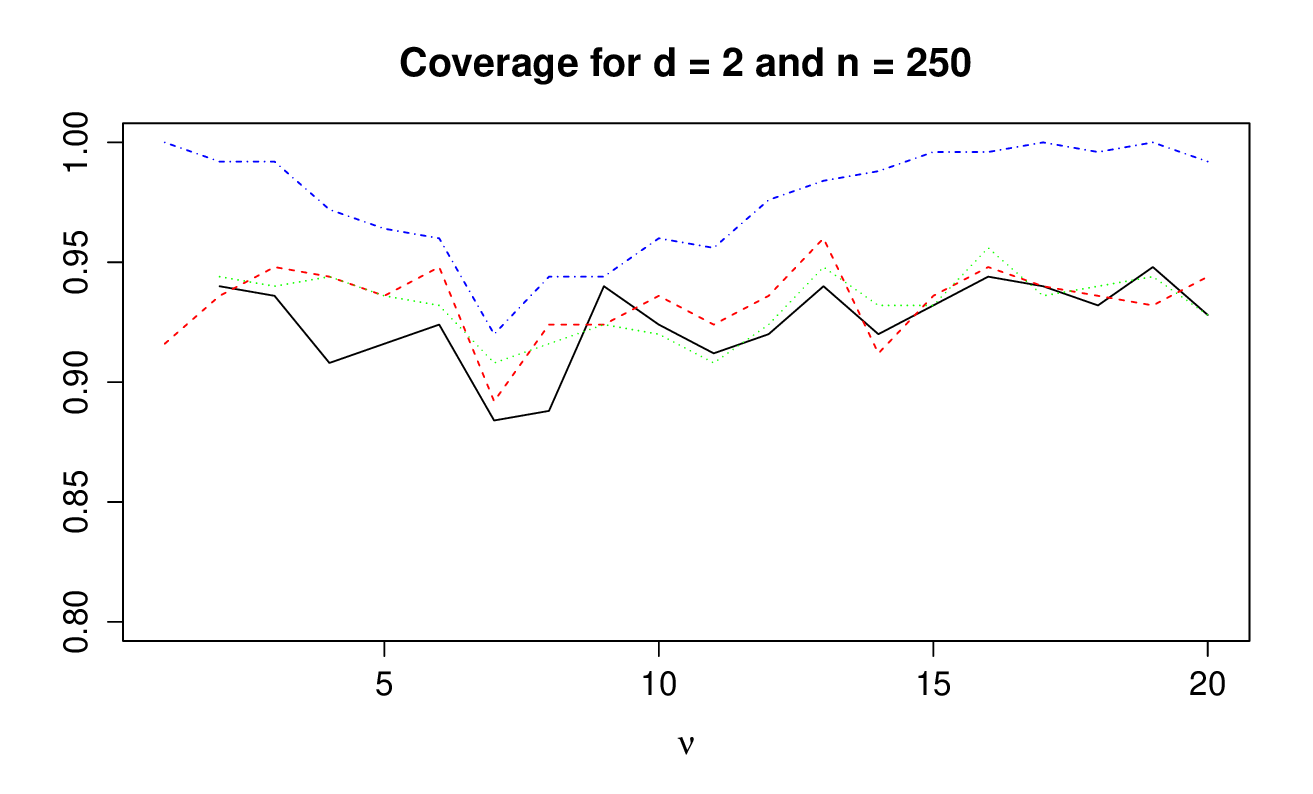}&
\includegraphics[width=6cm, height=5cm]{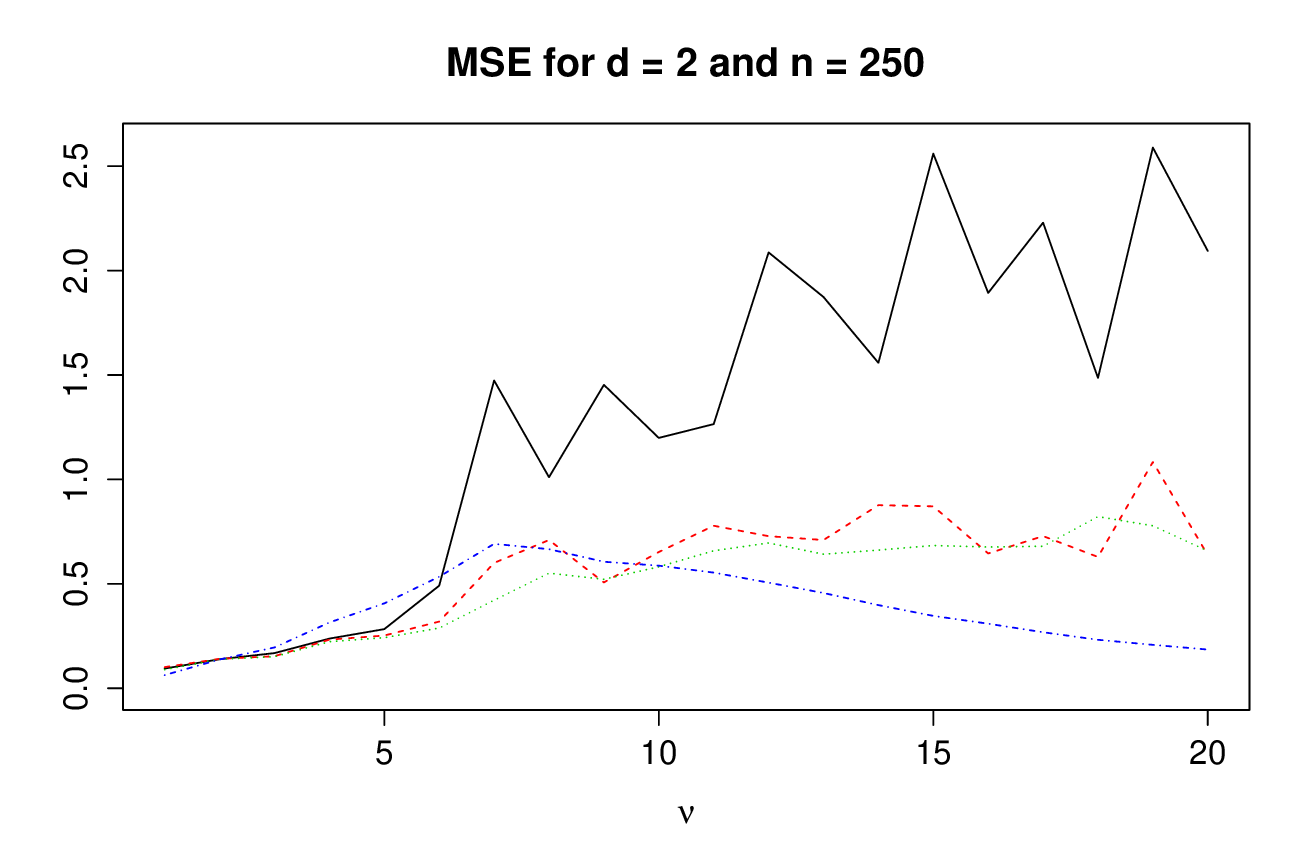}\\
(e) & (f)\\
\end{tabular}
\end{center}
\caption{Frequentist analysis of the multivariate $t$ of dimension $d=2$: (a)-(b) Coverage and MSE for $n=50$; (c)-(d) Coverage and MSE for $n=100$; (e)-(f) Coverage and MSE for $n=250$. We have considered four prior distributions for $\nu$: Anscombe prior (black continuous), Jeffreys prior (red dashed), Relles \& Rogers prior (green dotted) and the loss-based prior (blue dashed-dotted).}
\label{fig:tsimd2}
\end{figure}

\begin{figure}[h!]
\begin{center}
\begin{tabular}{c c}
\includegraphics[width=6cm, height=5cm]{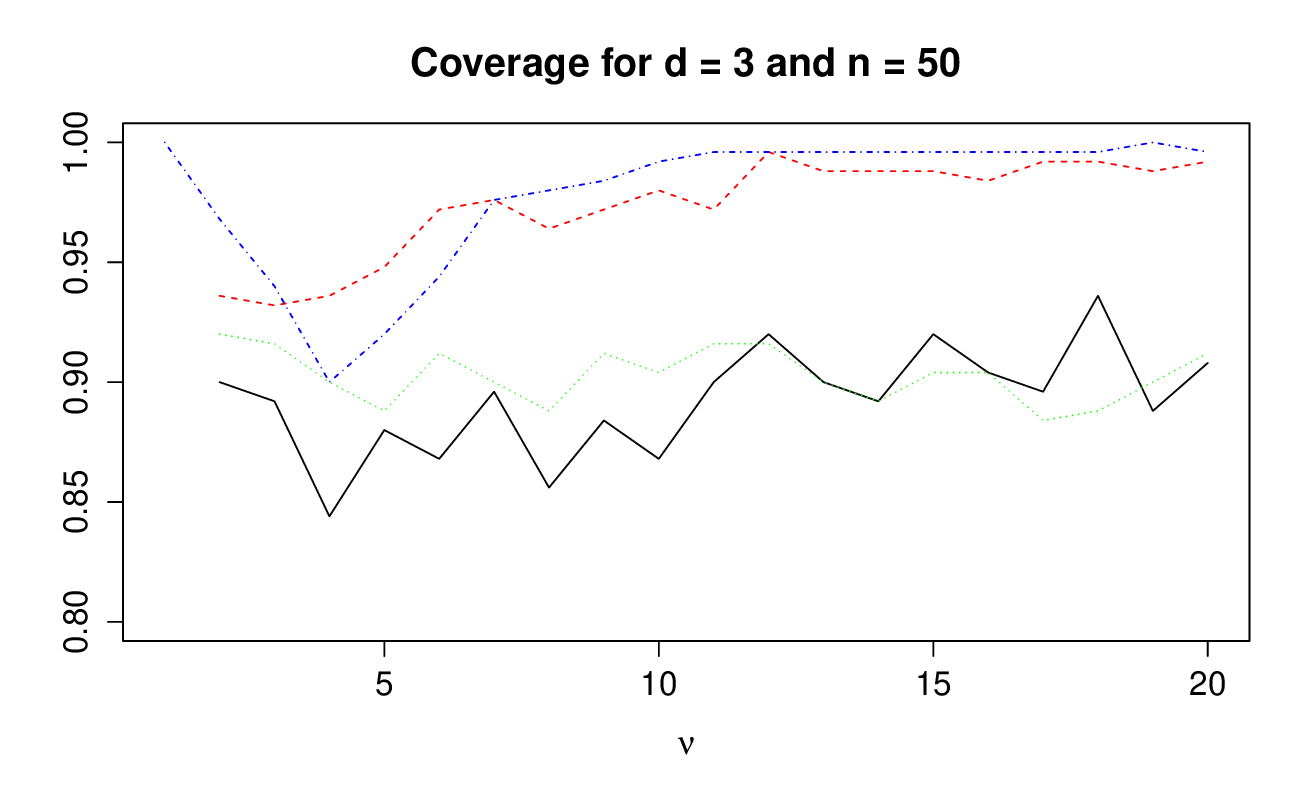}&
\includegraphics[width=6cm, height=5cm]{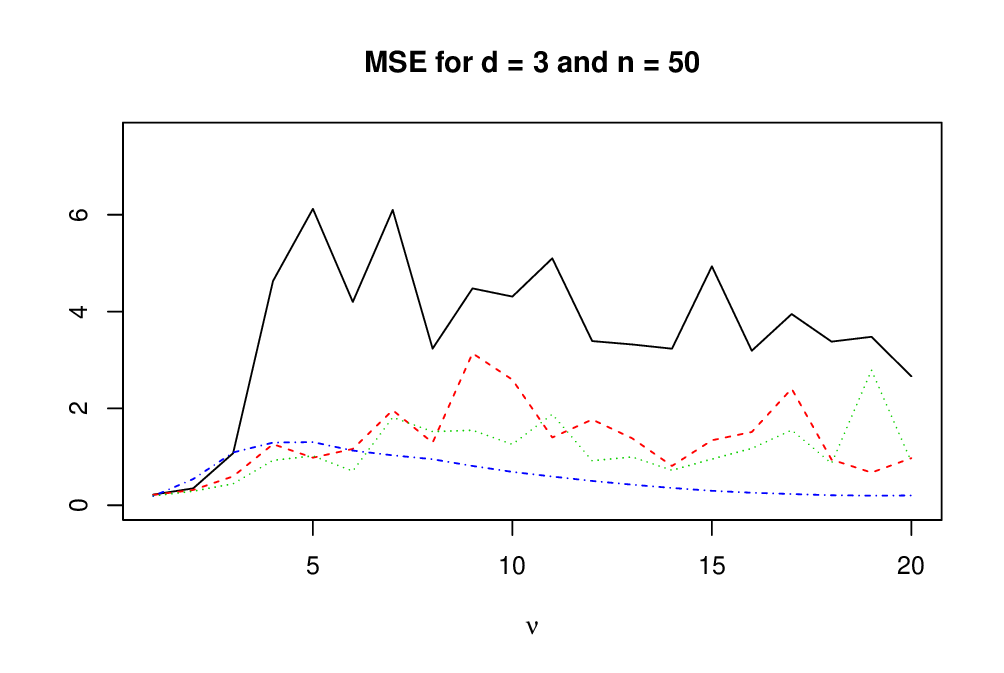}\\
(a) & (b) \\
\includegraphics[width=6cm, height=5cm]{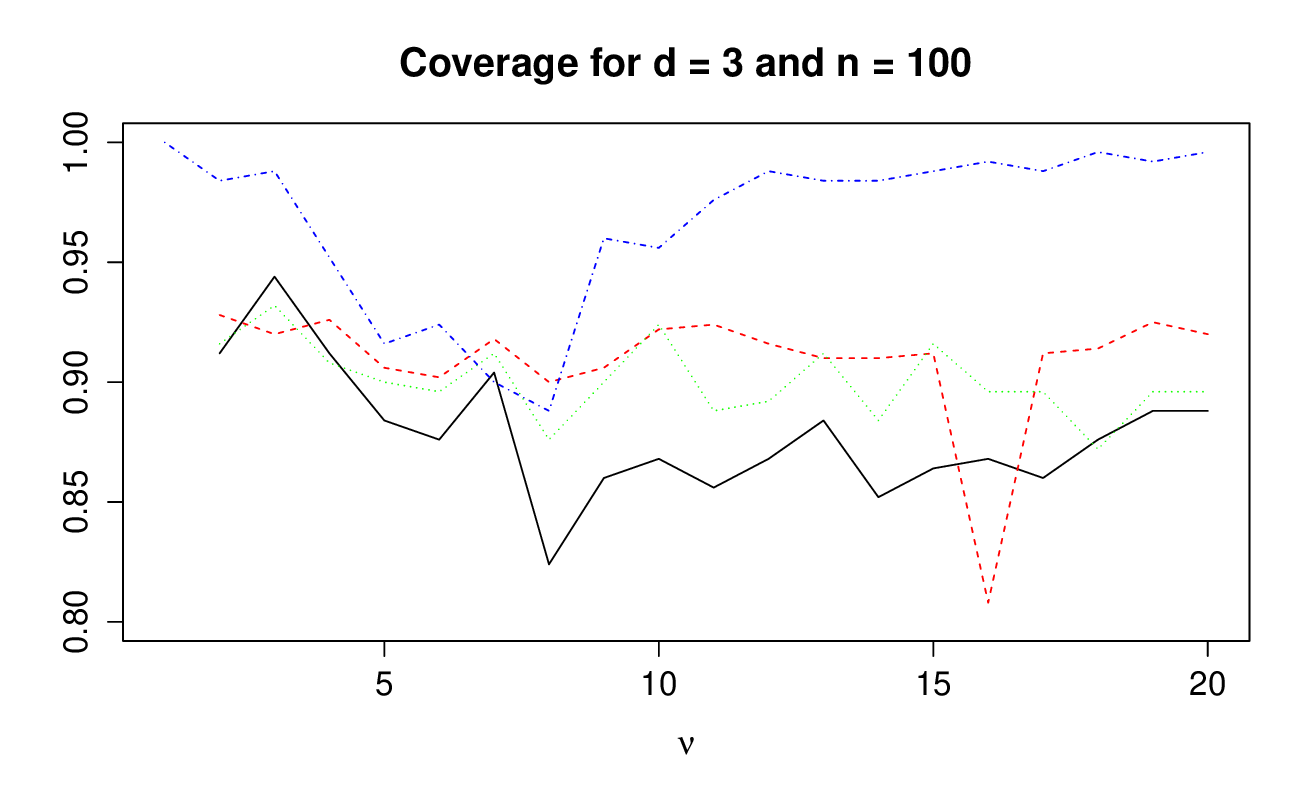}&
\includegraphics[width=6cm, height=5cm]{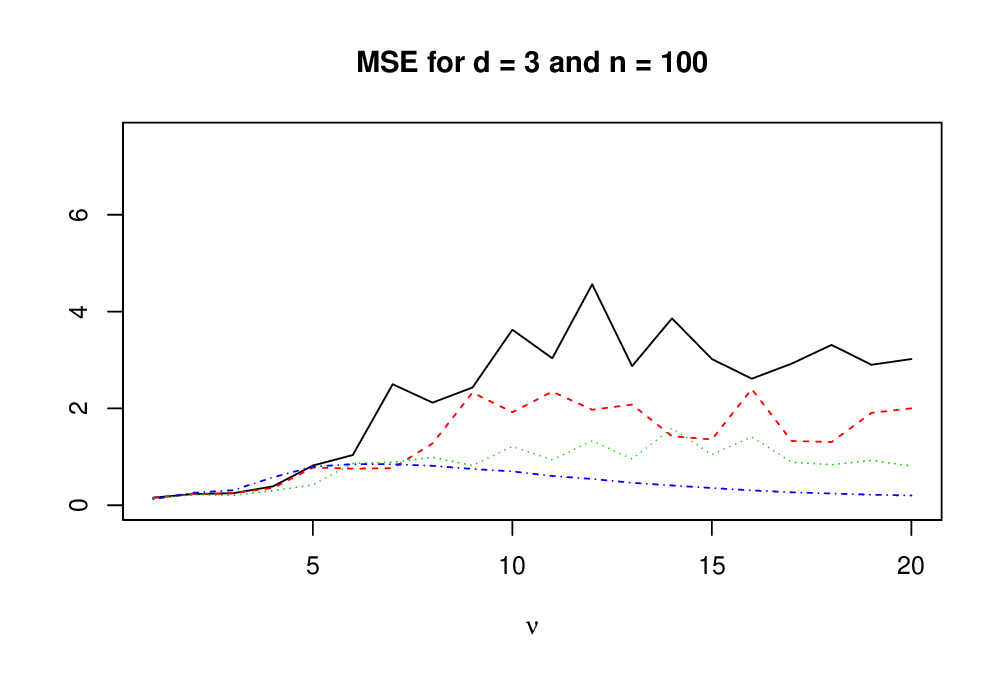}\\
(c) & (d)\\
\includegraphics[width=6cm, height=5cm]{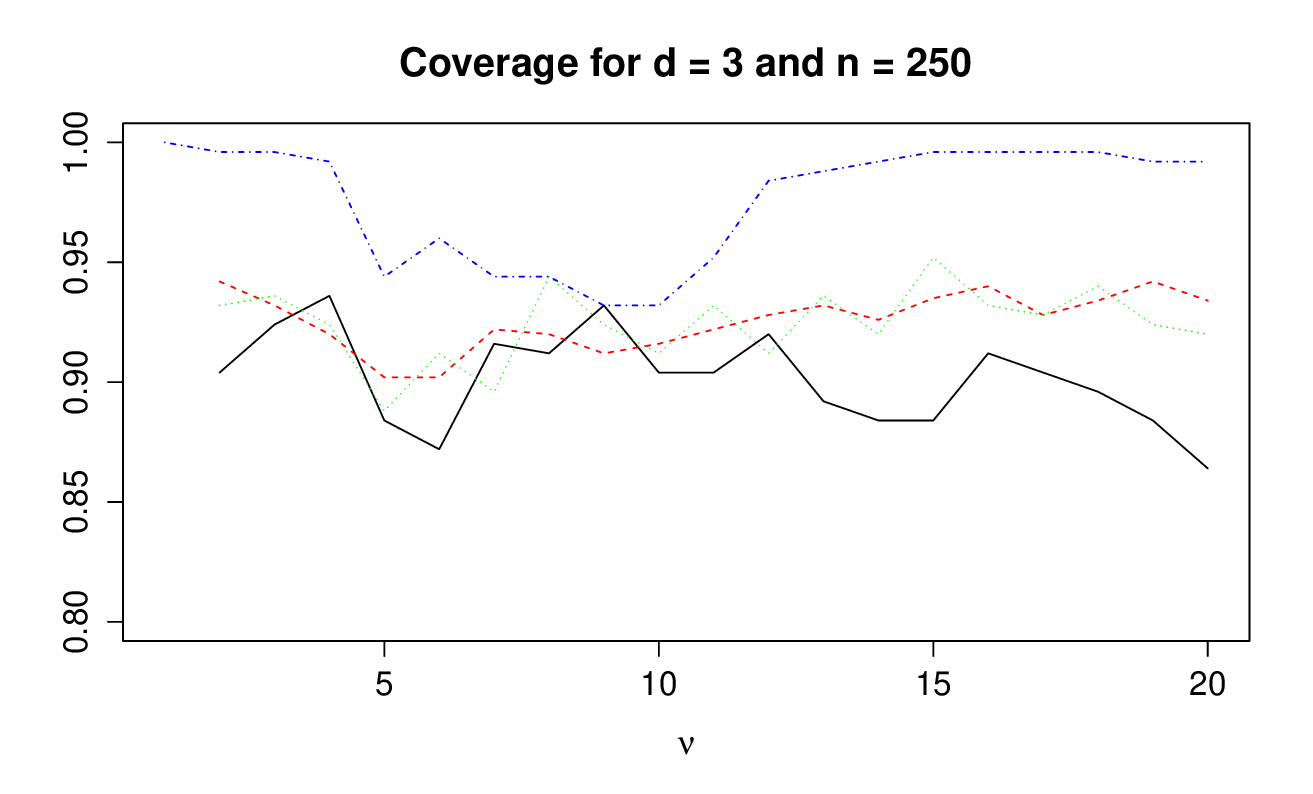}&
\includegraphics[width=6cm, height=5cm]{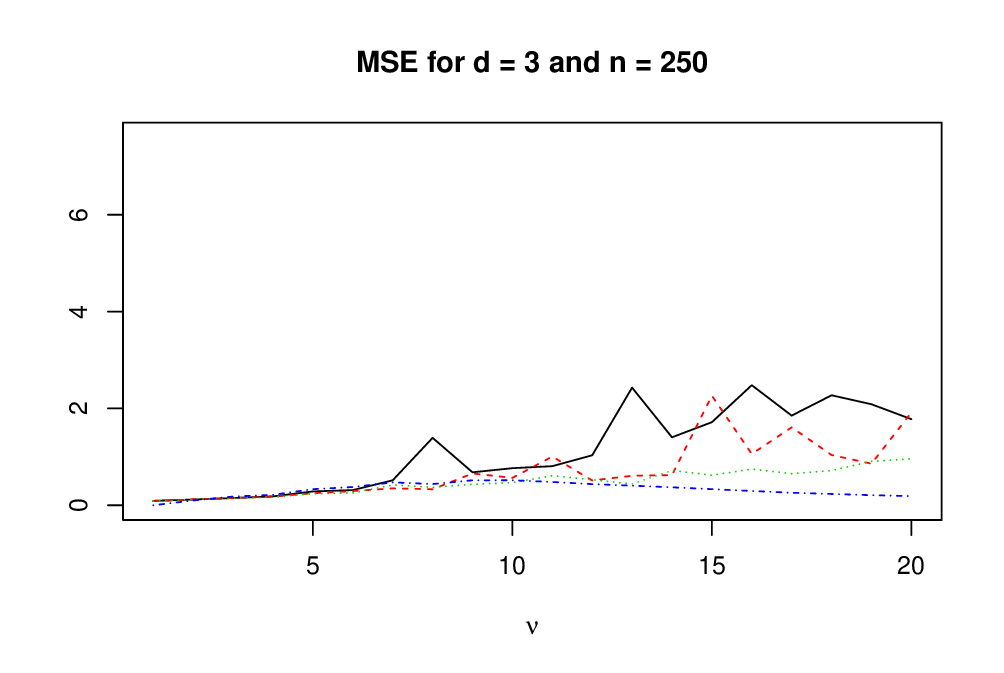}\\
(e) & (f)\\
\end{tabular}
\end{center}
\caption{Frequentist analysis of the multivariate $t$ of dimension $d=3$: (a)-(b) Coverage and MSE for $n=50$; (c)-(d) Coverage and MSE for $n=100$; (e)-(f) Coverage and MSE for $n=250$. We have considered four prior distributions for $\nu$: Anscombe prior (black continuous), Jeffreys prior (red dashed), Relles \& Rogers prior (green dotted) and the loss-based prior (blue dashed-dotted).}
\label{fig:tsimd3}
\end{figure}

Figure \ref{fig:tsimd2} shows the results for $d=2$, where we have the coverage (left column) and the MSE (right column) for the three sample sizes considered. The Anscombe prior appears to have the overall worst performance. In particular, the MSE, with the exception of the very low end of the parameter space, is always above the MSE obtained by employing any of the other priors. Also, for large values of $\nu$, the sample size appears to have little effect. As expected, the Jeffreys prior and the Relles \& Roger prior have similar performance, in particular for relatively large values of the number of degrees of freedom. The proposed prior, in terms of MSE, appears to be the most influenced by the data, \textit{i.e.}~the sample size. In fact, the value in its higher region noticeably decreases as $n$ increases. Furthermore, it has the best performance for relatively large values of $\nu$. If we consider the coverage, we note similar frequentist performances of the four priors for relatively small values of $\nu$, with slightly higher coverage of the proposed prior (which is explained by the discrete nature of this prior). Both Anscombe prior and the loss-based prior tend to 100\% as $\nu$ approaches 20, while the remaining two priors appear to ``under-cover'' the credible interval. This is more prominent for $n=50$ and for $n=100$.
The simulation results for the case $d=3$ are presented in Figure \ref{fig:tsimd3}. We note that the Anscombe prior is affected by the increase in the dimensionality of the $t$ distribution, in particular for small sample sizes. Although in a more confined way, both Jeffreys and Relles \& Rogers prior are affected as well. The increase in $d$ appears not to have any appreciable effect on the proposed loss-based prior. For what it concerns the coverage, the only noticeable difference from the case $d=2$ is in the tendency of the Anscombe prior to lie below the nominal value of 95\%, for any sample size.
An interesting aspect to highlight is the ``bumpiness'' of the MSE for the three priors we compare the loss-based to. This is particularly prominent for the Anscombe prior. The reason of the behaviour can be sought in the difficulty in sampling from models where heavy-tailed distributions are combined to heavy-tailed priors \citep{JarRob07}. Due to the truncated nature of the loss-based prior, which exhibits a relatively light tail, the effect is not noticeable, making it a good candidate to be used in the absence of sufficient prior information about the true number of degrees of freedom.

The uncertainty of the estimated mean coverages reported in Figures \ref{fig:tsimd2} and \ref{fig:tsimd3} is presented in the Appendix C. For each of the four considered priors we have computed the 95\% confidence intervals for the mean coverage estimated on the basis of the repeated 250 simulations per scenario. Also, to have a better understanding of the relationship between the width of the posterior 95\% credible intervals and the coverage, we have reported the posterior mean credible intervals and their width in the Appendix B.

\subsubsection{Non-integer values of $\nu$}\label{sec:nonintegerprior}
In this section, we study the behaviour of the loss-based prior when the true multivariate-$t$ distribution has non-integer degrees of freedom. We focus our simulations on the region of the parameter space where the densities are more different, that is for $\nu=2.5, 3.5, 4.5, 5.5$. We compare, for $d=2, 3$ and for $n=100, 250$ the frequentist performance of the loss-based prior and the three other objective priors: Anscombe prior (AP), Jeffreys prior (JP) and Relles \& Rogers prior (RRP). The MSE and the mean posterior 95\% credible intervals for the four scenarios are reported in Tables \ref{Tab:speciald2} and \ref{Tab:speciald3}. The loss-based prior, in terms of MSE, suffers a worsening when compared to the case where the data is simulated from multivariate-$t$ densities with integer degrees of freedom. This appears to be the base for both $d=2$ and $d=3$; although it decreases when we consider $n=250$ instead of $n=100$. There are however some considerations to be made in support of the loss-based prior. First, the performance, in terms of MSE, is indeed worse as it support does not contain the true value of $\nu$, but the difference from the other priors is, in general, minimal. Second, the width of the posterior mean credible intervals remains stable, meaning that the accuracy of the estimates of $\nu$ is not sensibly affected by the non-discrete nature of the true parameter.
\begin{table}[h!]
\centering
\begin{tabular}{cccccc}
\hline
DF & AP & JP & RRP & LBP & $\mbox{LBP}^*$ \\
\hline
2.5 & 0.28 & 0.26 & 0.25 & 0.40 & 0.23 \\
 & (1.70, 4.38) & (1.67, 4.24) & (1.63, 4.12) & (2.04, 5.03) & (1.65, 4.02) \\
3.5 & 0.46 & 0.37 & 0.39 & 0.77 & 0.28 \\
 & (2.27, 8.71) & (2.18, 7.47) & (2.10, 7.64) & (2.53, 10.90) & (2.10, 6.51)\\
4.5 & 1.17 & 0.66 & 0.48 & 0.98 & 0.29 \\
 & (2.82, 34.11) & (2.64, 27.67) & (2.54, 14.90) & (3.16, 17.63) & (2.46, 9.35) \\
5.5 & 1.68 & 0.82 & 0.65 & 1.13 & 0.31 \\
 & (3.38, 161.77) & (3.15, 29.37) & (2.99, 27.37) & (3.88, 22.09) & (2.76, 12.10) \\
\hline
\hline
2.5 & 0.17 & 0.17 & 0.16 & 0.24 & 0.16\\
 & (1.95, 3.40) & (1.93, 3.37) & (1.90, 3.33) & (2.11, 3.37) & (1.98, 3.28)\\
3.5 & 0.23 & 0.22 & 0.21 & 0.28 & 0.21\\
 & (2.63, 5.25) & (2.60, 5.19) & (2.56, 5.09) & (2.90, 5.71) & (2.57, 5.07)\\
4.5 & 0.61 & 0.30 & 0.31 & 0.45 & 0.23\\
 & (3.27, 11.84) & (3.22, 7.84) & (3.16, 7.64) & (3.51, 9.22) & (3.11, 7.20)\\
5.5 & 0.37 & 0.35 & 0.31 & 0.55 & 0.24\\
 & (3.83, 12.11) & (3.80, 11.03) & (3.71, 10.72) & (4.18, 13.67) & (3.58, 9.53)\\
\hline
\end{tabular}
\caption{Posterior MSE and mean 95\% credible interval for the multivariate$t$ with $d=2$ and $n=100$ (top table) and $n=250$ (bottom table). The samples have been drawn from densities with non-integer number of degrees of freedom: $\nu=2.5, 3.5, 4.5, 5.5$. The last column to the right contains the results obtained by applying the loss based prior ($\mbox{LBP}^*$) with the support of half integers:$\{1,1.5,2,2.5,\ldots,29,29.5,30\}$.}
\label{Tab:speciald2}
\end{table}
\begin{table}[h!]
\centering
\begin{tabular}{cccccc}
\hline
DF & AP & JP & RRP & LBP & $\mbox{LBP}^*$ \\
\hline
2.5 & 0.30 & 0.21 & 0.24 & 0.39 & 0.23 \\
 & (1.78, 4.34) & (1.73, 3.88) & (1.71, 3.85) & (2.06, 4.27) & (2.01, 3.65) \\
3.5 & 0.29 & 0.72 & 0.27 & 0.44 & 0.24 \\
 & (2.34, 6.29) & (2.29, 31.91) & (2.21, 5.91) & (2.54, 7.52) & (2.32, 5.69) \\
4.5 & 1.77 & 0.54 & 0.38 & 0.67 & 0.27 \\
 & (2.96, 14.95) & (2.85, 19.66) & (2.76, 17.60) & (3.27, 12.98) & (2.79, 8.14) \\
5.5 & 5.65 & 0.67 & 0.50 & 0.87 & 0.30 \\
 & (3.56, 50.59) & (3.18, 65.56) & (3.28, 22.29) & (3.94, 17.68) & (3.15, 10.71) \\
\hline
\hline
2.5 & 0.15 & 0.15 & 0.15 & 0.22 & 0.20 \\
 & (2.01, 3.27) & (2.01, 3.25) & (1.98, 3.22) & (2.18  3.17) & (2.14, 3.07) \\
3.5 & 0.17 & 0.17 & 0.16 & 0.19 & 0.17 \\
 & (2.73, 4.84) & (2.68, 4.76) & (2.64, 4.70) & (3.00, 4.95) & (2.92, 4.64) \\
4.5 & 0.23 & 0.23 & 0.22 & 0.28 & 0.19 \\
 & (3.40, 6.74) & (3.33, 6.67) & (3.30, 6.63) & (3.58, 7.30) & (3.39, 6.30) \\
5.5 & 0.27 & 0.27 & 0.26 & 0.37 & 0.21 \\
 & (4.06, 9.35) & (4.02, 9.38) & (3.96, 9.32) & (4.31, 10.63) & (4.02, 8.82) \\
\hline
\end{tabular}
\caption{Posterior MSE and mean 95\% credible interval for the multivariate$t$ with $d=3$ and $n=100$ (top table) and $n=250$ (bottom table). The samples have been drawn from densities with non-integer number of degrees of freedom: $\nu=2.5, 3.5, 4.5, 5.5$. The last column to the right contains the results obtained by applying the loss based prior ($\mbox{LBP}^*$) with the support of half integers:$\{1,1.5,2,2.5,\ldots,29,29.5,30\}$.}
\label{Tab:speciald3}
\end{table}

The construction of the loss-based prior is sufficiently flexible to accommodate discretisation of the parameter space for $\nu$ on a denser set (\textit{e.g.}~including non-integer values of $\nu$). By applying the definition of the loss-based prior as given in Section \ref{sec:priors}, and considering the support of the number of degrees of freedom as $\{1,1.5,2,2.5,\ldots,29,29.5,30\}$, we obtain a prior distribution for $\nu$ with the same properties as the prior on the integers. In Tables \ref{Tab:speciald2} and \ref{Tab:speciald3}, we show the simulations results for the non-integer degrees of freedom considered above using the loss-based prior with a support on the half-integers ($\mbox{LBP}^*$). We note that both the MSE and the mean posterior credible intervals give results that are in-line to the ones of the other priors and, in some circumstances, provide better performances.

\subsection{$t$-copula}\label{sec:simulcopula}
For the $t$-copula we have considered the following simulation scenarios. The sample sizes were $n=50$, $n=100$ and $n=250$, while for the correlation coefficient we have chosen $\rho=0.25$, $\rho=0.50$ and $\rho=0.75$. We have limited our study to the bivariate case, \textit{i.e.}~$d=2$, as the extension to any dimension is straightforward (see also Section \ref{sec:discussion}). For the marginals, without loss of generality, we have chosen equal location and scale parameters, that is $\mu_1=\mu_2=0$, $\sigma_1=\sigma_2=1$, and $\nu_1=\nu_2=3$. For the priors on the parameters other than $\nu$, as discussed in Section \ref{sec:otherprios}, we have chosen minimally informative priors. Samples from the posterior distributions are obtained using a MCMC algorithm where continuous parameters are sampled using a Random Walk Metropolis with Normal proposals, while the discrete parameters (the degrees of freedom of the copula and the degrees of freedom of the marginals) are sampled directly using the corresponding posterior probabilities in each iteration. In all the simulation scenarios, $N=500$ posterior samples are obtained using a burn-in period of $1000$ iterations, and a thinning period of $10$ iterations ($6000$ iterations in total).

\begin{figure}[h!]
\begin{center}
\begin{tabular}{c c}
\includegraphics[width=6cm, height=5cm]{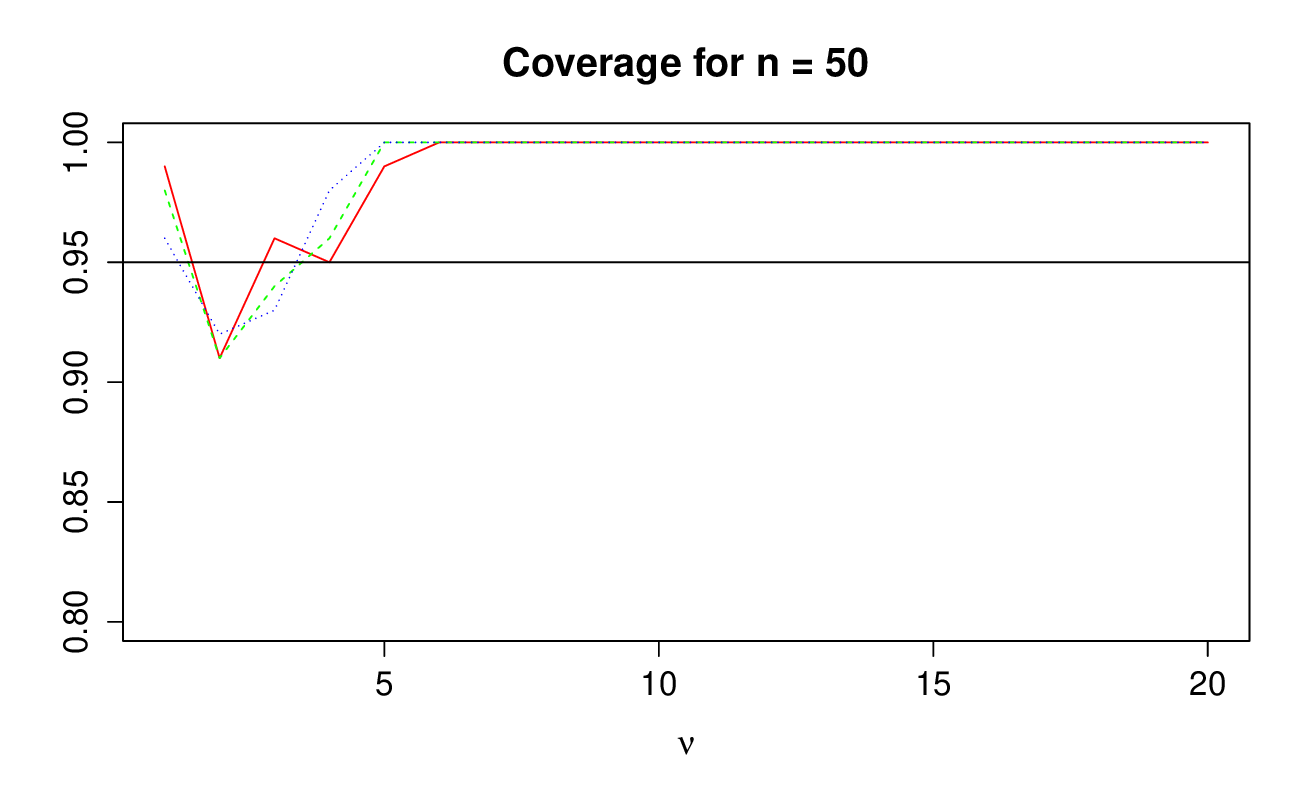}&
\includegraphics[width=6cm, height=5cm]{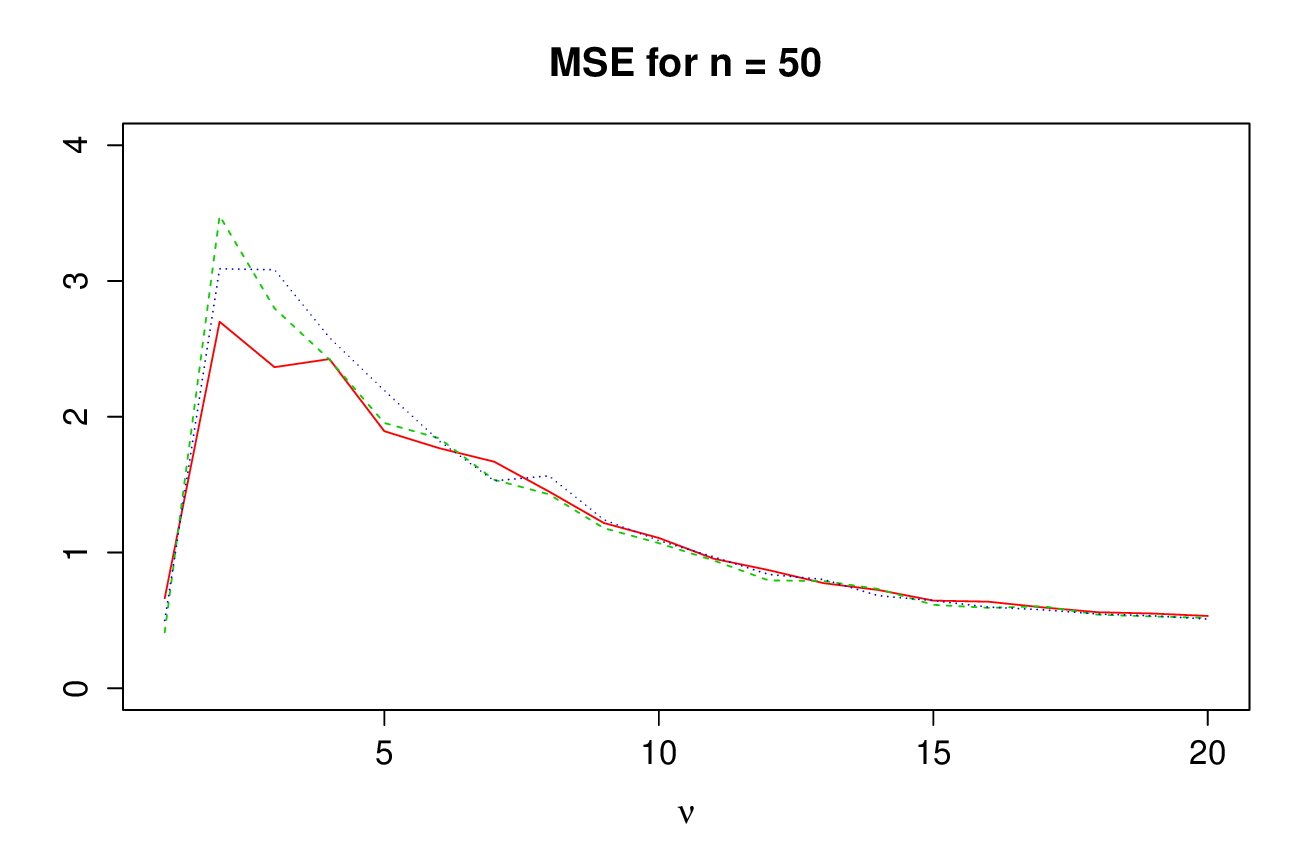}\\
(a) & (b) \\
\includegraphics[width=6cm, height=5cm]{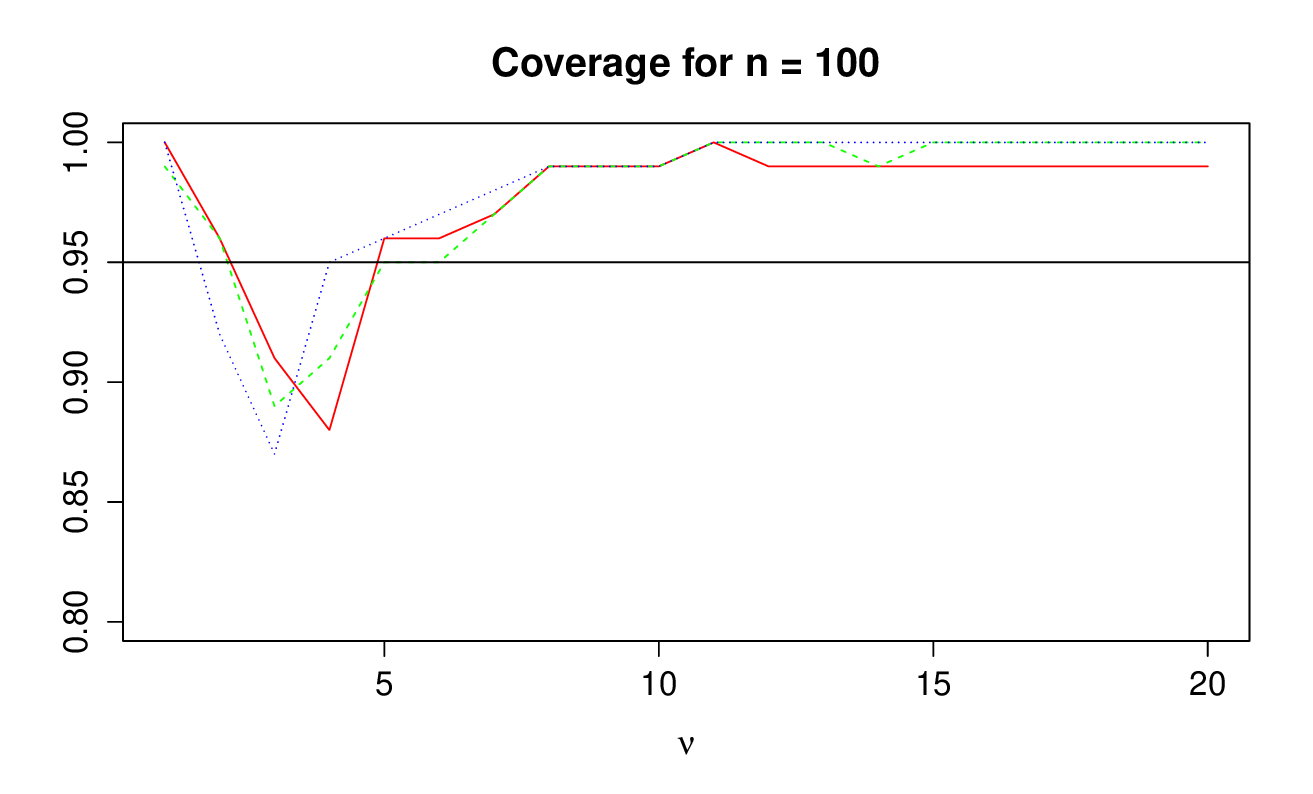}&
\includegraphics[width=6cm, height=5cm]{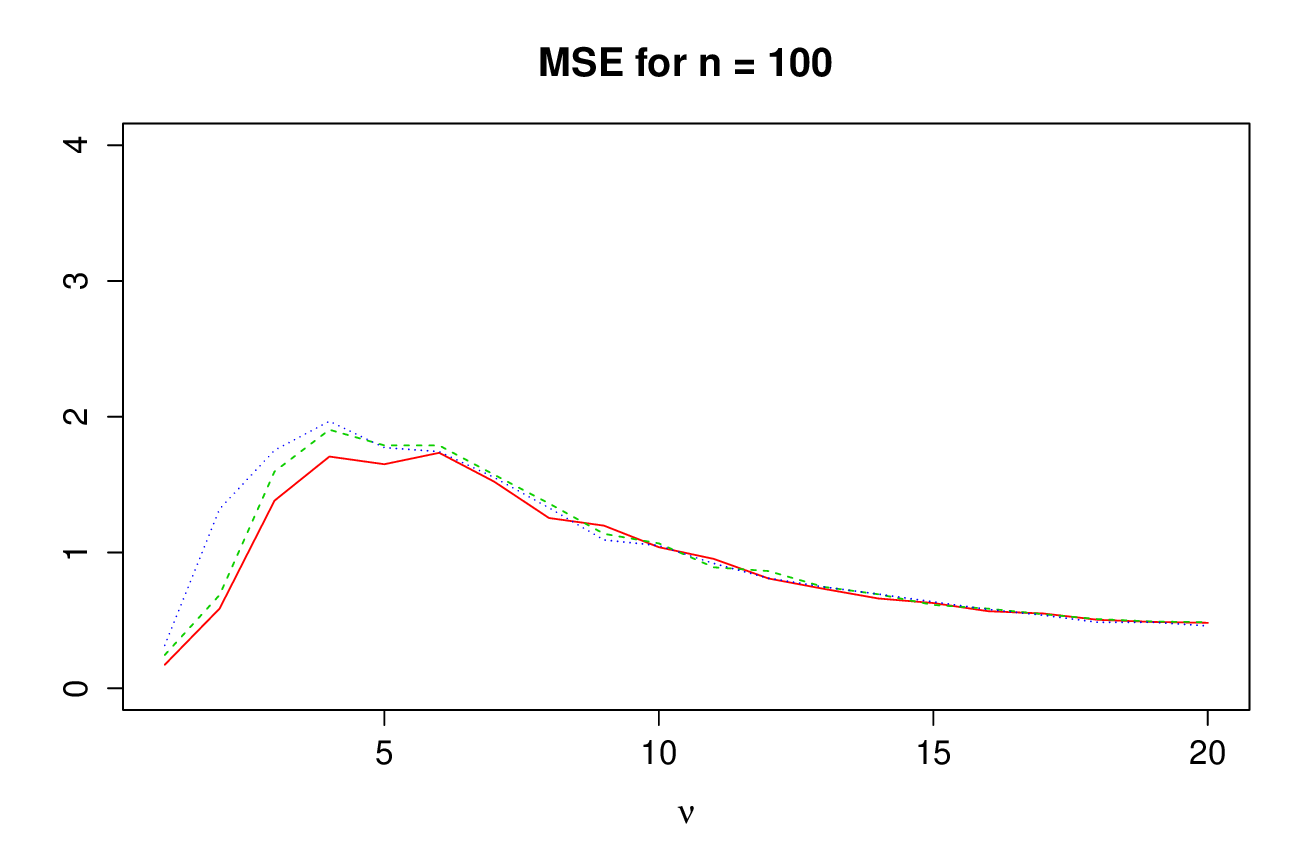}\\
(c) & (d)\\
\includegraphics[width=6cm, height=5cm]{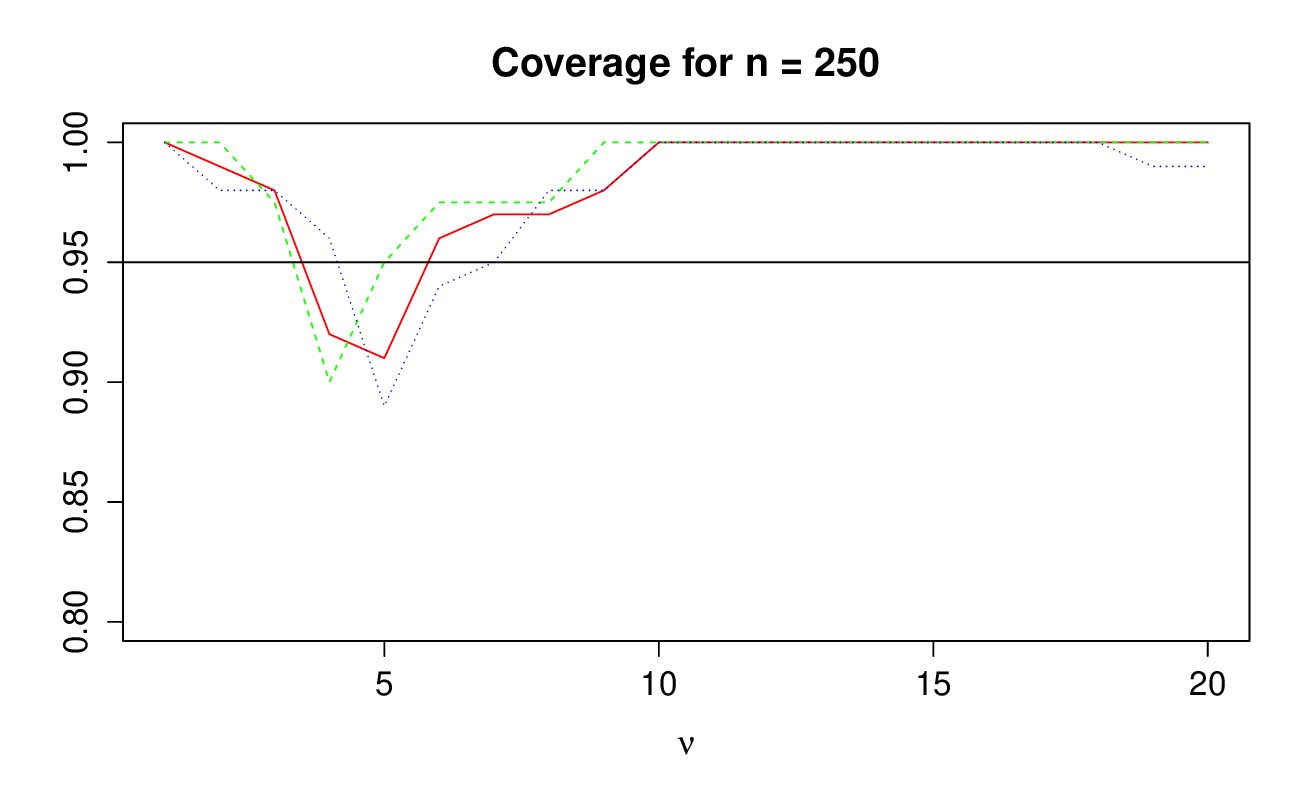}&
\includegraphics[width=6cm, height=5cm]{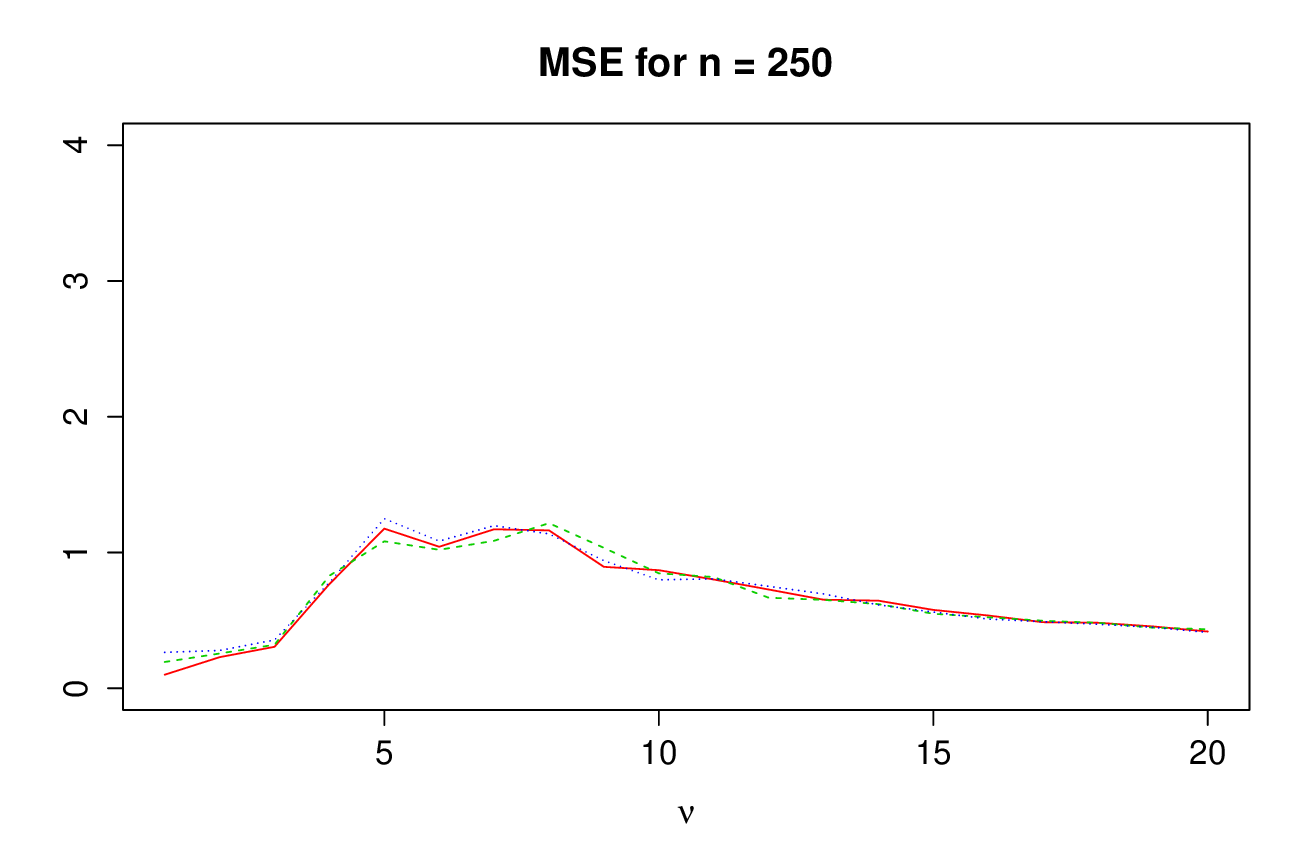}\\
(e) & (f)\\
\end{tabular}
\end{center}
\caption{Frequentist analysis of the $t$-copula: (a)-(b) Coverage and MSE for $n=50$; (c)-(d) Coverage and MSE for $n=100$; (e)-(f) Coverage and MSE for $n=250$. We have considered $\rho=0.25$ (continuous red line), $\rho=0.50$ (dashed green line) and $\rho=0.75$ (dotted blue line).}
\label{fig:copulasim}
\end{figure}

We have then generated 250 \textit{i.i.d.} samples for $\nu=1,\ldots,20$ for each scenario. The results obtained by applying the prior for $\nu$ described in Section \ref{sec:tcoppriors}, are summarised in Figure \ref{fig:copulasim}.
In particular, we note the following. The effect of $\rho$ appears to be minimal, appreciable only in the MSE for $n=50$ and for a number of degrees of freedom between $\nu=3$ and $\nu=5$.  As one would expect, the larger the sample size the higher is the accuracy of the estimate; feature noticeable by inspecting the MSE curves. For what it concerns the coverage, the performance of the loss-based prior is in line with the one for the number of degrees of freedom of a $t$ density, either in the univariate case \citep{VW14} or in the multivariate case (see Section \ref{sec:simulmultit}). In particular, we note a tendency to cover the 100\% of samples for $\nu$ approaching the maximum value, and this is more obvious for relatively small sample sizes. Similarities with the univariate and multivariate case can be seen in the MSE from the median as well. In fact, there is a peak in the relatively lower region of the parameter space, with a curve that rapidly decreases and $\nu$ increases.

Similarly as for the multivariate $t$, we have computed the 95\% confidence intervals of the estimated mean coverage based on the 250 samples per scenario. These intervals are reported in the Appendix C.
\pagebreak

\section{Applications}\label{sec:applications}
In this section we illustrate two financial applications in the context of modelling bivariate daily logarithm returns using the multivariate $t$ distribution and the $t$-copula with Student-$t$ marginals. In the first application, we compare the inference obtained with the proposed loss-based prior for the multivariate $t$ distribution with that of three alternative priors (see Section \ref{sec:priors}). Simulations from the posterior distribution associated to the proposed prior are obtained using an iterative MCMC algorithm (Metropolis within Gibbs) in which we employ a random walk Metropolis for the continuous parameters, using Normal proposal distributions, while the posterior of the degrees of freedom parameter (which are discrete and bounded) are directly sampled using their corresponding probabilities. The variance of the Normal proposals are chosen in order to obtain around 30\% acceptance rates. For the three alternative models, we employ the t-walk algorithm \citep{CF10}, which is implemented in the R package `Rtwalk'. In the second application, which illustrates the use of the proposed loss-based prior for the $t$-copula, simulations from the posterior distribution are again obtained using an iterative MCMC method composed by a random walk Metropolis for the continuous parameters and direct sampling for the discrete parameters. For each of these models, we obtained $N=5000$ samples from the posterior distribution after a burn-in period of $5000$ iterations and a thinning period of $50$ iterations (this is, $255000$ iterations in total). This configuration produced stable traceplots of the MCMC posterior samples and the log-posterior. R codes used here are available under request.

\subsection{Multivariate $t$: Bivariate log-returns}\label{example:multivariatet}
\begin{figure}[h!]
\begin{center}
\begin{tabular}{c c}
\includegraphics[width=6cm, height=5cm]{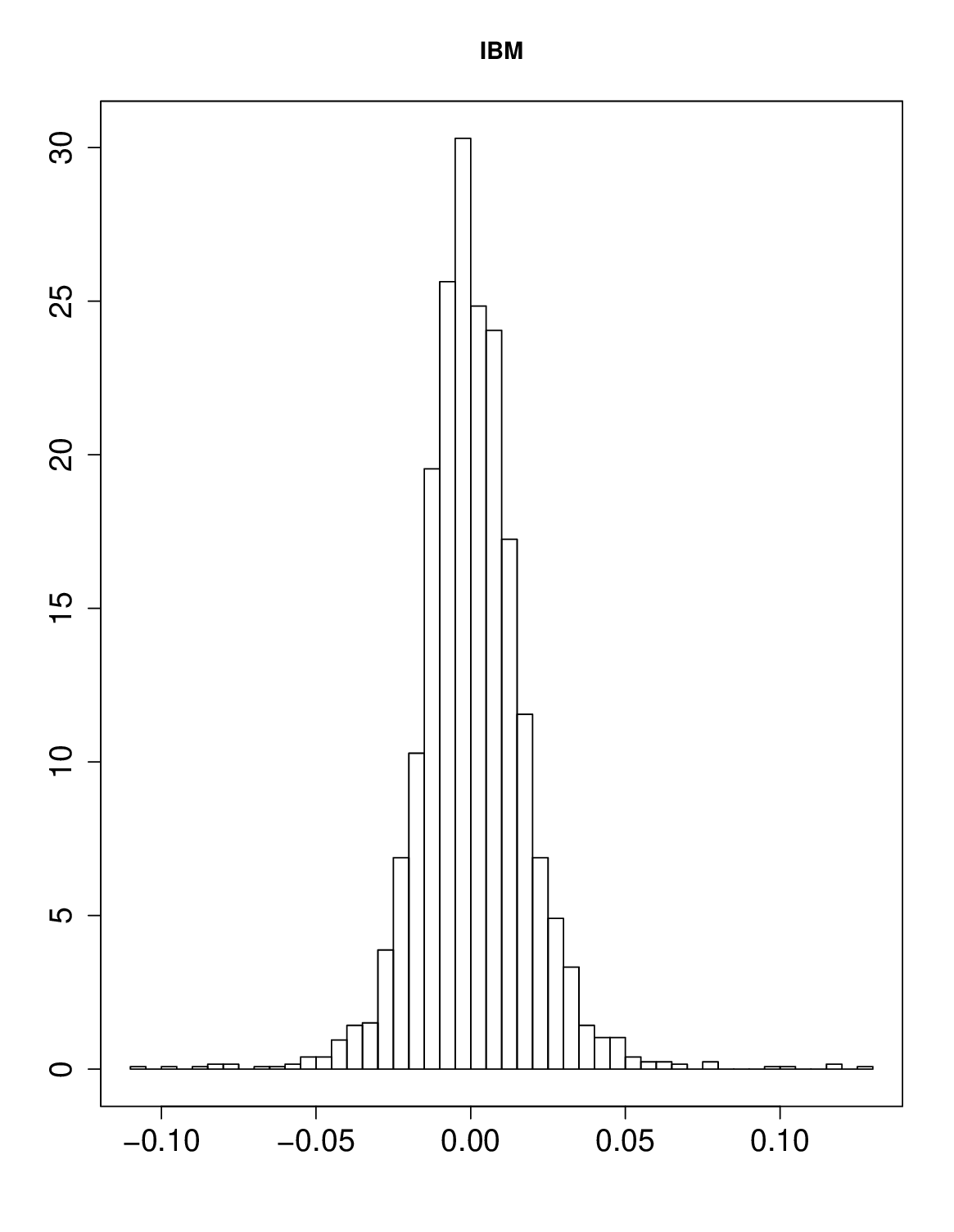}&
\includegraphics[width=6cm, height=5cm]{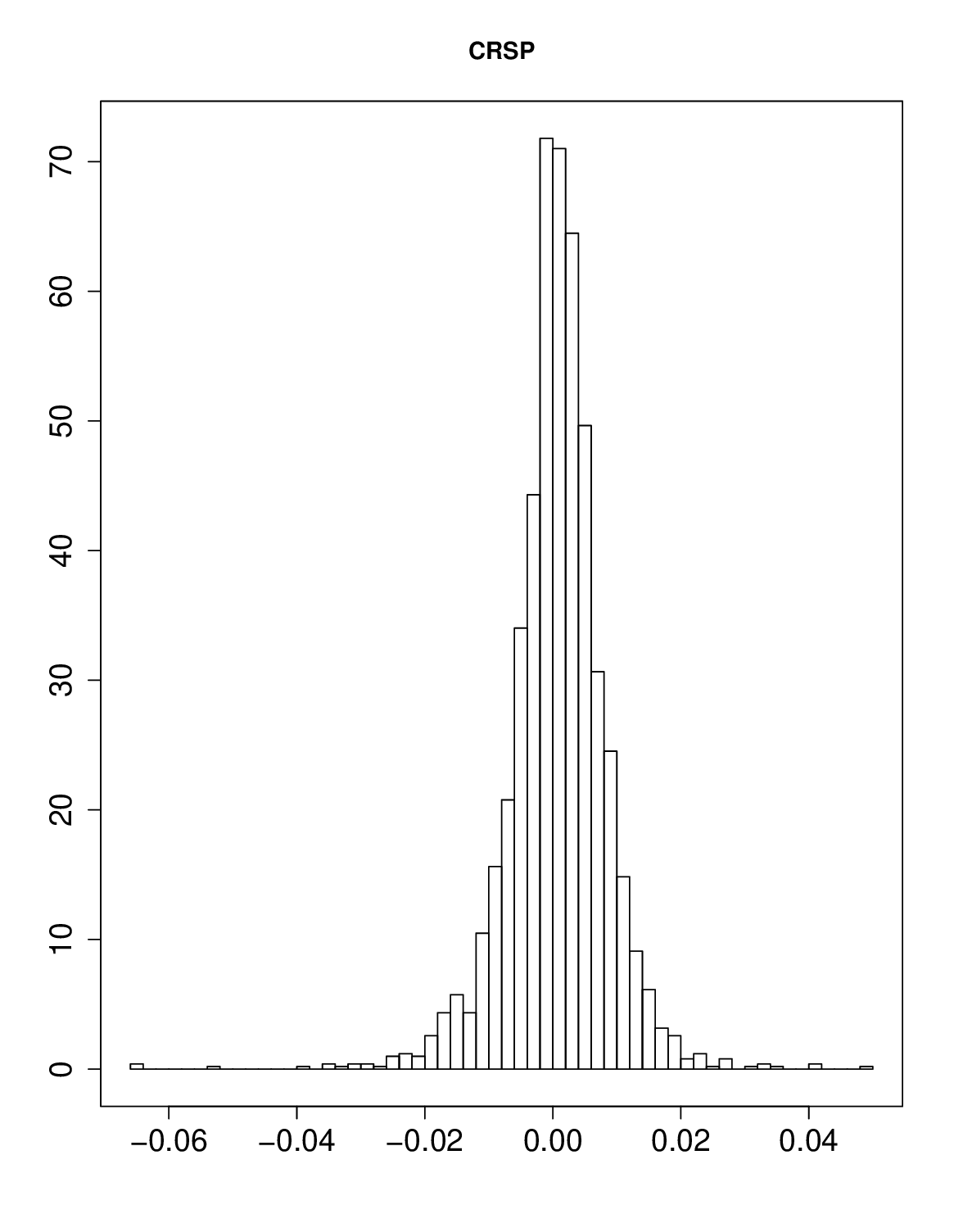}\\
(a) & (b) \\
\includegraphics[width=6cm, height=5cm]{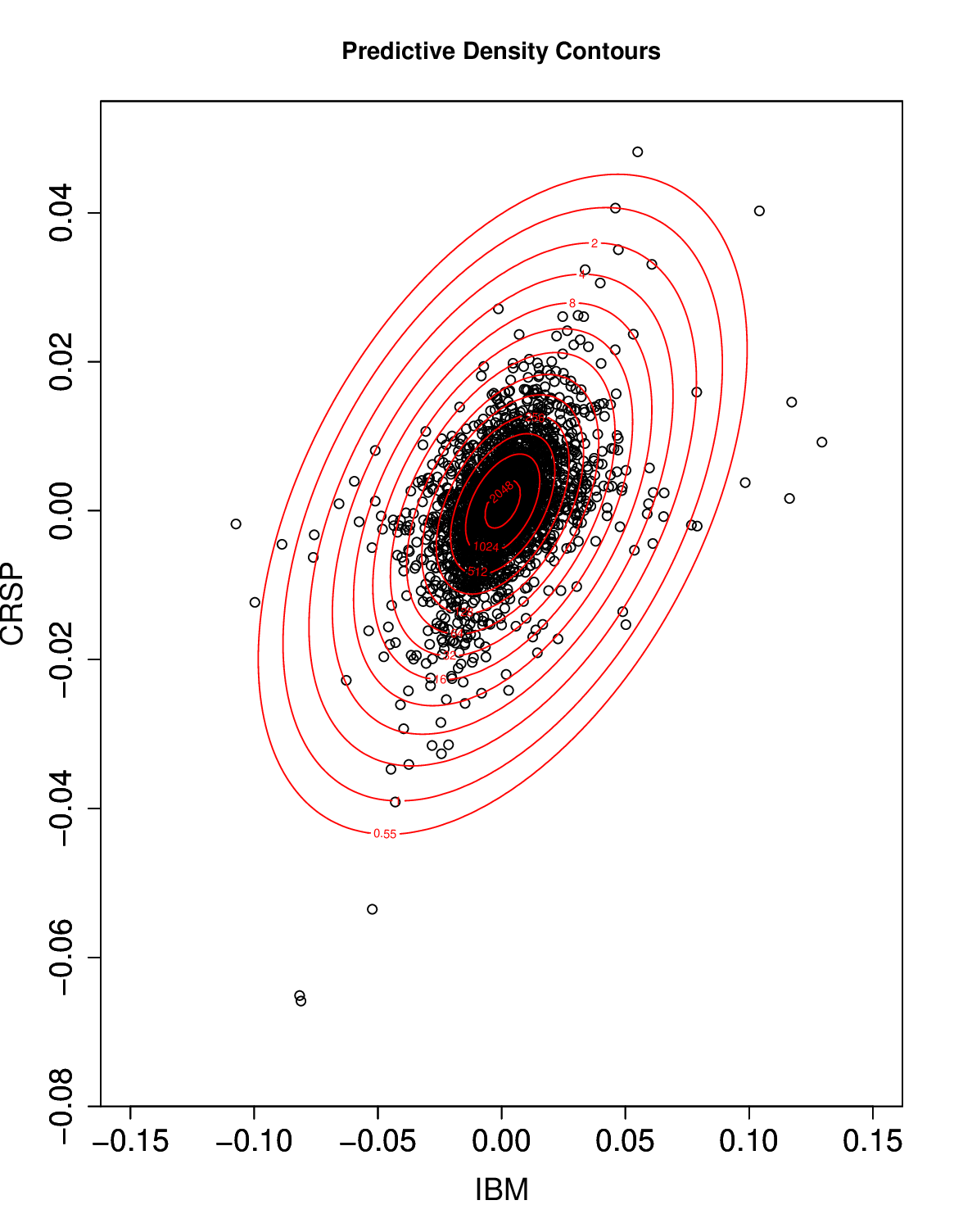} & \\
(c) &
\end{tabular}
\end{center}
\caption{ IBM returns \emph{vs.} CRSP returns data: (a) Histogram of IBM data; (b) Histogram of CRSP data; and (c) Predictive contour plots associated to the LBP and levels = (0.55,1,2,4,8,16,32,64,128,256,512,1024,2048).}
\label{fig:PredLr}
\end{figure}

We present an application of the bivariate $t$ distribution in the context of modelling daily log-returns from the Center for Research in Security Prices (CRSP) Database. The data contains $n=2528$ observations corresponding to the daily log-returns of IBM (Permno 12490) and CRSP (the return for the CRSP value-weighted index, including dividends) of the period from the 3\textsuperscript{rd} of January 1969 to the 31\textsuperscript{st} of December 1998. The data are available from the `Ecdat' R package \citep{Ecdat} and has been analysed using a bivariate $t$ distribution, using likelihood estimation, in \cite{R11}. We analyse these data using also a bivariate $t$ distribution in a Bayesian framework. We adopt the prior structure:
 \begin{eqnarray*}
\pi(\bm{\mu},\bm{\Sigma},\nu) = \dfrac{1}{\vert \bm{\Sigma}\vert^{\frac{3}{2}}}\pi(\nu),
 \end{eqnarray*}
where $\pi(\nu)$ represents the objective prior on the degrees of freedom of the bivariate $t$ distribution proposed in Section \ref{sec:tprior}. Table \ref{table:MVtExample} shows the maximum likelihood estimators (MLE) of the parameters as well as the posterior median estimators associated to the 4 priors choices: the loss-based prior (LBP), the Anscombe prior (AP), the Jeffreys prior (JP) and the Relles \& Rogers prior (RRP). This table also presents the 95\% Bootstrap confidence intervals (based on 1000 Bootstrap samples) and the 95\% credible intervals associated to each model. The maximum a posteriori (MAP) is reported for $\nu$ in the LBP case. In this example we obtained similar estimators with all the different approaches due to the large sample size. The fit of the predictive distribution associated to the LBP is illustrated in Figure \ref{fig:PredLr} for different contour plot levels.
\begin{table}[ht]
\begin{center}
$\begin{array}{cccccc}
\hline
\text{Parameter} & \text{MLE}  & \text{LBP} & \text{AP} & \text{JP} & \text{RRP}\\
\hline
\mu_1 & 5.00 & 4.33 & 4.34 & 4.34 & 4.35 \\
 \times 10^{-4}   & (-1.04,10.6)  &  (-1.28,10.0) &  (-1.48,10.1) & (-1.87,10.3)  & (-1.42,10.3)  \\
\mu_2 & 8.41 & 8.54 & 8.58 & 8.47 & 8.58\\
 \times 10^{-4}   &  (6.10,11.3)  & (6.01,11.1)  & (5.69,11.2)  & (5.60,11.4)  & (5.70,11.3)  \\
\sigma_1^2 & 1.58 & 1.54 & 1.56 & 1.56 & 1.56 \\
 \times 10^{-4}   &  (1.44,1.69)  & (1.44,1.66)  & (1.43,1.71)  & (1.43,1.71)  & (1.43,1.70)  \\
\sigma_2^2 & 3.15 & 3.11 & 3.14 & 3.15 & 3.14\\
  \times 10^{-5}  &  (2.85,3.51)  & (2.83,3.42)  & (2.79,3.55)  & (2.81,3.55)  & (2.79,3.54)  \\
\sigma_{12} & 3.34 & 3.26 & 3.29 & 3.31 & 3.30 \\
\times 10^{-5}   &  (2.90,3.78)  & (2.87,3.70)  & (2.82,3.83)  & (2.84,3.82)  & (2.82,3.81)  \\
\nu & 4.19 & 4 &  4.12 & 4.15 & 4.12 \\
    &  (3.75,4.71)  & \{4\}  & (3.65,4.69)  & (3.66,4.70)  & (3.65,4.72)  \\
\hline
\end{array}$
\caption{ IBM returns \emph{vs.} CRSP returns data: MLE, 95\% Bootstrap intervals, Bayesian estimators, and 95\% credible intervals.}
\label{table:MVtExample}
\end{center}
\end{table}

%

\pagebreak
\subsection{$t$-copula: Bivariate log-returns}\label{example:copula}
\begin{figure}[h!]
\begin{center}
\begin{tabular}{c c}
\includegraphics[width=6cm, height=5cm]{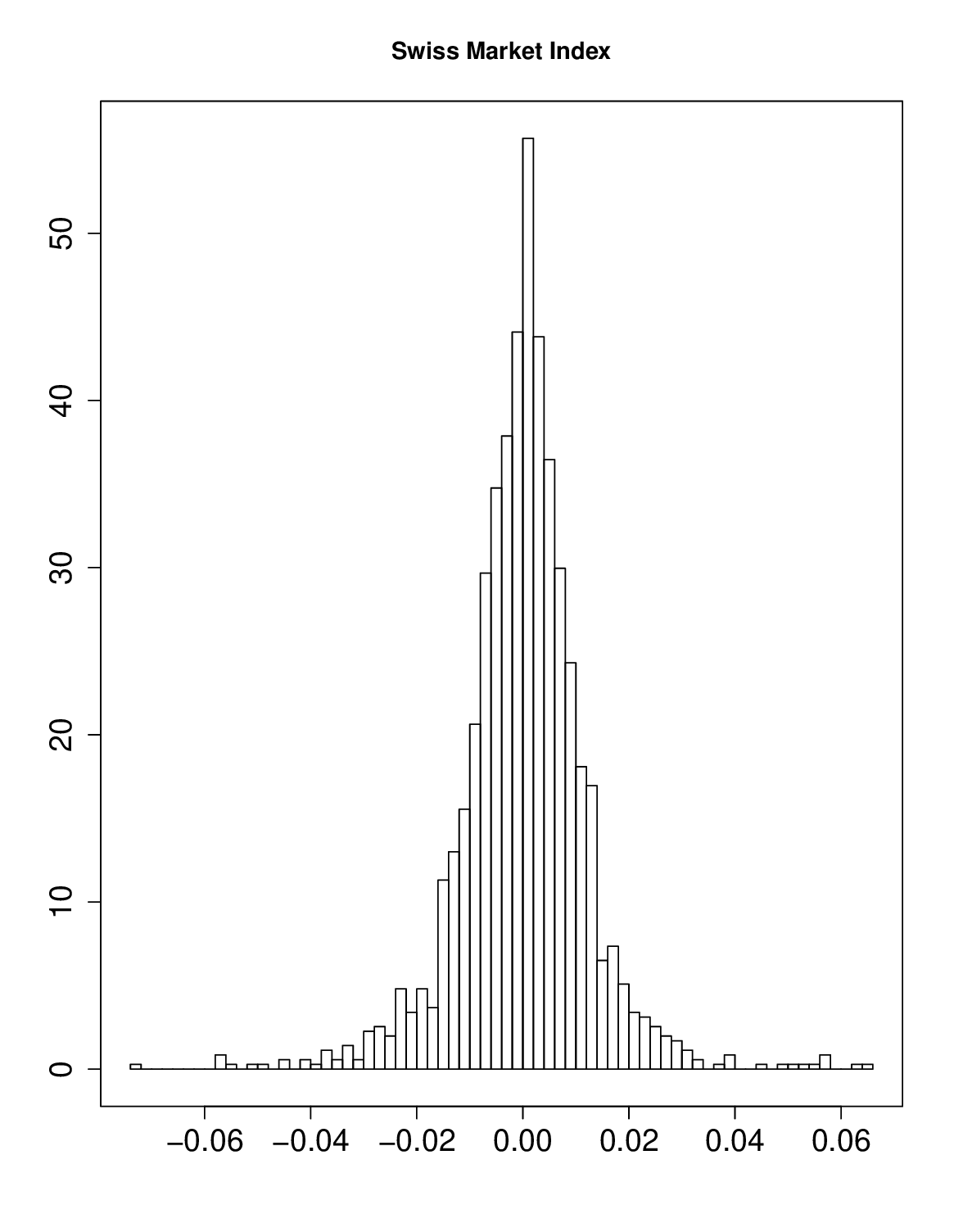}&
\includegraphics[width=6cm, height=5cm]{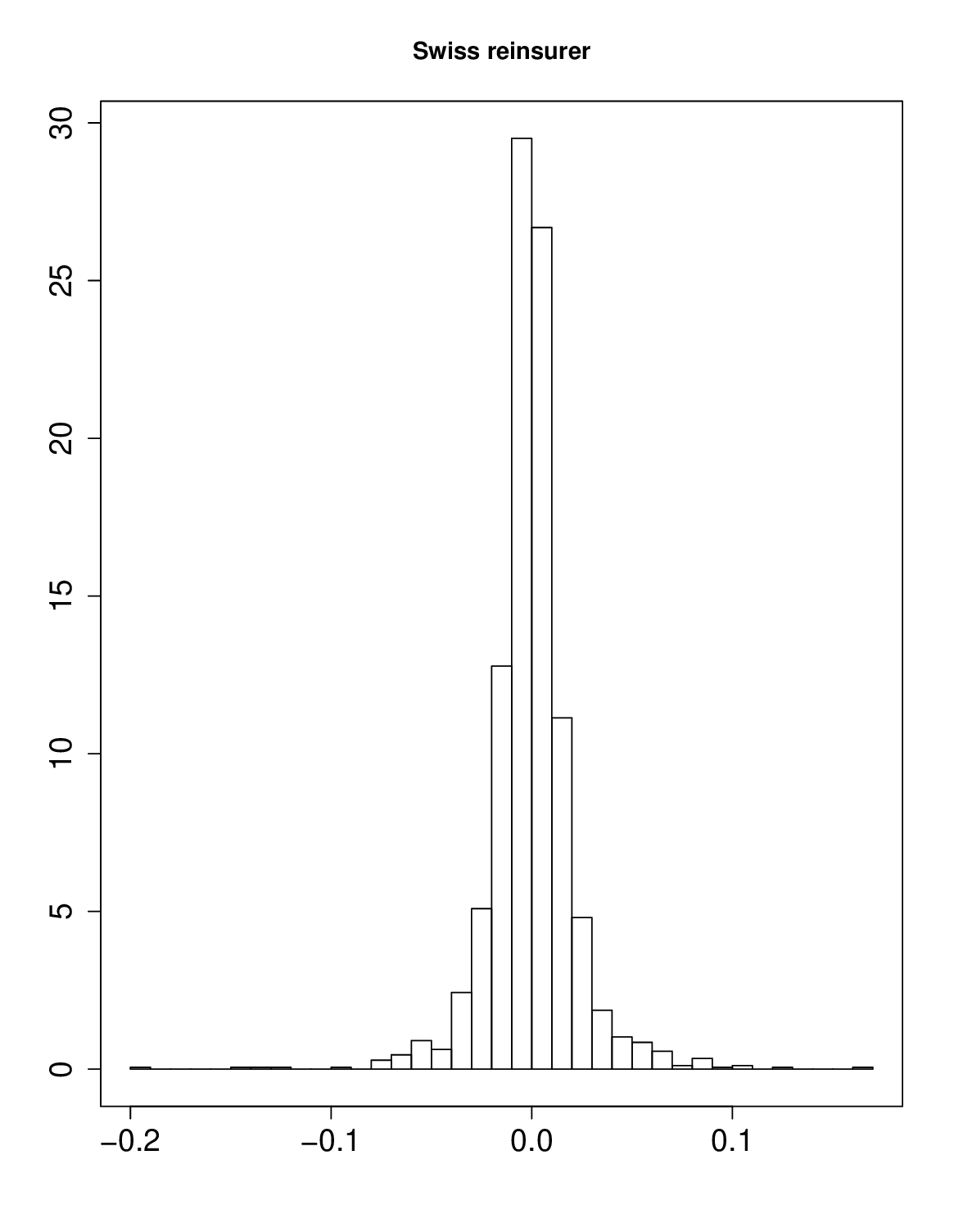}\\
(a) & (b) \\
\includegraphics[width=6cm, height=5cm]{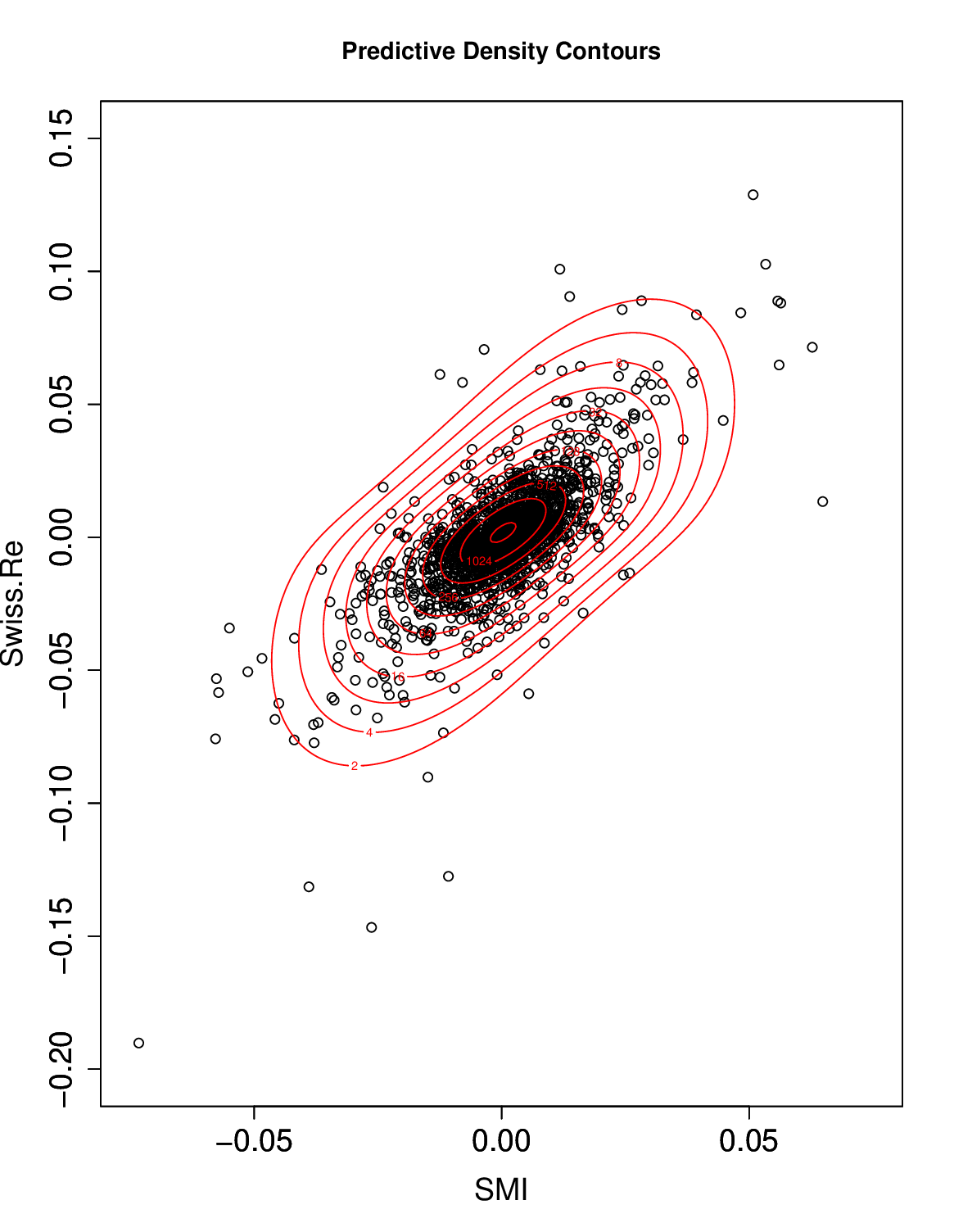} & \\
(c) &
\end{tabular}
\end{center}
\caption{ Swiss Market Index vs. Swiss reinsurer data: (a) Histogram of Swiss Market Index data; (b) Histogram of Swiss reinsurer data data; and (c) Predictive contour plots associated to the LBP and levels = (2,4,8,16,32,64,128,256,512,1024,2048)).}
\label{fig:PredSwiss}
\end{figure}

We model jointly the daily log-returns for the Swiss Market Index (SMI) and Swiss reinsurer (Swiss.Re). The data are available from the R package `ghyp' \citep{ghyp} and contain $n=1769$ observations corresponding to the period January 2000 to January 2007. We model these data using a bivariate $t$-copula with Student-$t$ marginals. This model can capture heavy tails of the marginals as well as tail dependence \citep{DM05}. We adopt the following prior structure, as introduced in Section \ref{sec:otherprios}:
 \begin{eqnarray*}
\pi(\mu_1,\mu_2,\sigma_1,\sigma_2,\nu_1,\nu_2,\nu,\rho) = \pi(\mu_1)\pi(\mu_2)\pi(\sigma_1)\pi(\sigma_2)\pi(\nu_1)\pi(\nu_2)\pi(\nu,\rho),
 \end{eqnarray*}
where $\pi(\mu_j)$, $j=1,2$, are Normal densities with mean zero and scale parameter $100$; $\pi(\sigma_j)$ are Cauchy densities (which reflect vague prior information, see \citealp{RS15});  $\pi(\nu_j)$ are the objective (loss-based) priors proposed in \cite{VW14}; and the joint prior $\pi(\nu,\rho)$ is decomposed as $\pi(\nu\vert\rho)\pi(\rho)$, where $\pi(\nu\vert\rho)$ is the LBP proposed in \ref{sec:tcoppriors} and $\pi(\rho)$ is a Beta density (on $(1+\rho)/2$) with shape parameters $(1/2,1/2)$. In order to simplify the implementation, we use $\pi(\nu\vert\rho) = \pi(\nu\vert\rho=0)$ as discussed in Section \ref{sec:tcoppriors}. Table \ref{table:BtCExample} shows the MLE of the parameters as well as the posterior median estimators associated to this prior structure. The MAP is reported for $\nu$. This table also presents the 95\% Bootstrap confidence intervals (based on 1000 Bootstrap samples) and the 95\% credible intervals. Figure \ref{fig:PredSwiss} illustrates the fit of the predictive contour plots.

In order to quantify the dependence between the marginals, we employ the coefficient of tail dependence and the Kendall's $\tau$ Rank Correlation, which are respectively given by \citep{DM05}:
\begin{eqnarray*}
\lambda &=& 2t_{\nu+1}\left(-\sqrt{\nu+1}\sqrt{1-\rho}/\sqrt{1+\rho}\right),\\
\tau &=& \dfrac{2}{\pi}\operatorname{arcsin}(\rho).
\end{eqnarray*}
The estimators of $\lambda$ and $\tau$ (reported in Table \ref{table:BtCExample}) indicate tail dependence of the marginals.
\begin{table}[ht]
\begin{center}
$\begin{array}{ccc}
\hline
\text{Parameter} & \text{MLE}  & \text{LBP}\\
\hline
\mu_1 & 3.16 & 3.17 \\
\times10^{-4} & (-1.39,7.39)  & (-1.18,7.53) \\
\mu_2 & -2.18 & -2.14\\
\times10^{-4} & (-8.85,4.29)  & (-8.11,4.32) \\
\sigma_1 & 7.94  & 8.13  \\
\times10^{-3} & (7.49,8.41)  & (7.80,8.48) \\
\sigma_2 & 1.12  & 1.18  \\
\times10^{-2} & (1.05,1.20)  & (1.13,1.23) \\
\rho & 0.69 & 0.69  \\
  & (0.66,0.72)  & (0.66,0.71) \\
\nu_1 & 3.45  & 4\\
 & (3.03,4.05)  &  \{4\} \\
\nu_2 & 2.52  & 3 \\
  & (2.26,2.85)  & \{3\} \\
\nu & 3.93  & 4\\
 & (3.12,5.34)  &  \{4,5,6\} \\
\lambda & 0.38  &  0.36 \\
 & (0.33,0.43)  &  (0.30,0.40) \\
\tau & 0.48  & 0.49 \\
 & (0.46,0.52)  &  (0.46,0.51) \\
\hline
\end{array}$
\caption{Swiss Market Index vs. Swiss reinsurer data: MLE, 95\% Bootstrap intervals, Bayesian estimators, and 95\% credible intervals.}
\label{table:BtCExample}
\end{center}
\end{table}

\section{Discussion}\label{sec:discussion}

The multivariate $t$ distribution and the $t$-copula are models of great importance in financial applications, among other areas. The multivariate $t$ distribution is typically used as a robust model to capture departures from normality in terms of heavy tails (outliers), while the $t$-copula is often employed to construct multivariate models that can capture a wider range of tail-dependence than that of the Normal copula \citep{Eetal01}.

We have proposed noninformative priors for the degrees of freedom in the multivariate $t$ distribution and the $t$-copula. These priors are built upon an objective criterion based on loss functions previously proposed in \cite{VW14}, and further generalised in \cite{VW15}. Thus, our work extends the prior proposed in \cite{VW14}, for the univariate $t$ distribution, to the multivariate case, while it represents the first objective prior for the degrees of freedom of the $t$-copula, to the best of our knowledge. Our simulation studies illustrate the good frequentist performance of the posterior distribution associated to the proposed objective priors. They also show that the posterior distribution associated to these priors is easier to sample from (due to the truncated and discrete nature of the prior), and lead to sensible inferences. For what it concerns the multivariate $t$ distribution, we have compared the frequentist properties of the proposed prior to three alternative options presented in literature. Overall, the loss-based prior appears to give better results, in particular for the larger dimension considered. Furthermore, its performance is more stable, in particular for relatively large values of $\nu$.

Although we have focused on low-dimensional scenarios in our applications and simulations, the extension of the proposed prior distributions to higher dimensions is immediate. For the multivariate $t$ case, the construction of the prior is scalable since it only requires unidimensional integration, regardless of the dimension of the multivariate $t$ distribution. The construction of the prior for the $t$-copula in higher dimensions is slightly more challenging. However, in the context of copula modelling, it has been largely advocated the use of the pair-copula decomposition, rather than a direct use of a multivariate copula, as a means to model complex patterns of tail dependence \citep{A04}. The pair-copula decomposition is used to construct multivariate distributions based on bivariate copulas associated to pairs of variables. Since we have fully addressed the construction of priors for the bivariate $t$-copula, our results may serve as a framework for modelling data in higher dimensions via the pair-copula construction. Alternatively, \cite{KKC17} discuss a few methods to compute the Kullback--Leibler divergence between copulas of relatively large dimensions, which result to be computationally efficient.

In the real data example presented in Section \ref{example:copula}, we have employed symmetric Student-$t$ marginals since they were appropriate in our context. However, given that the proposed prior does not depend on the choice of the marginals, it is possible to employ more flexible marginal distributions, such as the two-piece Student-$t$ (see \citealp{RS15} for an extensive discussion of the family of two-piece distributions), in order to capture skewness and heavy tails. \cite{L17} proposed an objective prior for the degrees of freedom parameter in the univariate two-piece Student-$t$ distribution, which is constructed using the loss-based principle discussed in Section \ref{sec:priors}. They show that this prior does not depend on the skewness parameter, and that it coincides with that proposed in \cite{VW14} for the univariate Student-$t$ distribution (see Section \ref{sec:priors}). For the skewness parameter, \cite{L17} employ the noninformative prior proposed in \cite{RS14}. Thus, the Bayesian model applied in Section \ref{example:copula} can be easily extended to capture skewness on the marginals by using these ideas.

\section*{Acknowledgements}
We thank two referees, an Associate Editor, and the Editor for helpful comments. Cristiano Villa was supported by the Royal Society's Research Grant RG150786.

\newpage
\section*{Appendix}
\section*{Appendix A - Sampling algorithms}\label{sec:appendixA}
In this appendix we describe the MCMC strategy to make approximate posterior inference for the case where the loss-based prior is employed. The algorithms have been implemented in R and run on a workstation with two Intel(R) Xeon(R) CPU with 2.10 GHz and 32 cores each.

The algorithms used to sample from the posterior distributions for the multivariate $t$ model and the $t$-copula model (see Section \ref{sec:posterior}) adopt the same strategy. In particular, the posterior sample for the number of degrees of freedom is obtained by direct sampling, while the posterior sample for the remaining parameters by means of a random walk Metropolis. We may remark that one can employ other samplers, such as Gibbs or Metropolis-Gibbs samplers instead.

\subsection*{Multivariate $t$}
For the multivariate $t$ model the parameters are the number of degrees of freedom $\nu$, the location vector $\boldsymbol\mu$ and the scale matrix $\boldsymbol\Sigma$. To illustrate, let us consider the case $d=2$ (the generalisation to any dimension is conceptually straightforward). If at iteration $i$ the chain is in position $\boldsymbol\theta^{(i)}=\left(\nu^{(i)},\mu_1^{(i)}\mu_2^{(i)},\sigma_1^{(i)},\sigma_2^{(i)},\sigma_{12}^{(i)}\right)$, then iteration $(i+1)$ is generated as follows.

\begin{algorithm}[h!]
\begin{algorithmic}[1]
\STATE Sampling $\nu$ -- Given that the parameter space for the number of degrees of freedom is finite, and bounded above by $\nu_{\max}$, we proceed as follows:
	\begin{itemize}
	\item compute $\pi(\nu_l^{(i)}\mid\mathbf{x},\boldsymbol{\mu},\boldsymbol{\Sigma})$, for $l=1,\ldots,\nu_{\max}$
	\item sample $\nu^{(i+1)}\sim \pi(\nu_l^{(i)}\mid\mathbf{x},\boldsymbol{\mu},\boldsymbol{\Sigma})$
	\end{itemize}
\STATE Sampling $\mu_j$, with $j=1,2$ -- Defining the proposed values as $\mu_j^*$, we have
	\begin{itemize}
	\item $\mu_j^* \sim N(\mu_j^{(i)},\phi)$, where $\phi$ is calibrated to get the desired acceptance rate
	\item $\mu_j^{(i+1)}=\mu_j^*$ with probability
	$$\alpha_{\mu_j} = \min\left\{1,\frac{\pi(\mu_j^*\mid\mathbf{x})}{\pi(\mu_j^{(i)}\mid\mathbf{x})}\right\}$$
	\end{itemize}
\STATE Sampling $\sigma_j$, with $j=\{1,2,12\}$ -- Defining the proposed values as $\sigma_j^*$, we have
	\begin{itemize}
	\item $\sigma_j^* \sim TN(\sigma_j^{(i)},\phi^\prime)$, where $TN$ is a truncated normal so to generate positive values only and $\phi^\prime$ is calibrated to get the desired acceptance rate
	\item $\sigma_j^{(i+1)}=\sigma_j^*$ with probability
	$$\alpha_{\sigma_j} = \min\left\{1,\frac{\pi(\sigma_j^*\mid\mathbf{x})}{\pi(\sigma_j^{(i)}\mid\mathbf{x})}\frac{TN(\sigma_j^{(i)},\phi^\prime)}{TN(\sigma_j^*,\phi^\prime)}\right\}$$
	\end{itemize}
\end{algorithmic}
\caption{Posterior sampler: Multivariate $t$ distribution.}\label{alg:MVT}
\end{algorithm}


\subsection*{$t$-copula}
The algorithm to sample from the $t$-copula is similar to the previous one. Let us assume that $d=2$; then, if the chain is $\boldsymbol\theta^{(i)}=\left(\nu_1^{(i)},\nu_2^{(i)},\nu^{(i)},\mu_1^{(i)}\mu_2^{(i)},\sigma_1^{(i)},\sigma_2^{(i)},\rho^{(i)}\right)$. The samples from the posteriors of the number of degrees of freedom, location and scale parameters for the marginal are sampled as in the algorithm for the multivariate $t$. For all parameters the iteration $(i+1)$ is generated as follows.

\begin{algorithm}[h!]
\begin{algorithmic}[1]
\STATE Sampling $\nu_1$, $\nu_2$ and $\nu$ -- Given that the parameter space for the number of degrees of freedom is finite, and bounded above by $\nu_{\max}$, we proceed as follows:
	\begin{itemize}
	\item compute $\pi(\nu_l^{(i)}\mid\mathbf{x},\mathbf{R})$, for $l=1,\ldots,\nu_{\max}$
	\item sample $\nu^{(i+1)}\sim\pi(\nu_l^{(i)}\mid\mathbf{x},\mathbf{R})$
	\end{itemize}
\STATE Sampling $\mu_j$, with $j=1,2$ -- Defining the proposed values as $\mu_j^*$, we have
	\begin{itemize}
	\item $\mu_j^* \sim N(\mu_j^{(i)},\phi)$, where $\phi$ is calibrated to get the desired acceptance rate
	\item $\mu_j^{(i+1)}=\mu_j^*$ with probability
	$$\alpha_{\mu_j} = \min\left\{1,\frac{\pi(\mu_j^*\mid\mathbf{x})}{\pi(\mu_j^{(i)}\mid\mathbf{x})}\right\}$$
	\end{itemize}
\STATE Sampling $\sigma_j$, with $j=\{1,2,12\}$ -- Defining the proposed values as $\sigma_j^*$, we have
	\begin{itemize}
	\item $\sigma_j^* \sim TN(\sigma_j^{(i)},\phi^\prime)$, where $TN$ is a truncated normal so to generate positive values only and $\phi^\prime$ is calibrated to get the desired acceptance rate
	\item $\sigma_j^{(i+1)}=\sigma_j^*$ with probability
	$$\alpha_{\sigma_j} = \min\left\{1,\frac{\pi(\sigma_j^*\mid\mathbf{x})}{\pi(\sigma_j^{(i)}\mid\mathbf{x})}\frac{TN(\sigma_j^{(i)},\phi^\prime)}{TN(\sigma_j^*,\phi^\prime)}\right\}$$
	\end{itemize}
\STATE Sampling $\rho$ -- We sample from a normal truncated in the interval $[-1,1]$:
	\begin{itemize}
	\item $\rho^* \sim TN(\rho^{(i)},\phi^{\prime\prime)}$, where $TN$ is a truncated normal and $\phi^{\prime\prime}$ is calibrated to get the desired acceptance rate
\item $\rho^{(i+1)}=\rho^*$ with probability
	$$\alpha_{\rho} = \min\left\{1,\frac{\pi(\rho^*\mid\mathbf{x})}{\pi(\rho^{(i)}\mid\mathbf{x})}\frac{TN(\rho^{(i)},\phi^{\prime\prime})}{TN(\rho^*,\phi^{\prime\prime})}\right\}$$
	\end{itemize}
\end{algorithmic}
\caption{Posterior sampler: $t$-copula.}\label{alg:MVTC}
\end{algorithm}



\section*{Appendix B - Mean credible intervals for the multivariate $t$}\label{sec:appendixB}
In this appendix we report the mean 95\% credible intervals for the simulation study of the multivariate $t$. In particular, we report the interval and its width based on the 250 samples considered under each simulation scenario. For the multivariate $t$ with $d=2$, we have Table \ref{tab:meanCId2n50} for the case $n=50$, Table \ref{tab:meanCId2n100} for the case $n=100$ and Table \ref{tab:meanCId2n250} for $n=250$. Tables \ref{tab:meanCId3n50}, \ref{tab:meanCId3n100} and \ref{tab:meanCId3n250} report the results for the multivariate $t$ with $d=3$ and, respectively, $n=50,100,250$.
\begin{table}[h!]
\centering
\footnotesize
\begin{tabular}{ccccccccc}
\hline 
 & \multicolumn{2}{c}{AP} & \multicolumn{2}{c}{JP} & \multicolumn{2}{c}{RRP} & \multicolumn{2}{c}{LBP} \\ 
$\nu$ & Width & C.I. & Width & C.I. & Width & C.I. & Width & C.I. \\
\hline
1 & 0.74 & (1.01, 1.76) & 0.90 & (0.65, 1.55) & 0.62 & (1.01, 1.64) & 0.38 & (1.00, 1.38) \\
2 & 18.53 & (1.25, 19.78) & 3.49 & (1.19, 4.68) & 3.06 & (1.21, 4.28) & 4.97 & (1.48, 6.44) \\
3 & 90.51 & (1.71, 92.22) & 12.14 & (1.61, 13.75) & 7.74 & (1.53, 9.26) & 12.40 & (2.11, 14.52) \\
4 & 175.43 & (2.07, 177.50) & 25.19 & (1.98, 27.17) & 18.30 & (1.85, 20.15) & 17.73 & (2.72, 20.45) \\
5 & 281.51 & (2.50, 284.01) & 31.72 & (2.21, 33.93) & 30.63 & (2.03, 32.65) & 21.07 & (3.03, 24.10) \\
6 & 339.30 & (2.70, 341.99) & 55.81 & (2.47, 58.28) & 33.52 & (2.23, 35.75) & 22.80 & (3.49, 26.29) \\
7 & 380.55 & (2.92, 383.47) & 65.63 & (2.64, 68.27) & 50.44 & (2.38, 52.81) & 23.61 & (3.83, 27.44) \\
8 & 428.15 & (3.13, 431.27) & 95.90 & (2.80, 98.70) & 63.75 & (2.56, 66.30) & 24.17 & (4.14, 28.31) \\
9 & 486.33 & (3.33, 489.67) & 79.02 & (2.86, 81.88) & 65.50 & (2.64, 68.14) & 24.19 & (4.29, 28.47) \\
10 & 470.21 & (3.32, 473.53) & 84.37 & (2.98, 87.34) & 74.99 & (2.70, 77.69) & 24.35 & (4.49, 28.83) \\
11 & 492.66 & (3.54, 496.20) & 94.60 & (3.09, 97.69) & 83.98 & (2.82, 86.80) & 24.41 & (4.63, 29.03) \\
12 & 527.02 & (3.53, 530.55) & 117.64 & (3.20, 120.84) & 91.41 & (2.82, 94.23) & 24.50 & (4.75, 29.25) \\
13 & 533.14 & (3.64, 536.78) & 111.33 & (3.20, 114.54) & 101.02 & (2.93, 103.95) & 24.46 & (4.85, 29.31) \\
14 & 551.22 & (3.66, 554.87) & 121.68 & (3.27, 124.94) & 97.68 & (2.97, 100.65) & 24.42 & (4.94, 29.36) \\
15 & 556.78 & (3.78, 560.57) & 129.72 & (3.27, 132.99) & 98.91 & (2.94, 101.86) & 24.37 & (5.05, 29.42) \\
16 & 591.35 & (3.86, 595.21) & 143.46 & (3.37, 146.84) & 108.63 & (2.99, 111.62) & 24.40 & (5.10, 29.50) \\
17 & 590.60 & (3.99, 594.59) & 141.79 & (3.42, 145.21) & 102.60 & (3.05, 105.65) & 24.37 & (5.14, 29.51) \\
18 & 602.05 & (3.92, 605.97) & 128.91 & (3.42, 132.32) & 95.38 & (3.09, 98.47) & 24.38 & (5.14, 29.51) \\
19 & 598.65 & (4.07, 602.73) & 126.95 & (3.42, 130.37) & 105.15 & (3.05, 108.21) & 24.35 & (5.25, 29.60) \\
20 & 602.86 & (4.00, 606.86) & 129.90 & (3.48, 133.38) & 96.06 & (3.15, 99.22) & 24.32 & (5.28, 29.60) \\
\hline 
\end{tabular}
\caption{Mean posterior 95\% credible interval and width for the multivariate $t$ with $d=2$ and $n=50$.}
\label{tab:meanCId2n50}
\end{table}

\begin{table}[h!]
\centering
\footnotesize
\begin{tabular}{ccccccccc}
\hline 
 & \multicolumn{2}{c}{AP} & \multicolumn{2}{c}{JP} & \multicolumn{2}{c}{RRP} & \multicolumn{2}{c}{LBP} \\ 
$\nu$ & Width & C.I. & Width & C.I. & Width & C.I. & Width & C.I. \\
\hline
1 & 0.45 & (1.01, 1.46) & 0.30 & (0.37, 0.67) & 0.39 & (1.01, 1.40) & 0.08 & (1.00, 1.08) \\
2 & 1.94 & (1.37, 3.31) & 0.89 & (0.70, 1.59) & 1.70 & (1.37, 3.08) & 1.64 & (1.79, 3.43) \\
3 & 16.25 & (1.92, 18.16) & 1.93 & (0.96, 2.89) & 3.77 & (1.87, 5.64) & 5.68 & (2.25, 7.93) \\
4 & 44.78 & (2.37, 47.15) & 7.74 & (1.24, 8.99) & 9.36 & (2.35, 11.71) & 11.51 & (2.92, 14.43) \\
5 & 107.15 & (2.86, 110.00) & 8.38 & (1.43, 9.81) & 15.06 & (2.77, 17.83) & 16.27 & (3.53, 19.80) \\
6 & 187.17 & (3.28, 190.45) & 19.31 & (1.66, 20.97) & 24.21 & (3.12, 27.32) & 19.10 & (4.10, 23.19) \\
7 & 260.59 & (3.58, 264.17) & 16.96 & (1.81, 18.77) & 46.24 & (3.42, 49.66) & 20.57 & (4.52, 25.10) \\
8 & 317.11 & (4.02, 321.13) & 41.09 & (1.96, 43.05) & 57.43 & (3.64, 61.07) & 21.72 & (4.93, 26.65) \\
9 & 358.66 & (4.29, 362.94) & 74.40 & (2.02, 76.42) & 62.99 & (3.86, 66.86) & 22.14 & (5.22, 27.36) \\
10 & 394.70 & (4.45, 399.15) & 69.57 & (2.15, 71.72) & 121.95 & (4.11, 126.07) & 22.53 & (5.64, 28.17) \\
11 & 425.74 & (4.70, 430.44) & 55.90 & (2.22, 58.12) & 101.56 & (4.21, 105.77) & 22.69 & (5.85, 28.54) \\
12 & 460.23 & (4.91, 465.14) & 63.71 & (2.33, 66.04) & 112.70 & (4.38, 117.08) & 22.71 & (6.03, 28.75) \\
13 & 492.25 & (5.25, 497.51) & 67.23 & (2.45, 69.68) & 128.91 & (4.52, 133.43) & 22.76 & (6.22, 28.99) \\
14 & 531.81 & (5.40, 537.21) & 67.01 & (2.46, 69.46) & 118.69 & (4.63, 123.32) & 22.78 & (6.31, 29.09) \\
15 & 542.53 & (5.55, 548.08) & 85.42 & (2.49, 87.91) & 117.96 & (4.71, 122.66) & 22.80 & (6.46, 29.26) \\
16 & 572.59 & (5.67, 578.26) & 70.96 & (2.55, 73.51) & 145.35 & (4.72, 150.07) & 22.78 & (6.55, 29.33) \\
17 & 579.06 & (5.69, 584.75) & 85.49 & (2.61, 88.10) & 159.69 & (4.87, 164.56) & 22.74 & (6.66, 29.40) \\
18 & 576.03 & (5.83, 581.86) & 97.01 & (2.61, 99.62) & 135.04 & (4.87, 139.90) & 22.70 & (6.77, 29.46) \\
19 & 600.30 & (6.03, 606.33) & 99.24 & (2.66, 101.90) & 203.34 & (5.01, 208.34) & 22.69 & (6.85, 29.54) \\
20 & 620.26 & (6.04, 626.30) & 90.47 & (2.67, 93.14) & 190.51 & (5.02, 195.52) & 22.67 & (6.92, 29.59) \\
\hline 
\end{tabular}
\caption{Mean posterior 95\% credible interval and width for the multivariate $t$ with $d=2$ and $n=100$.}
\label{tab:meanCId2n100}
\end{table}

\begin{table}[h!]
\centering
\footnotesize
\begin{tabular}{ccccccccc}
\hline 
 & \multicolumn{2}{c}{AP} & \multicolumn{2}{c}{JP} & \multicolumn{2}{c}{RRP} & \multicolumn{2}{c}{LBP} \\ 
$\nu$ & Width & C.I. & Width & C.I. & Width & C.I. & Width & C.I. \\
\hline
1 & 0.26 & (1.01, 1.26) & 0.37 & (0.83, 1.20) & 0.23 & (1.01, 1.24) & 0.00 & (1.00, 1.00) \\
2 & 1.09 & (1.55, 2.65) & 1.01 & (1.59, 2.60) & 1.00 & (1.57, 2.57) & 0.41 & (2.00, 2.41) \\
3 & 2.16 & (2.21, 4.37) & 1.95 & (2.26, 4.21) & 1.91 & (2.24, 4.15) & 1.88 & (2.53, 4.41) \\
4 & 3.98 & (2.84, 6.83) & 3.36 & (2.91, 6.26) & 3.26 & (2.87, 6.13) & 4.00 & (3.21, 7.20) \\
5 & 11.14 & (3.41, 14.55) & 5.35 & (3.49, 8.84) & 5.05 & (3.39, 8.44) & 7.34 & (3.81, 11.15) \\
6 & 43.17 & (3.96, 47.12) & 9.33 & (3.97, 13.31) & 12.24 & (3.93, 16.16) & 11.07 & (4.42, 15.49) \\
7 & 63.53 & (4.38, 67.91) & 19.59 & (4.56, 24.14) & 16.34 & (4.39, 20.73) & 14.45 & (5.16, 19.61) \\
8 & 107.44 & (4.88, 112.32) & 25.17 & (4.92, 30.09) & 21.81 & (4.78, 26.59) & 16.70 & (5.64, 22.34) \\
9 & 173.96 & (5.36, 179.32) & 34.16 & (5.27, 39.43) & 28.93 & (5.15, 34.08) & 18.07 & (6.16, 24.23) \\
10 & 220.61 & (5.84, 226.45) & 42.92 & (5.61, 48.53) & 32.74 & (5.48, 38.22) & 19.07 & (6.57, 25.64) \\
11 & 265.56 & (6.25, 271.82) & 57.50 & (5.94, 63.44) & 47.32 & (5.80, 53.12) & 19.77 & (6.98, 26.75) \\
12 & 340.43 & (6.68, 347.12) & 91.92 & (6.28, 98.20) & 66.65 & (6.16, 72.81) & 20.11 & (7.40, 27.51) \\
13 & 379.35 & (7.13, 386.48) & 88.72 & (6.59, 95.30) & 81.22 & (6.31, 87.53) & 20.21 & (7.73, 27.94) \\
14 & 414.60 & (7.35, 421.95) & 121.08 & (6.86, 127.94) & 93.10 & (6.63, 99.73) & 20.41 & (7.96, 28.37) \\
15 & 452.50 & (7.71, 460.20) & 168.56 & (7.04, 175.60) & 107.73 & (6.82, 114.55) & 20.45 & (8.27, 28.72) \\
16 & 480.05 & (8.12, 488.16) & 117.83 & (7.34, 125.17) & 144.20 & (7.03, 151.24) & 20.40 & (8.56, 28.96) \\
17 & 491.86 & (8.19, 500.05) & 142.88 & (7.52, 150.40) & 192.27 & (7.29, 199.55) & 20.32 & (8.74, 29.06) \\
18 & 524.31 & (8.36, 532.67) & 144.22 & (7.69, 151.91) & 166.88 & (7.49, 174.37) & 20.36 & (8.87, 29.23) \\
19 & 542.43 & (8.71, 551.13) & 236.55 & (7.86, 244.41) & 186.97 & (7.65, 194.62) & 20.21 & (9.08, 29.29) \\
20 & 545.08 & (8.86, 553.94) & 177.33 & (7.97, 185.29) & 161.82 & (7.68, 169.50) & 20.21 & (9.23, 29.44) \\

\hline 
\end{tabular}
\caption{Mean posterior 95\% credible interval and width for the multivariate $t$ with $d=2$ and $n=250$.}
\label{tab:meanCId2n250}
\end{table}

\begin{table}[h!]
\centering
\footnotesize
\begin{tabular}{ccccccccc}
\hline 
 & \multicolumn{2}{c}{AP} & \multicolumn{2}{c}{JP} & \multicolumn{2}{c}{RRP} & \multicolumn{2}{c}{LBP} \\ 
$\nu$ & Width & C.I. & Width & C.I. & Width & C.I. & Width & C.I. \\
\hline
1 & 0.60 & (1.02, 1.62) & 0.60 & (1.01, 1.62) & 0.53 & (1.01, 1.55) & 0.24 & (1.00, 1.25) \\
2 & 2.49 & (1.34, 3.84) & 2.80 & (1.27, 4.07) & 2.49 & (1.26, 3.75) & 3.11 & (1.54, 4.66) \\
3 & 19.91 & (1.86, 21.77) & 23.90 & (1.72, 25.62) & 6.06 & (1.67, 7.73) & 8.27 & (2.20, 10.48) \\
4 & 365.49 & (3.12, 368.61) & 56.91 & (2.13, 59.04) & 406.74 & (2.20, 408.94) & 14.01 & (2.77, 16.78) \\
5 & 250.68 & (2.97, 253.65) & 107.61 & (2.56, 110.17) & 51.97 & (2.41, 54.38) & 18.11 & (3.28, 21.40) \\
6 & 7523.21 & (17.67, 7540.88) & 157.08 & (2.77, 159.85) & 134.65 & (2.65, 137.30) & 20.42 & (3.70, 24.11) \\
7 & 5495.75 & (3.69, 5499.44) & 186.06 & (3.06, 189.11) & 716.81 & (2.93, 719.75) & 22.20 & (4.13, 26.33) \\
8 & 932.00 & (4.84, 936.84) & 231.79 & (3.38, 235.17) & 2403.52 & (3.62, 2407.14) & 22.75 & (4.55, 27.31) \\
9 & 500.04 & (4.71, 504.75) & 249.08 & (3.53, 252.62) & 154.45 & (3.46, 157.90) & 23.16 & (4.80, 27.95) \\
10 & 4655.09 & (7.37, 4662.47) & 280.20 & (3.66, 283.87) & 207.99 & (3.56, 211.55) & 23.36 & (4.96, 28.32) \\
11 & 1653.89 & (5.76, 1659.65) & 324.57 & (3.76, 328.33) & 330.24 & (3.61, 333.85) & 23.54 & (5.14, 28.68) \\
12 & 3129.56 & (5.79, 3135.35) & 319.44 & (3.95, 323.39) & 166.70 & (3.73, 170.44) & 23.64 & (5.29, 28.93) \\
13 & 17313.08 & (6.33, 17319.42) & 356.76 & (4.01, 360.77) & 250.59 & (3.93, 254.52) & 23.68 & (5.45, 29.13) \\
14 & 5059.18 & (7.40, 5066.58) & 365.50 & (4.15, 369.64) & 183.16 & (3.94, 187.10) & 23.69 & (5.59, 29.28) \\
15 & 3439.42 & (8.11, 3447.53) & 368.47 & (4.19, 372.66) & 197.37 & (4.21, 201.58) & 23.64 & (5.74, 29.38) \\
16 & 62553.59 & (18.35, 62571.94) & 374.86 & (4.29, 379.15) & 245.07 & (4.20, 249.27) & 23.66 & (5.80, 29.46) \\
17 & 7835.78 & (10.14, 7845.93) & 397.39 & (4.35, 401.73) & 197.49 & (4.13, 201.61) & 23.58 & (5.94, 29.51) \\
18 & 4060.36 & (8.32, 4068.68) & 409.59 & (4.39, 413.99) & 262.63 & (4.65, 267.27) & 23.57 & (6.01, 29.58) \\
19 & 5973.94 & (9.38, 5983.32) & 413.07 & (4.55, 417.62) & 204.88 & (4.15, 209.03) & 23.49 & (6.09, 29.58) \\
20 & 7067.15 & (15.15, 7082.30) & 73.08 & (4.30, 77.38) & 578.11 & (4.52, 582.63) & 23.52 & (6.10, 29.62) \\
\hline 
\end{tabular}
\caption{Mean posterior 95\% credible interval and width for the multivariate $t$ with $d=3$ and $n=50$.}
\label{tab:meanCId3n50}
\end{table}

\begin{table}[h!]
\centering
\footnotesize
\begin{tabular}{ccccccccc}
\hline 
 & \multicolumn{2}{c}{AP} & \multicolumn{2}{c}{JP} & \multicolumn{2}{c}{RRP} & \multicolumn{2}{c}{LBP} \\ 
$\nu$ & Width & C.I. & Width & C.I. & Width & C.I. & Width & C.I. \\
\hline
1 & 0.41 & (1.01, 1.42) & 0.40 & (1.01, 1.41) & 0.38 & (1.01, 1.39) & 0.06 & (1, 1.07) \\
2 & 1.53 & (1.46, 2.99) & 1.50 & (1.46, 2.96) & 1.48 & (1.41, 2.89) & 1.13 & (1.87, 3) \\
3 & 2.91 & (2.05, 4.96) & 2.94 & (2.04, 4.98) & 2.61 & (1.99, 4.6) & 3.51 & (2.29, 5.8) \\
4 & 6.62 & (2.68, 9.29) & 6.55 & (2.59, 9.13) & 5.32 & (2.51, 7.83) & 7.29 & (3, 10.28) \\
5 & 23.09 & (3.27, 26.36) & 11.61 & (3.16, 14.77) & 9.63 & (3.03, 12.66) & 11.94 & (3.63, 15.58) \\
6 & 67.97 & (3.95, 71.91) & 23.2 & (3.63, 26.83) & 21.89 & (3.48, 25.37) & 15.25 & (4.25, 19.5) \\
7 & 247.29 & (4.36, 251.65) & 42.96 & (4.06, 47.02) & 54.92 & (3.89, 58.81) & 18.01 & (4.79, 22.8) \\
8 & 1064.37 & (4.90, 1069.27) & 56.42 & (4.44, 60.86) & 50.72 & (4.2, 54.93) & 19.62 & (5.31, 24.93) \\
9 & 217.11 & (5.22, 222.33) & 64.55 & (4.71, 69.26) & 58.26 & (4.64, 62.9) & 20.51 & (5.7, 26.21) \\
10 & 492.91 & (6.67, 499.58) & 97.17 & (5.05, 102.22) & 96.45 & (4.89, 101.34) & 21.05 & (6.13, 27.17) \\
11 & 507.53 & (6.40, 513.93) & 131.78 & (5.31, 137.1) & 190.91 & (5.16, 196.07) & 21.32 & (6.39, 27.71) \\
12 & 2036.46 & (7.50, 2043.96) & 218.06 & (5.62, 223.68) & 106.9 & (5.28, 112.19) & 21.59 & (6.67, 28.26) \\
13 & 831.60 & (7.19, 838.80) & 165.13 & (5.9, 171.03) & 123.79 & (5.54, 129.33) & 21.6 & (6.93, 28.53) \\
14 & 2112.50 & (8.50, 2121.00) & 212.01 & (6.16, 218.17) & 350.13 & (6.05, 356.18) & 21.6 & (7.18, 28.78) \\
15 & 1568.83 & (8.92, 1577.75) & 227.92 & (6.33, 234.25) & 149.51 & (6.03, 155.54) & 21.58 & (7.37, 28.95) \\
16 & 2498.57 & (8.09, 2506.66) & 351.65 & (5.9, 357.55) & 175.7 & (6.15, 181.85) & 21.6 & (7.56, 29.16) \\
17 & 3285.01 & (9.31, 3294.31) & 234.06 & (6.64, 240.7) & 166.53 & (6.36, 172.89) & 21.52 & (7.71, 29.24) \\
18 & 2697.02 & (8.86, 2705.88) & 231.83 & (6.88, 238.71) & 264.3 & (6.57, 270.87) & 21.47 & (7.87, 29.33) \\
19 & 9678.97 & (9.86, 9688.83) & 28.75 & (0.74, 29.49) & 277.23 & (6.58, 283.81) & 21.51 & (7.98, 29.49) \\
20 & 6897.81 & (10.22, 6908.02) & 83.49 & (2.95, 86.44) & 175.66 & (6.58, 182.24) & 21.43 & (8.13, 29.56) \\

\hline 
\end{tabular}
\caption{Mean posterior 95\% credible interval and width for the multivariate $t$ with $d=3$ and $n=100$.}
\label{tab:meanCId3n100}
\end{table}

\begin{table}[h!]
\centering
\footnotesize
\begin{tabular}{ccccccccc}
\hline 
 & \multicolumn{2}{c}{AP} & \multicolumn{2}{c}{JP} & \multicolumn{2}{c}{RRP} & \multicolumn{2}{c}{LBP} \\ 
$\nu$ & Width & C.I. & Width & C.I. & Width & C.I. & Width & C.I. \\
\hline
1 & 0.23 & (1.01, 1.24) & 0.22 & (1.01, 1.23) & 0.22 & (1.01, 1.23) & 0.00 & (1.00, 1.00) \\
2 & 0.89 & (1.63, 2.52) & 0.88 & (1.62, 2.49) & 0.88 & (1.60, 2.49) & 0.26 & (2.00, 2.26) \\
3 & 1.61 & (2.37, 3.99) & 1.64 & (2.34, 3.98) & 1.62 & (2.32, 3.94) & 1.40 & (2.65, 4.05) \\
4 & 2.75 & (3.06, 5.81) & 2.62 & (3.00, 5.62) & 2.58 & (2.98, 5.57) & 2.72 & (3.25, 5.97) \\
5 & 4.48 & (3.71, 8.18) & 4.14 & (3.67, 7.80) & 4.08 & (3.62, 7.70) & 4.84 & (3.96, 8.80) \\
6 & 6.73 & (4.32, 11.04) & 6.06 & (4.28, 10.34) & 5.84 & (4.18, 10.02) & 7.38 & (4.63, 12.01) \\
7 & 12.86 & (4.88, 17.75) & 12.08 & (4.85, 16.94) & 9.09 & (4.79, 13.89) & 10.40 & (5.27, 15.66) \\
8 & 32.60 & (5.38, 37.98) & 13.49 & (5.31, 18.80) & 15.07 & (5.20, 20.27) & 12.87 & (5.84, 18.71) \\
9 & 92.38 & (6.07, 98.45) & 23.03 & (5.88, 28.91) & 22.51 & (5.77, 28.28) & 15.12 & (6.58, 21.69) \\
10 & 76.60 & (6.75, 83.34) & 31.65 & (6.41, 38.05) & 27.14 & (6.23, 33.37) & 16.82 & (7.12, 23.94) \\
11 & 140.17 & (7.05, 147.22) & 35.76 & (6.85, 42.61) & 27.00 & (6.70, 33.70) & 17.76 & (7.59, 25.35) \\
12 & 121.34 & (7.48, 128.82) & 41.83 & (7.18, 49.01) & 41.75 & (6.98, 48.73) & 18.56 & (8.03, 26.59) \\
13 & 135.07 & (8.02, 143.09) & 60.12 & (7.55, 67.67) & 52.83 & (7.35, 60.18) & 18.99 & (8.49, 27.48) \\
14 & 404.39 & (8.69, 413.08) & 69.23 & (8.02, 77.25) & 62.69 & (7.68, 70.37) & 19.20 & (8.87, 28.06) \\
15 & 442.43 & (9.14, 451.56) & 10.53 & (1.00, 11.53) & 57.04 & (8.04, 65.08) & 19.28 & (9.24, 28.51) \\
16 & 290.49 & (9.52, 300.01) & 144.85 & (8.61, 153.45) & 94.60 & (8.40, 103.00) & 19.28 & (9.51, 28.79) \\
17 & 610.36 & (10.42, 620.77) & 202.83 & (9.00, 211.83) & 232.48 & (8.75, 241.22) & 19.27 & (9.77, 29.04) \\
18 & 740.49 & (11.13, 751.62) & 123.17 & (9.24, 132.41) & 90.40 & (8.80, 99.20) & 19.25 & (9.96, 29.21) \\
19 & 894.99 & (11.06, 906.05) & 164.16 & (9.46, 173.62) & 154.89 & (9.22, 164.11) & 19.10 & (10.22, 29.32) \\
20 & 1122.68 & (11.66, 1134.34) & 302.44 & (9.85, 312.29) & 139.62 & (9.75, 149.37) & 18.97 & (10.47, 29.44) \\

\hline 
\end{tabular}
\caption{Mean posterior 95\% credible interval and width for the multivariate $t$ with $d=3$ and $n=250$.}
\label{tab:meanCId3n250}
\end{table}

\section*{Appendix C - 95\% Interval for the mean coverage}\label{sec:appendixC}
In this appendix we present the 95\% intervals for the mean coverage for the multivariate $t$ frequentist analysis. As the coverage, in most of the cases, is close to 1, we used the Wilson approximation \citep{wilson1927}. That is:
$$\frac{\hat{p}+\dfrac{z^2}{2w}}{1+\dfrac{z^2}{w}} \pm \frac{z}{1+\dfrac{z^2}{w}}\sqrt{\frac{\hat{p}(1-\hat{p})}{w}+\frac{z^2}{4w^2}},$$
where $\hat{p}$ is the estimated coverage, $w=250$ is the the number of samples used to estimate $p$ and $z=1.96$ is the 0.975-quantile of the standard normal distribution. Table \ref{tab:coverageintervalsn50} reports the intervals for the simulation with a sample size of $n=50$, Table \ref{tab:coverageintervalsn100} reports the results for the simulation with $n=100$ and, finally, Table \ref{tab:coverageintervalsn250} shows the intervals for the scenario with $n=250$.
\begin{table}
\centering
\footnotesize
\begin{tabular}{ccccccccc}
\hline 
 & \multicolumn{4}{c}{$d=2$} & \multicolumn{4}{c}{$d=3$} \\ 
$\nu$ & AP & JP & RRP & LBP & AP & JP & RRP & LBP \\
\hline
1 & NA & (0.87,0.94) & NA & (0.98,1) & NA & NA & NA & (0.98,1) \\
2 & (0.93,0.98) & (0.91,0.97) & (0.90,0.96) & (0.94,0.99) & (0.86,0.93) & (0.90,0.96) & (0.88,0.95) & (0.94,0.98) \\
3 & (0.91,0.97) & (0.87,0.94) & (0.88,0.95) & (0.92,0.98) & (0.85,0.93) & (0.89,0.96) & (0.88,0.94) & (0.90,0.96) \\
4 & (0.91,0.97) & (0.90,0.96) & (0.85,0.93) & (0.86,0.93) & (0.79,0.88) & (0.90,0.96) & (0.86,0.93) & (0.86,0.93) \\
5 & (0.86,0.93) & (0.90,0.96) & (0.86,0.94) & (0.89,0.96) & (0.83,0.92) & (0.91,0.97) & (0.84,0.92) & (0.88,0.95) \\
6 & (0.91,0.97) & (0.92,0.97) & (0.87,0.94) & (0.91,0.97) & (0.82,0.90) & (0.94,0.99) & (0.87,0.94) & (0.91,0.97) \\
7 & (0.94,0.98) & (0.91,0.97) & (0.90,0.95) & (0.95,0.99) & (0.85,0.93) & (0.95,0.98) & (0.86,0.93) & (0.94,0.98) \\
8 & (0.95,0.99) & (0.91,0.97) & (0.88,0.95) & (0.97,0.99) & (0.81,0.89) & (0.93,0.98) & (0.84,0.92) & (0.95,0.99) \\
9 & (0.96,0.99) & (0.89,0.95) & (0.86,0.94) & (0.97,0.99) & (0.84,0.92) & (0.94,0.99) & (0.87,0.94) & (0.96,0.99) \\
10 & (0.97,1) & (0.89,0.95) & (0.87,0.94) & (0.97,1) & (0.82,0.90) & (0.95,0.99) & (0.86,0.94) & (0.97,0.99) \\
11 & (0.97,1) & (0.88,0.94) & (0.84,0.92) & (0.97,1) & (0.86,0.93) & (0.94,0.99) & (0.88,0.94) & (0.97,1) \\
12 & (0.97,1) & (0.89,0.95) & (0.85,0.93) & (0.97,1) & (0.88,0.95) & (0.97,1) & (0.88,0.94) & (0.97,1) \\
13 & (0.97,1) & (0.88,0.95) & (0.82,0.90) & (0.97,1) & (0.86,0.93) & (0.97,1) & (0.86,0.93) & (0.97,1) \\
14 & (0.97,1) & (0.84,0.92) & (0.83,0.91) & (0.97,1) & (0.85,0.93) & (0.97,1) & (0.85,0.93) & (0.97,1) \\
15 & (0.96,0.99) & (0.87,0.94) & (0.83,0.91) & (0.97,1) & (0.88,0.95) & (0.97,1) & (0.86,0.94) & (0.97,1) \\
16 & (0.96,0.99) & (0.83,0.92) & (0.79,0.88) & (0.97,1) & (0.86,0.94) & (0.96,0.99) & (0.86,0.94) & (0.97,1) \\
17 & (0.96,0.99) & (0.87,0.94) & (0.84,0.92) & (0.97,1) & (0.85,0.93) & (0.97,1) & (0.84,0.92) & (0.97,1) \\
18 & (0.96,0.99) & (0.84,0.92) & (0.79,0.88) & (0.97,1) & (0.90,0.96) & (0.97,1) & (0.84,0.92) & (0.97,1) \\
19 & (0.95,0.99) & (0.86,0.94) & (0.78,0.87) & (0.97,1)& (0.84,0.92) & (0.97,1) & (0.86,0.93) & (0.97,1) \\
20 & (0.97,1) & (0.85,0.93) & (0.82,0.90) & (0.97,1) & (0.87,0.94) & (0.97,1) & (0.87,0.94) & (0.97,1) \\
\hline 
\end{tabular}
\caption{95\% intervals for the mean coverage of the multivariate $t$ with $d=2,3$ and $n=50$.}
\label{tab:coverageintervalsn50}
\end{table}
\begin{table}
\centering
\footnotesize
\begin{tabular}{ccccccccc}
\hline 
 & \multicolumn{4}{c}{$d=2$} & \multicolumn{4}{c}{$d=3$} \\ 
$\nu$ & AP & JP & RRP & LBP & AP & JP & RRP & LBP \\
\hline
1 & NA & (0.89,0.95) & NA & (0.99,1) & NA & NA & NA & (0.99,1) \\
2 & (0.87,0.94) & (0.89,0.95) & (0.88,0.94) & (0.96,0.99) & (0.87,0.94) & (0.89,0.95) & (0.88,0.94) & (0.96,0.99) \\
3 & (0.88,0.95) & (0.89,0.96) & (0.88,0.95) & (0.94,0.98) & (0.91,0.97) & (0.88,0.95) & (0.89,0.96) & (0.97,1) \\
4 & (0.88,0.94) & (0.87,0.94) & (0.88,0.95) & (0.88,0.94) & (0.87,0.94) & (0.89,0.95) & (0.87,0.94) & (0.92,0.97) \\
5 & (0.88,0.94) & (0.89,0.95) & (0.86,0.94) & (0.87,0.94) & (0.84,0.92) & (0.86,0.94) & (0.86,0.93) & (0.88,0.94) \\
6 & (0.86,0.94) & (0.89,0.96) & (0.89,0.95) & (0.87,0.94) & (0.83,0.91) & (0.86,0.93) & (0.85,0.93) & (0.88,0.95) \\
7 & (0.83,0.92) & (0.88,0.95) & (0.89,0.95) & (0.89,0.96) & (0.86,0.94) & (0.88,0.95) & (0.87,0.94) & (0.86,0.93) \\
8 & (0.86,0.93) & (0.88,0.95) & (0.88,0.95) & (0.94,0.98) & (0.77,0.87) & (0.86,0.93) & (0.83,0.91) & (0.84,0.92) \\
9 & (0.90,0.96) & (0.88,0.94) & (0.87,0.94) & (0.95,0.99) & (0.81,0.90) & (0.86,0.94) & (0.86,0.93) & (0.93,0.98) \\
10 & (0.88,0.95) & (0.90,0.96) & (0.87,0.94) & (0.97,1) & (0.82,0.90) & (0.88,0.95) & (0.88,0.95) & (0.92,0.98) \\
11 & (0.89,0.95) & (0.90,0.96) & (0.86,0.94) & (0.98,1) & (0.81,0.89) & (0.88,0.95) & (0.84,0.92) & (0.95,0.99) \\
12 & (0.92,0.98) & (0.89,0.95) & (0.87,0.94) & (0.98,1) & (0.82,0.90) & (0.88,0.94) & (0.85,0.93) & (0.97,1) \\
13 & (0.91,0.97) & (0.88,0.95) & (0.88,0.95) & (0.98,1) & (0.84,0.92) & (0.87,0.94) & (0.87,0.94) & (0.96,0.99) \\
14 & (0.88,0.95) & (0.90,0.96) & (0.85,0.93) & (0.98,1) & (0.80,0.89) & (0.87,0.94) & (0.84,0.92) & (0.96,0.99) \\
15 & (0.90,0.96) & (0.90,0.96) & (0.88,0.94) & (0.98,1) & (0.82,0.90) & (0.87,0.94) & (0.88,0.94) & (0.97,1) \\
16 & (0.92,0.98) & (0.87,0.94) & (0.86,0.93) & (0.98,1) & (0.82,0.90) & (0.76,0.85) & (0.85,0.93) & (0.97,1) \\
17 & (0.91,0.97) & (0.88,0.94) & (0.84,0.92) & (0.98,1) & (0.81,0.90) & (0.87,0.94) & (0.85,0.93) & (0.97,1) \\
18 & (0.91,0.97) & (0.86,0.93) & (0.84,0.92) & (0.98,1) & (0.83,0.91) & (0.87,0.94) & (0.83,0.91) & (0.98,1) \\
19 & (0.90,0.96) & (0.88,0.95) & (0.87,0.94) & (0.98,1) & (0.84,0.92) & (0.88,0.94) & (0.85,0.93) & (0.97,1) \\
20 & (0.91,0.97) & (0.88,0.95) & (0.83,0.91) & (0.98,1) & (0.84,0.92) & (0.87,0.94) & (0.85,0.93) & (0.98,1) \\
\hline 
\end{tabular}
\caption{95\% intervals for the mean coverage of the multivariate $t$ with $d=2,3$ and $n=100$.}
\label{tab:coverageintervalsn100}
\end{table}
\begin{table}
\centering
\footnotesize
\begin{tabular}{ccccccccc}
\hline 
 & \multicolumn{4}{c}{$d=2$} & \multicolumn{4}{c}{$d=3$} \\ 
$\nu$ & AP & JP & RRP & LBP & AP & JP & RRP & LBP \\
\hline
1 & NA & (0.88,0.94) & NA & (0.99,1) & NA & NA & NA & (0.99,1) \\
2 & (0.90,0.96) & (0.90,0.96) & (0.91,0.97) & (0.97,1) & (0.86,0.94) & (0.91,0.97) & (0.89,0.96) & (0.98,1) \\
3 & (0.90,0.96) & (0.91,0.97) & (0.90,0.96) & (0.97,1) & (0.88,0.95) & (0.89,0.96) & (0.90,0.96) & (0.98,1) \\
4 & (0.87,0.94) & (0.91,0.97) & (0.91,0.97) & (0.94,0.99) & (0.90,0.96) & (0.88,0.95) & (0.88,0.95) & (0.97,1) \\
5 & (0.88,0.94) & (0.90,0.96) & (0.90,0.96) & (0.93,0.98) & (0.84,0.92) & (0.86,0.93) & (0.84,0.92) & (0.91,0.97) \\
6 & (0.88,0.95) & (0.91,0.97) & (0.89,0.96) & (0.93,0.98) & (0.83,0.91) & (0.86,0.93) & (0.87,0.94) & (0.93,0.98) \\
7 & (0.84,0.92) & (0.85,0.93) & (0.87,0.94) & (0.88,0.95) & (0.88,0.94) & (0.88,0.95) & (0.85,0.93) & (0.91,0.97) \\
8 & (0.84,0.92) & (0.88,0.95) & (0.88,0.94) & (0.91,0.97) & (0.87,0.94) & (0.88,0.95) & (0.91,0.97) & (0.91,0.97) \\
9 & (0.90,0.96) & (0.88,0.95) & (0.88,0.95) & (0.91,0.97) & (0.89,0.96) & (0.87,0.94) & (0.88,0.95) & (0.89,0.96) \\
10 & (0.88,0.95) & (0.90,0.96) & (0.88,0.95) & (0.93,0.98) & (0.86,0.94) & (0.88,0.94) & (0.87,0.94) & (0.89,0.96) \\
11 & (0.87,0.94) & (0.88,0.95) & (0.87,0.94) & (0.92,0.98) & (0.86,0.94) & (0.88,0.95) & (0.89,0.96) & (0.92,0.97) \\
12 & (0.88,0.95) & (0.90,0.96) & (0.88,0.95) & (0.95,0.99) & (0.88,0.95) & (0.89,0.95) & (0.87,0.94) & (0.96,0.99) \\
13 & (0.90,0.96) & (0.93,0.98) & (0.91,0.97) & (0.96,0.99) & (0.85,0.93) & (0.89,0.96) & (0.90,0.96) & (0.97,1) \\
14 & (0.88,0.95) & (0.87,0.94) & (0.89,0.96) & (0.97,1) & (0.84,0.92) & (0.89,0.95) & (0.88,0.95) & (0.97,1) \\
15 & (0.89,0.96) & (0.90,0.96) & (0.89,0.96) & (0.98,1) & (0.84,0.92) & (0.88,0.94) & (0.92,0.97) & (0.98,1) \\
16 & (0.91,0.97) & (0.91,0.97) & (0.92,0.98) & (0.98,1) & (0.87,0.94) & (0.90,0.96) & (0.89,0.96) & (0.98,1) \\
17 & (0.90,0.96) & (0.90,0.96) & (0.90,0.96) & (0.99,1) & (0.86,0.94) & (0.89,0.95) & (0.89,0.95) & (0.98,1) \\
18 & (0.89,0.96) & (0.90,0.96) & (0.90,0.96) & (0.98,1) & (0.85,0.93) & (0.90,0.96) & (0.90,0.96) & (0.98,1) \\
19 & (0.91,0.97) & (0.89,0.96) & (0.91,0.97) & (0.99,1) & (0.84,0.92) & (0.91,0.97) & (0.88,0.95) & (0.97,1) \\
20 & (0.89,0.95) & (0.91,0.97) & (0.89,0.95) & (0.97,1) & (0.82,0.90) & (0.90,0.96) & (0.88,0.95) & (0.97,1) \\
\hline 
\end{tabular}
\caption{95\% intervals for the mean coverage of the multivariate $t$ with $d=2,3$ and $n=250$.}
\label{tab:coverageintervalsn250}
\end{table}

\begin{landscape}
\begin{table}
\centering
\footnotesize
\begin{tabular}{cccccccccc}
\hline 
 & \multicolumn{3}{c}{$n=50$} & \multicolumn{3}{c}{$n=100$} & \multicolumn{3}{c}{$n=250$} \\ 
$\nu$ & $\rho=0.25$ & $\rho=0.50$ & $\rho=0.75$ & $\rho=0.25$ & $\rho=0.50$ & $\rho=0.75$ & $\rho=0.25$ & $\rho=0.50$ & $\rho=0.75$ \\
\hline
1 & (0.97,1) & (0.95,0.99) & (0.93,0.98) & (0.99,1) & (0.97,1) & (0.99,1) & (0.99,1) & (0.99,1) & (0.99,1) \\
2 & (0.87,0.94) & (0.87,0.94) & (0.88,0.95) & (0.93,0.98) & (0.93,0.98) & (0.88,0.95) & (0.97,1) & (0.99,1) & (0.95,0.99) \\
3 & (0.93,0.98) & (0.90,0.96) & (0.89,0.96) & (0.87,0.94) & (0.85,0.92) & (0.82,0.91) & (0.95,0.99) & (0.95,0.99) & (0.95,0.99) \\
4 & (0.92,0.97) & (0.93,0.98) & (0.95,0.99) & (0.83,0.92) & (0.87,0.94) & (0.92,0.97) & (0.88,0.95) & (0.86,0.93) & (0.93,0.98) \\
5 & (0.97,1) & (0.98,1) & (0.98,1) & (0.93,0.98) & (0.92,0.97) & (0.93,0.98) & (0.87,0.94) & (0.92,0.97) & (0.85,0.92) \\
6 & (0.98,1) & (0.98,1) & (0.98,1) & (0.93,0.98) & (0.92,0.97) & (0.94,0.99) & (0.93,0.98) & (0.95,0.99) & (0.90,0.96) \\
7 & (0.98,1) & (0.98,1) & (0.98,1) & (0.94,0.99) & (0.94,0.99) & (0.95,0.99) & (0.94,0.99) & (0.95,0.99) & (0.92,0.97) \\
8 & (0.98,1) & (0.98,1) & (0.98,1) & (0.97,1) & (0.97,1) & (0.97,1) & (0.94,0.99) & (0.95,0.99) & (0.95,0.99) \\
9 & (0.98,1) & (0.98,1) & (0.98,1) & (0.97,1) & (0.97,1) & (0.97,1) & (0.95,0.99) & (0.99,1) & (0.95,0.99) \\
10 & (0.98,1) & (0.98,1) & (0.98,1) & (0.97,1) & (0.97,1) & (0.97,1) & (0.98,1) & (0.98,1) & (0.98,1) \\
11 & (0.98,1) & (0.98,1) & (0.98,1) & (0.99,1) & (0.98,1) & (0.98,1) & (0.98,1) & (0.98,1) & (0.98,1) \\
12 & (0.98,1) & (0.98,1) & (0.98,1) & (0.97,1) & (0.98,1) & (0.98,1) & (0.98,1) & (0.98,1) & (0.98,1) \\
13 & (0.98,1) & (0.98,1) & (0.98,1) & (0.97,1) & (0.98,1) & (0.98,1) & (0.98,1) & (0.98,1) & (0.98,1) \\
14 & (0.98,1) & (0.98,1) & (0.98,1) & (0.97,1) & (0.98,1) & (0.98,1) & (0.98,1) & (0.98,1) & (0.98,1) \\
15 & (0.98,1) & (0.98,1) & (0.98,1) & (0.97,1) & (0.98,1) & (0.98,1) & (0.98,1) & (0.98,1) & (0.98,1) \\
16 & (0.98,1) & (0.98,1) & (0.98,1) & (0.97,1) & (0.98,1) & (0.98,1) & (0.98,1) & (0.98,1) & (0.98,1) \\
17 & (0.98,1) & (0.98,1) & (0.98,1) & (0.97,1) & (0.98,1) & (0.98,1) & (0.98,1) & (0.98,1) & (0.98,1) \\
18 & (0.98,1) & (0.98,1) & (0.98,1) & (0.97,1) & (0.98,1) & (0.98,1) & (0.98,1) & (0.98,1) & (0.98,1) \\
19 & (0.98,1) & (0.98,1) & (0.98,1) & (0.97,1) & (0.98,1) & (0.98,1) & (0.98,1) & (0.98,1) & (0.98,1) \\
20 & (0.98,1) & (0.98,1) & (0.98,1) & (0.97,1) & (0.98,1) & (0.98,1) & (0.98,1) & (0.98,1) & (0.98,1) \\
\hline 
\end{tabular}
\caption{95\% intervals for the mean coverage of the bivariate $t$-copula with $\rho=0.25,0.50,0.75$ and sample sizes $n=50,100,250$.}
\label{tab:coverageintervalscopula}
\end{table}
\end{landscape}

\bibliographystyle{plainnat}
\bibliography{references}

\end{document}